\documentclass[useAMS,usenatbib,usegraphicx,twocolumn]{mn2e}
  
\usepackage{float}
\usepackage{aas_macros}
\usepackage{threeparttable}
\usepackage{grffile}
\usepackage{amsmath, amssymb}
\usepackage{multirow}
\usepackage{wallpaper}
\usepackage{mathbbol}
\usepackage{rotating}
\usepackage{blindtext}
\usepackage[T1]{fontenc}

\usepackage{tocloft}
\newcommand{\fat}{\textbf}
\newcommand{\ita}{\textit}

\newcommand{\beq}{\begin{equation}}
\newcommand{\eeq}{\end{equation}}  
\newcommand{\RNum}[1]{\uppercase\expandafter{\romannumeral #1\relax}}

  \title[Deuterium fractionation in filaments]{Fast deuterium fractionation in magnetized and turbulent filaments}
  
  \author[B.~K\"ortgen]{B.~K\"ortgen$^1$, S.~Bovino$^1$, D.R.G.~Schleicher$^2$, A.~Stutz$^2$, R.~Banerjee$^1$,
  \newauthor{A.~Giannetti$^3$ and S.~Leurini$^4$}\\
  $^{1}$ Hamburger Sternwarte, Universit\"at Hamburg, Gojenbergsweg 112, D-21029 Hamburg, Germany \\
  $^{2}$ Departamento de Astronom\'{i}a, Facultad Ciencias F\'{i}sicas y Matem\'{a}ticas, Universidad de Concepci\'{o}n, \\
  \  \ Av. Esteban Iturra s/n Barrio Universitario, Casilla 160--C, Concepci\'{o}n, Chile\\
  $^{3}$ INAF - Istituto di Radioastronomia \& Italian ALMA Regional Centre, Via P. Gobetti 101, I-40129 Bologna, Italy\\
  $^{4}$ INAF Osservatorio Astronomico di Cagliari, Via della Scienza 5, I-09047 Selargius, Italy
  }

\date{Released 2016}

\pagerange{\pageref{firstpage}--\pageref{lastpage}} \pubyear{2016}

\begin{document}

\label{firstpage}
\maketitle

\begin{abstract}
Deuterium fractionation is considered as an important process to infer the chemical ages of prestellar cores in filaments. We present here the first magneto-hydrodynamical simulations including a chemical network to study deuterium fractionation in magnetized and turbulent filaments and their substructures. The filaments typically show widespread deuterium fractionation with average values $\gtrsim0.01$. For individual cores of similar age, we observe the 
deuteration fraction to increase with time, but also to be independent of their average properties such as density, virial or mass-to-magnetic flux ratio. We further find a correlation of the deuteration fraction with core mass, average H$_2$ density and virial parameter only at late evolutionary stages of the filament and attribute this to the lifetime of the individual cores. Specifically, chemically 
old cores reveal higher deuteration fractions.
Within the radial profiles of selected cores, we notice differences in the structure of the deuteration fraction or surface density, which we can attribute to their different turbulent properties. High deuteration fractions of the order $0.01-0.1$ may be reached within approximately $200$~kyrs, corresponding to two free-fall times, as defined for cylindrical systems, of the filaments. 
\end{abstract}

\section{Introduction}
Filamentary structures have been known since the time of \citet{Barnard05} through the discovery on photographic plate data. Their ubiquitous presence inside molecular clouds has been most recently revealed by surveys such as the Herschel Gould Belt Survey \citep{Andre10}, the Herschel Hi-GAL Milky Way survey \citep{Molinari10} or the characterization of filaments in IC~5147 \citep{Arzoumanian2011}. The physics of such filaments under conditions of equilibrium was initially described by \citet{Chandrasekhar53}, taking into account gravity, thermal pressure, rotation and magnetic fields. These investigations have been extended by various authors, including \citet{Nagasawa87}, \citet{Fiege00}, \citet{Tomisaka14} and \citet{Toci15}. As most of star formation occurs in filaments \citep{Andre14,Stutz16}, it is thus particularly important to understand their general properties as well as their stability and how they first fragment into larger clumps and subsequently into cores.\\ 
The observed star formation rate and efficiency have been measured for many filaments with various properties and within different environments \cite[see e.g. review by][]{Andre14} and the question was posed how filaments with suppressed star formation can be stabilized against (radial) contraction. In this 
sense, the impact of different types of magnetic field structures was predominantly explored via linear stability analysis, with very early studies initially focusing on axial or helical fields \citep{Nakamura93, Matsumoto94, Fiege00}, while more recently also perpendicular field structures were considered by \citet[][]{Hanawa15}. As a central result, they found that filaments cannot be stabilized against fragmentation by axial magnetic fields, while perpendicular ones can have a relevant stabilizing contribution.\\ Polarimetry observations suggest a bi-modal distribution of magnetic field alignment with the filament \citep{Li13}, and more recent results suggest a lower specific star formation rate in the case of perpendicular magnetic fields \citep{Li17}. The result is thus in line with expectations based on linear theory. Numerical simulations by \citet{Seifried15} however show that axial fields can at least stabilize the filaments along the radial direction, leading to a reduced number of stars in their simulations. More recently, \citet{Seifried16} carried out 
high-resolution numerical simulations including a complex chemical network to study the evolution of filaments. They find strong chemical differentiation (i.e. chemical gradients) throughout the filament. Post-processing these data and generating synthetic observations, \citet{Seifried17} 
found that general filament properties (such as their mass or width) can be probed accurately via dust emission, but line emission observations might fail to reproduce the true properties. The authors state that e.g. the assumption of a constant excitation 
temperature introduces errors as it does not account for strong variations in the gas temperature along the line of sight.\\ 
The collapse and fragmentation of filaments have been studied intensively from a numerical point of view. \cite{Burkert04} have shown via 2D simulations that there appears to be a preferred edge-on (starting at the end points) fragmentation of filaments, which is independent of the 
initial conditions. They further emphasize that a degree of inhomogeneity of the filament surface may indeed help to drive turbulence within the filament. More recently \cite{Clarke15} have shown that this edge-on mode of fragmentation happens for all 
filaments with aspect ratios $\gtrsim2$ and that the resulting fragments greatly influence the gas inside the filaments. A similar study by \cite{Seifried15} reveals that filament fragmentation can be either edge-on, centralized or both, primarily depending on 
the initial ratio of the filament line-mass to the critical one.\\
Observationally, the fragmentation of filaments was analyzed by \citet[][see also \citealt{Hacar13}]{Hacar11}, finding that pc-scale filaments fragment into small-scale, velocity-coherent structures, which can 
then undergo fragmentation to form individual cores. These 'fibres' are closely linked to the vorticity in the filament, as has been suggested recently by \cite{Clarke17}. In addition, velocity gradients across cores have been found in the infrared dark cloud G351.77-051 by \citet{Leurini2011}. These gradients were assigned to CO outflows, which indicate past fragmentation and active star formation. Furthermore, in the Orion A cloud, where the combination of polarimetry and Zeeman measurements shows evidence for a helical field \citep{Heiles87, Bally89, Heiles97, Pattle17}, recent measurements suggest approximate equipartition between gravitational and magnetic energy on 1~pc scales, as well as potential oscillations and instabilities induced by magnetic fields \citep{Stutz16}, giving rise to the possible influence of large-scale environmental effects on the fragmentation of filaments. The potential oscillation is consistent with analytic estimates regarding the impact of helical fields \citep{Schleicher17} 
as well as with stellar dynamical calculations \citep{Boekholt17}.\\
Strong controversies are however still present regarding the way how stars form in gravitationally unstable cores, in particular high-mass stars. Whether it happens on approximately a free-fall time, as suggested in the competitive accretion scenario by \citet{Bonnell01}, or rather slowly, implying at least several free-fall times, in the core accretion model by \citet{McKee2003}, which assumes that turbulence and/or magnetic fields provide a major stabilizing contribution. It is therefore of central importance to obtain independent constraints on the timescale of star formation. A potentially powerful tool that was suggested in the literature is the measurement of deuteration fractions, which may be translated into timescales via chemical models \citep{Caselli02, Fontani11, Pagani11, Kong15, Lackington2016,Barnes16}. In relation to filaments, a timescale estimate has recently been achieved by \citet{Lackington2016} for the infrared dark cloud L332. They found deuteration ratios N$_2$D$^+/$N$_2$H$^+$ in the 
range $0.003-0.14$. Based on the chemical models by \citet{Kong15}, 
the authors deduced timescales for various cores within L332 of the order of several free-fall times to match the observed deuteration fractions, indicating rather old cores. However, the authors emphasize that a change in the CO depletion factor reduces the timescale to only a free-fall time, 
which points towards (dynamically) young objects. In addition, \citet{Barnes16} deduce a chemical age of about eight free-fall times for the dark cloud G035.39-00.33 to fit the observed average deuteration fraction of $0.04\pm0.01$, which they attribute to support of the filament by magnetic fields and turbulence. Furthermore, these authors find that deuteration is widespread over the filament rather than concentrated in individual cores, in agreement with recent findings by \citet{Pillai2012} who report 
widespread H$_2$D$^+$ emission in Cygnus X. \\
To test how reliably deuterium fractionation can be used as a chemical clock requires numerical simulations, which investigate the build-up of high deuteration fractions under realistic conditions. A first set of 3D magneto-hydrodynamical simulations of collapsing cores has been pursued by \citet{Goodson16}, who determined effective formation rates of deuterated species from detailed one-zone models, finding that observed fractions of up to $0.1$ can be reproduced after 3-4 free-fall times. Employing the chemical network by \citet{Walmsley2004}, which has been fully coupled to the 3D magnetohydrodynamics code \textsc{flash}, \citet{Koertgen17} have shown that rapid deuteration occurs within roughly one free-fall time. These models were pursued under the assumption of full depletion, which is expected above densities of about $10^4$~cm$^{-3}$ \citep[e.g.][]{Hocuk14}, and known from observational measurements \citep{Chen11, Hernandez11, Giannetti14}.\\
However, to further test this scenario, deuterium fractionation must be explored in a broader context. Particularly, models that follow the collapse of individual cores need pre-determined initial conditions, which are not known from observations. This is particularly true for the ortho-to-para ratio of molecular hydrogen, which plays a crucial role in the deuterium fractionation \citep[e.g.][]{Walmsley2004, Bovino17}. Here, we therefore investigate deuterium fractionation on the scale of a filament, focusing on the global chemical evolution as well as the properties of individual cores within the filament. Our numerical method and initial conditions are described in section~\ref{method}, and the results are outlined in section~\ref{results}. A discussion with our main conclusions is given in section~\ref{discussion}.

\section{Numerical Model}\label{method}
We describe here the numerical method employed, the filament setup as well as the chemical network and its initial conditions.
\subsection{Method}
We use the \textsc{flash} code (v4.2.2) to solve the magneto-hydrodynamic (MHD) equations as well as the Poisson equation for self-gravity \citep{Fryxell00,Dubey08}. The (magneto-)hydrodynamical system is solved with a HLL5R-solver, which preserves positivity of density and internal energy \citep{Bouchut10, Waagan11}, while we use a tree-solver for the self-gravity \citep[optimized for GPU,][]{Lukat16}. We further employ outflow boundary conditions for the hydrodynamics and isolated ones for the self-gravity. 
Gravitationally collapsing regions with increasing density are replaced by a Lagrangian sink particle to avoid violation of the True\-love criterion \citep[][]{Truelove97}. Besides other tests of gravitational stability \citep[see][]{Federrath10}, the grid cells have to exceed a 
threshold density of $n_\mathrm{thresh}=8\times10^7\,\mathrm{cm}^{-3}$ before a sink particle is formed. We anticipate here that no sink particles form in the simulations as most of the formation criteria are not fulfilled. \\
The size of the numerical domain is \mbox{$L_\mathrm{box}=6\,\mathrm{pc}$} in each direction. We allow a maximum refinement level of 13, corresponding to a minimum cell size of \mbox{$\Delta x=147\,\mathrm{AU}$}. The numerical grid is 
refined once the local Jeans length is resolved with less than 16 cells. In addition, the Jeans length at the threshold density is still refined with 8 grid cells.\\
The \textsc{flash} code is coupled to the astrochemistry package \textsc{krome}\footnote{Webpage KROME: http://www.kromepackage.org/} \citep{Grassi2013}, employing the high-order solver DLSODES for solving the chemical rate equations. The chemical initial conditions and the network are described further below.

\subsection{Initial filament properties and dynamics}
Recent observational surveys (with e.g. \ita{Herschel}) have revealed that interstellar filaments are surprisingly similar in their properties, such as e.g. their spatial distribution of the gas in the radial direction. As was shown by 
\citet[][]{Arzoumanian2011}, the radial density profile of observed filaments is best fitted with a Plummer--like distribution:
\beq
\rho\left(R\right)=\frac{\rho_\mathrm{ridge}}{\left\{1+\left(R/R_\mathrm{flat}\right)^2\right\}^{p/2}}.
\eeq
Here $\rho_\mathrm{ridge}$ is the central density, $R$ is the radius, $R_\mathrm{flat}$ the characteristic width of the flat inner plateau of the filament and $p$ is the typical exponent of the profile, respectively. \citet[][]{Arzoumanian2011} give values $1.5<p<2.5$ for the profile exponent and a typical filament width $\sim3\times R_\mathrm{flat}$ with $0.01$~pc $\leq R_\mathrm{flat} \leq 0.06$~pc. For the setup used here, we employ $p=2$ and $R_\mathrm{flat}=0.03$~pc \citep[like in e.g.][]{Seifried15} and we do not 
apply any cut--off radius or density. Hence, the radial density profile extends essentially to infinity.\\
The filament has a length of $L_\mathrm{fil}=2$ pc, thus ranging from $-1\,\mathrm{pc}$ to $+1\,\mathrm{pc}$. In order to avoid a pressure--jump at the outer edges of the filament we let the density decay exponentially for distances larger/smaller than $\pm1\,\mathrm{pc}$ along the major axis.\\
Following \citet[][]{Ostriker1964} there exists a critical filament mass per unit length above which the filament is unstable to perturbations and will undergo gravitational collapse. This critical \ita{thermal line--mass} is given by 
\beq
\left(\frac{M}{L}\right)_\mathrm{crit}=\frac{2c_\mathrm{s}^2}{G},
\eeq
where $G$ is Newton's constant and $c_\mathrm{s}$ is the sound speed. The simulations are isothermal with $T=15\,\mathrm{K}$, which results in a critical line--mass of $\left(M/L\right)_\mathrm{crit, thermal}=24\,\mathrm{M}_\odot/\mathrm{pc}$. The central density $\varrho_\mathrm{ridge}$ is varied 
to give filaments with line--masses of \mbox{$\left(M/L\right)_\mathrm{fil}\sim1.6\times\left(M/L\right)_\mathrm{crit, thermal}$} as well as of \mbox{$\left(M/L\right)_\mathrm{fil}\sim0.8\times\left(M/L\right)_\mathrm{crit}$}, where only gas within \mbox{$R=0.1\,\mathrm{pc}$} 
is taken into account.\\
Observations have revealed that the interstellar medium and molecular clouds -- and hence filaments -- are turbulent and magnetized \citep[see e.g.][and references therein]{MacLow04,Elmegreen04,McKee07, Crutcher10, Crutcher12,Schmidt13}. The turbulence is modeled by employing a turbulent velocity field, which is being generated in Fourier space with
\beq
E\left(k\right)\propto\left\{
\begin{array}{c}
 k^{10}\quad k<k_\mathrm{int},\\
 k^{-2}\quad k>k_\mathrm{int}
 \end{array}
 \right.
 \eeq
where the slope at $k>k_\mathrm{int}$ is typical for supersonic/shock--dominated turbulence and $k_\mathrm{int}$ being the integral scale. Using a box length of $L_\mathrm{box}\sim6\,\mathrm{pc}$ in each spatial direction, the integral scale is $\mathcal{L}_\mathrm{int}\sim L_\mathrm{box}/k_\mathrm{int}\sim1\,\mathrm{pc}\sim30\times R_\mathrm{flat}$. The turbulence is not driven and decays after approximately a crossing--time.\\
Magnetic fields in molecular clouds have been shown to be approximately constant as a function of density up to a threshold density of $n_\mathrm{thresh}\sim300\,\mathrm{cm}^{-3}$ \citep{Crutcher10,Crutcher12}. For $n>n_\mathrm{thresh}$, the magnetic field 
scales with density as $B\propto n^{\alpha}$ with $\alpha\sim0.5-0.65$ \citep{Crutcher12}. However, the scatter in these Zeeman observations is rather large and we follow the approach by \citet{Seifried15} and set $B=40\,\mu\mathrm{G}=\mathrm{const}$. As 
these authors note, this value for the magnetic field strength is rather low but still in agreement with observational data. 
The simulations are initialized with two different orientations of the magnetic field, namely a) $\fat{B}$ being perpendicular to the filament's major axis and b) $\fat{B}$ being parallel to the filament's major axis. An overview of the initial conditions explored here is given in Table~\ref{ini}.

\begin{table*}
	\caption{Overview of the initial conditions for the performed simulations. All listed quantities are measured within a radius of $R=0.1\,\mathrm{pc}$.  Column 5 depicts the ratio of the line-mass to the critical value when taking the magnetic field and 
	turbulence into account. Note that in the latter case the filaments are stable against gravitational collapse. The Alfv\'en speed $\mathcal{M}_\mathrm{a}$ is calculated by using the average density within the radius $R$. The last two columns 
	denote the free-fall times as calculated for a sphere or a cylinder \citep[][]{Toala12} with average density that is given in column 2.}
	\begin{tabular}{lcccccccccc}
	\hline
	\hline
	\fat{Run name}	&$\varrho_\mathrm{ridge}$ &$\left<n\right>$	&$\frac{\left(M/L\right)}{\left(M/L\right)_\mathrm{crit,th}}$ &$\frac{\left(M/L\right)}{\left(M/L\right)_\mathrm{crit,tot}}$	&$B$	&\fat{Orientation}	&$\mathcal{M}_\mathrm{rms}$	&$\mathcal{M}_\mathrm{a}$ &$t_\mathrm{ff,sph}$	&$t_\mathrm{ff,cyl}$\\
				&$\left[\mathrm{g\,cm}^{-3}\right]$	&$\left[\mathrm{cm}^{-3}\right]$	&	&	&$\left[\mu\mathrm{G}\right]$	&	&	&	&$\left[\mathrm{kyr}\right]$	&$\left[\mathrm{kyr}\right]$\\
	\hline
	ML1.6-M0.6-Perp	&$4\times10^{-19}$&$3.4\times10^5$	&1.6	&0.44	&40	&perpendicular	&0.6	&0.38	&57	&92\\
	ML1.6-M2.0-Perp	&$4\times10^{-19}$&$3.4\times10^5$	&1.6	&0.24	&40	&perpendicular	&2.0	&1.28	&	57&92\\
	ML1.6-M2.0-Para	&$4\times10^{-19}$&$3.4\times10^5$	&1.6	&0.24	&40	&parallel		&2.0	&1.28	&57	&92\\
	ML1.6-M4.0-Para	&$4\times10^{-19}$&$3.4\times10^5$	&1.6	&0.09	&40	&parallel		&4.0	&2.56	&	57&92\\
	ML1.6-M6.0-Perp	&$4\times10^{-19}$&$3.4\times10^5$	&1.6	&0.06	&40	&perpendicular	&6.0	&3.84	& 57	&92\\
	\hline
	ML0.8-M2.0-Perp	&$2\times10^{-19}$&$1.7\times10^5$	&0.8	&0.09	&40	&perpendicular	&2.0	&0.90	&81	&129\\
	ML0.8-M6.0-Perp	&$2\times10^{-19}$&$1.7\times10^5$	&0.8	&0.02	&40	&perpendicular	&6.0	&2.71	&81	&129\\
	\hline
	\hline
	\end{tabular}\label{ini}
\end{table*}

\subsection{Chemistry}
Deuterium fractionation is followed together with the hydrodynamical equations by employing the chemistry package \textsc{krome}.  We specifically emphasize that the network distinguishes between the ortho and para states of H$_2$, H$_3^+$ and H$_2$D$^+$, where H$_2$D$^+$ is the main tracer of the deuteration fraction employed in the simulation \citep{Walmsley2004}. We assume the gas to be fully depleted, an assumption which becomes accurate above number densities of $10^4$~cm$^{-3}$ and then applies to our initial conditions. The details and the uncertainties of the network have been largely discussed in \citet{Koertgen17}. We include the formation of molecules on the surface of grains but we do not take into account potential grain-surface processes as the H$_2$ ortho-to-para conversion on dust which can potentially accelerate the deuteration process \citep{Bovino17}. The latter would introduce in fact additional free parameters into the calculation, and here our main interest is to establish a lower limit based on a standard network. The initial H$_2$ ortho-to-para ratio is 3 and the initial cosmic--ray flux \mbox{$\zeta = 1.3 \times 10^{-17}$ s$^{-1}$}. We note that the cosmic--ray flux together with the initial H$_2$ ortho-to-para ratio can strongly affect the deuteration process \citep[see e.g.][]{Kong15}. However, the uncertainties of these quantities are still very large and a quantitative study on the effect of these parameters on our simulations would make our analysis more complex.

\section{Results}\label{results}
In the following, we will first discuss the global results for our reference run ML1.6-M4.0-Para, and subsequently assess the dependence of the deuteration fraction on the properties of the cores in different simulations. We then address the time evolution of typical cores as well as their radial profiles.

\subsection{Reference run}
To get an impression on the global dynamics, we first focus on our reference run ML1.6-M4.0-Para, with an average filament density of $3\times10^5$~cm$^{-3}$.  The line-mass corresponds to 1.6 times the thermal critical line mass, or $0.09$ times the critical line mass when considering the magnetic field and turbulent fluctuations. The magnetic and turbulent pressure thus play a crucial role in stabilizing the filament against gravitational collapse. The magnetic field  strength is $40$~$\mu$G, and the field is parallel to the filament major axis. The free-fall time of the system, considering the cylindrical geometry \citep[$t_\mathrm{ff,cyl}=\sqrt{R_\mathrm{fil}/2L_\mathrm{fil}G\left<\rho\right>}$,][]{Toala12}, corresponds to $92$~kyrs. The system of course is not going through free-fall due to the stabilizing effect of the magnetic field and turbulence, but it provides a characteristic timescale for larger-scale motions within the filament. 
\begin{figure*}
	\begin{tabular}{cc}
		\includegraphics[width=0.51\textwidth]{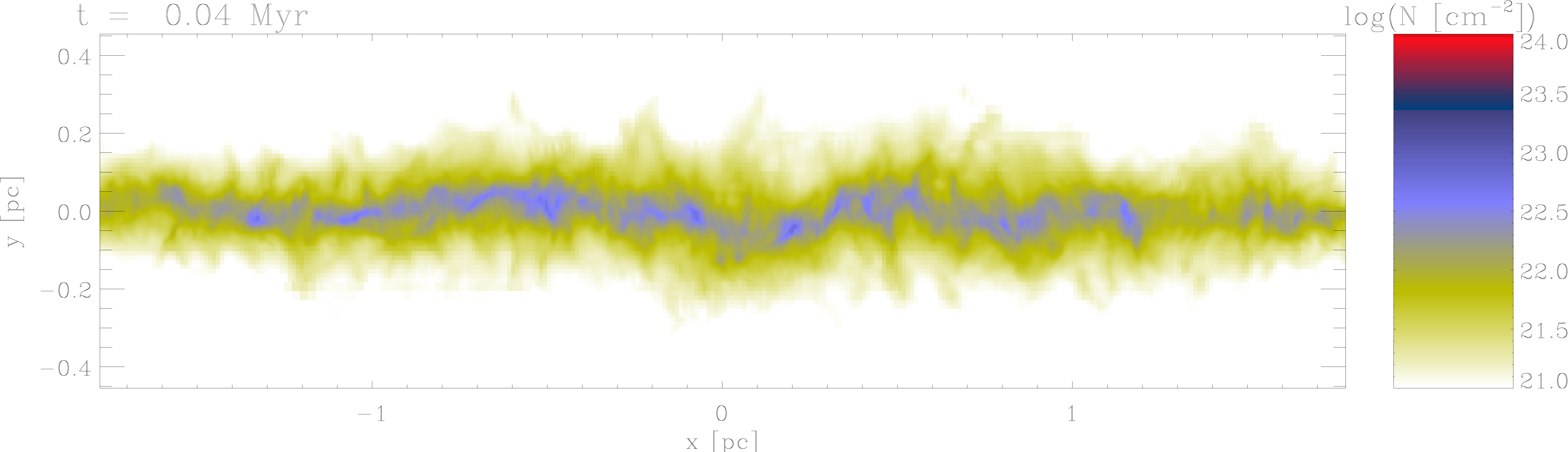}&\includegraphics[width=0.51\textwidth]{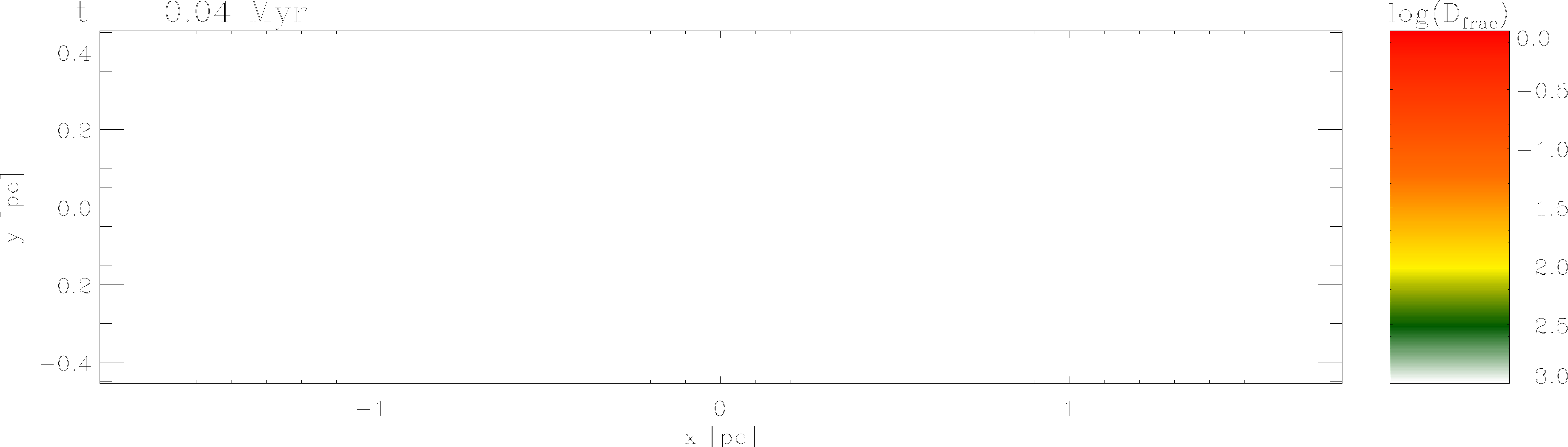}\\
		\includegraphics[width=0.51\textwidth]{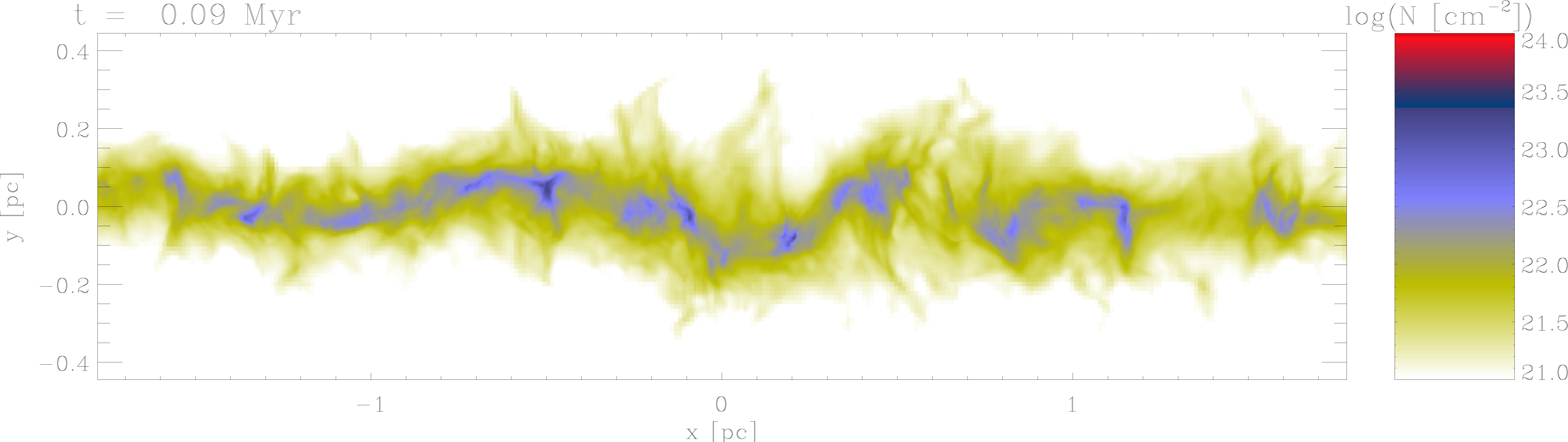}&\includegraphics[width=0.51\textwidth]{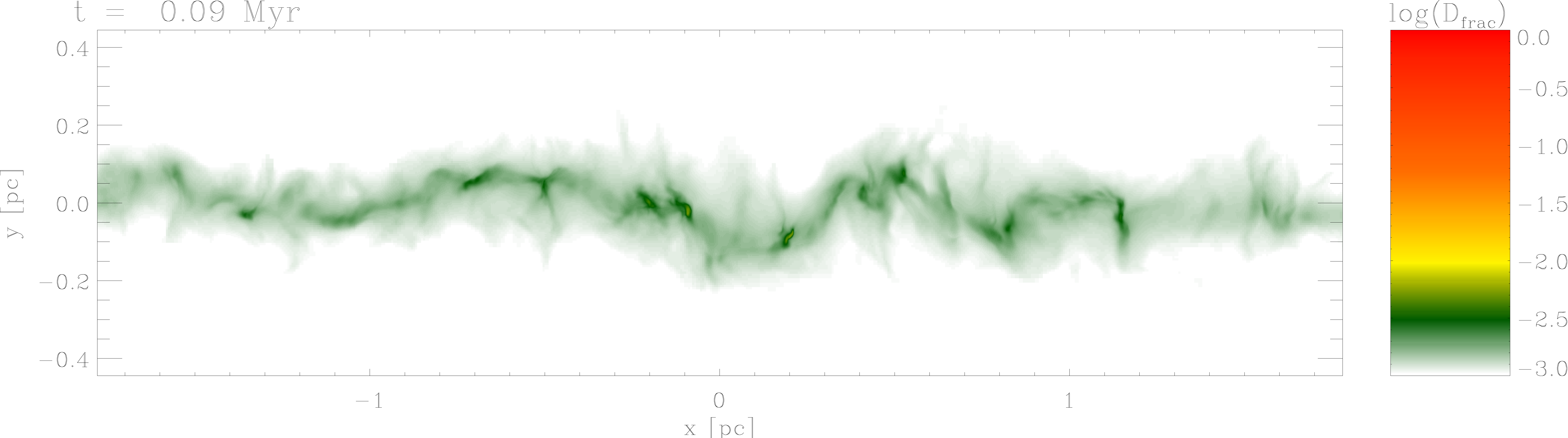}\\
		\includegraphics[width=0.51\textwidth]{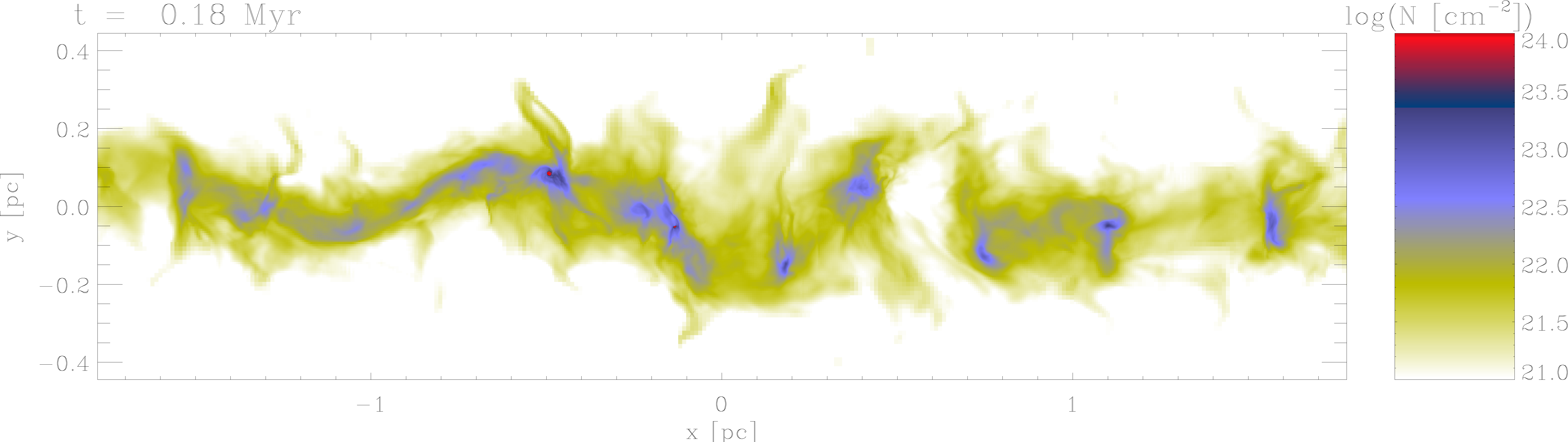}&\includegraphics[width=0.51\textwidth]{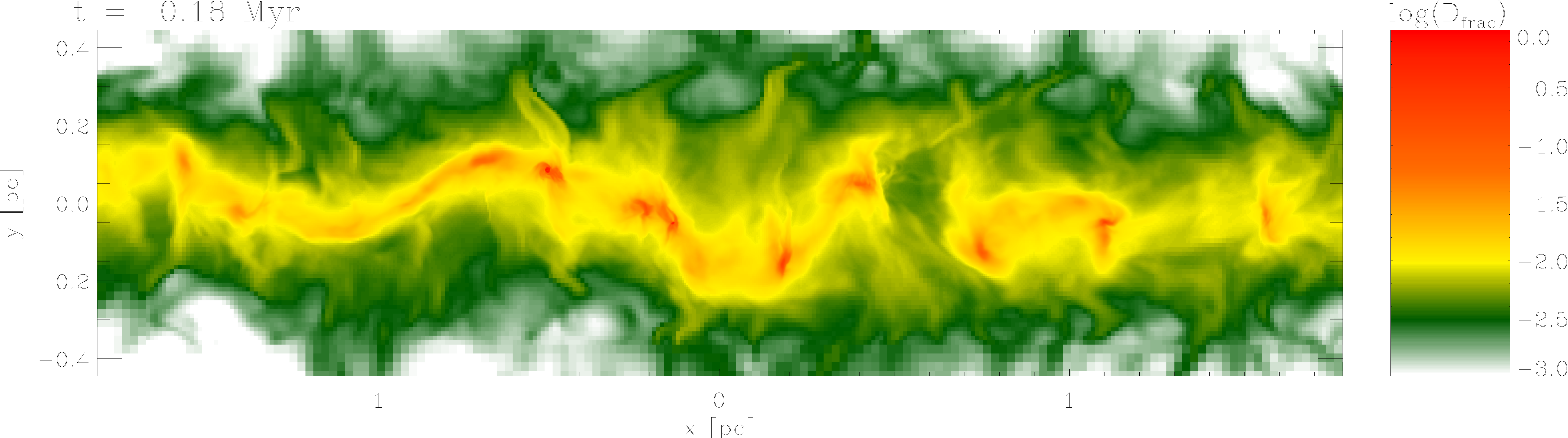}\\
		\includegraphics[width=0.51\textwidth]{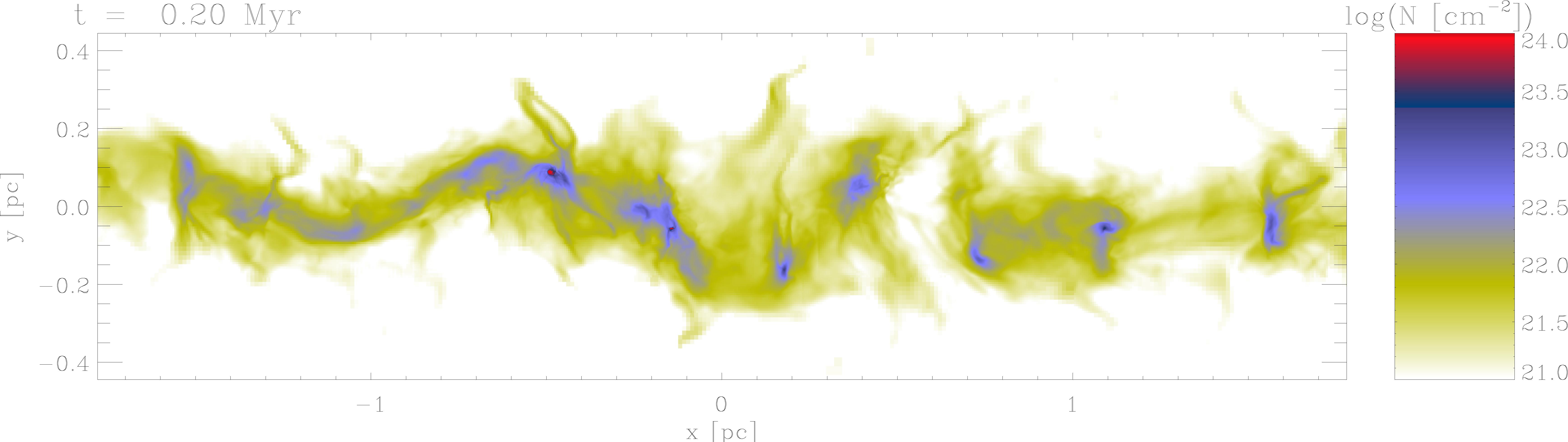}&\includegraphics[width=0.51\textwidth]{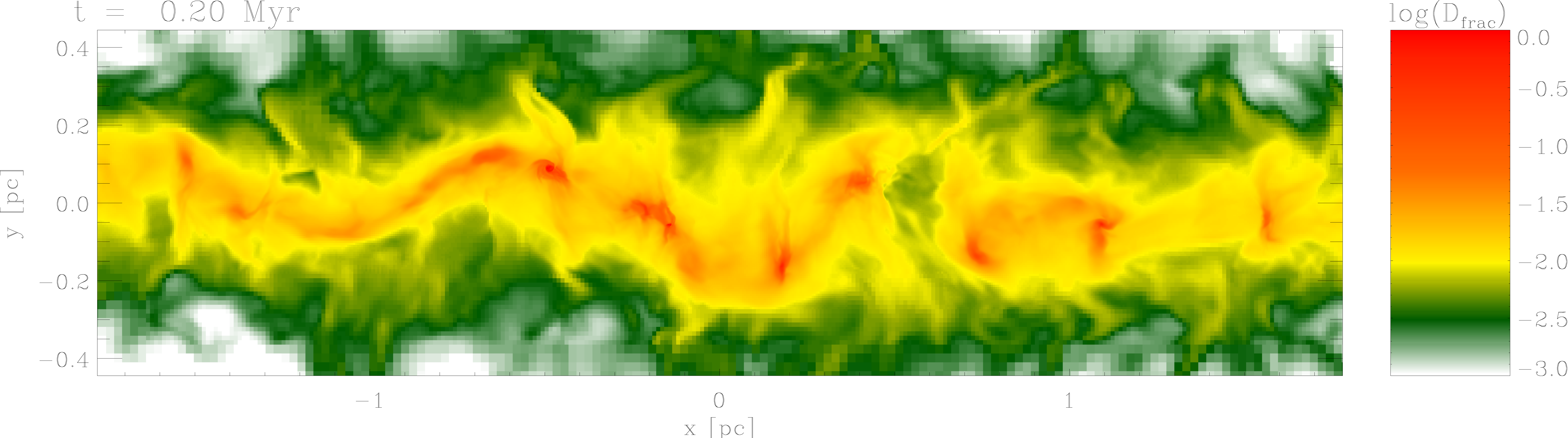}\\	
	\end{tabular}
	\caption{Time evolution of the filament in run ML1.6-M4.0-Para. \ita{Left:} Top-down view of column density for the filament. \ita{Right:} Same view for density-weighted deuteration fraction. The initially strong turbulence leads to a heavily bent and partially disrupted filament. Individual cores form due to compression. The times shown refer to 0.5, 1, 2, and 2.2 free-fall times. The integration length is $l=0.2\,\mathrm{pc}$.}	
\label{filstruct}
\end{figure*}
In Fig.~\ref{filstruct} we show snapshots of the column density and the density-weighted deuteration fraction 
\beq
D_\mathrm{frac}=\frac{\int{\frac{\left[\mathrm{H}_2\mathrm{D}^+\right]}{\left[\mathrm{H}_3^+\right]}\rho dz}}{\int{\rho dz}},
\label{eqDfrac}
\eeq 
in the x-y plane, that is, perpendicular to the initial magnetic field. In Eq. \ref{eqDfrac} $\rho$ is the density in each cell, $dz$ denotes the path-element (essentially the cell-size) along the z-direction and $\left[X\right]$ denotes the mass-fraction of species $X$ in that cell. The different maps correspond to 0.5, 1, 2, and 2.2 times the free-fall time (from top to bottom). 
At early times, $t\sim0.5\,t_\mathrm{ff,cyl}$, the innermost part near the ridge of the filament has already accreted mass and thereby enhanced column density. This increase is, however, not uniform along the major axis, but rather scattered throughout the filament 
due to compression and dilatation of individual gas parcels by turbulent motions. At the same time, the filament is observed to be slightly bent on larger scales due to the ambient turbulence in the filament itself as 
well as in the more diffuse gas surrounding it. At about one free-fall time ($t\sim90\,\mathrm{kyr}$), individual overdense regions can readily be identified. At the same time, the filament breaks up at $x\sim0.6\,$pc and $x\sim1.3\,$pc, a feature that can be associated with dispersion 
due to turbulence and material flowing along the magnetic field lines towards deeper gravitational potential wells. Interestingly, there appear striations in the diffuse gas, which are more pronounced near overdense regions within the filament. These features 
are generated via gravitational focusing due to the overdense structures within the filament.\\
The further evolution of the filament reveals stronger bending of the whole structure as well as ongoing fragmentation within the filament. The amplitude of the bending is $\sim0.1-0.2\,\mathrm{pc}$. The fragments show signs of ongoing accretion (observed by their increased column density), but some also seem to 
dissolve. The striations in the surroundings of the filament have now also become denser and, at least for the structure at $x\sim-0.5\,$pc, they appear to be tightly connected to the inner part of the filament. From $180$~kyr to the end of the simulation at 
$200$~kyr, the filaments do not evolve very much, but differences are seen for some of the overdensities. At the same time, the previously mentioned break-up of the filament proceeds, revealing regions with a decrease in 
column density by about an order of magnitude.\\
On the right side of Fig.~\ref{filstruct} the deuteration fraction is shown for the same temporal stages as discussed above for the column density. In general it is observed that this ratio traces the density 
quite well. This is to be expected, as chemical timescales depend on the density, and also the deuteration process is faster at higher densities. Initially the deuteration ratio is rather small of the order of $\lesssim10^{-3}$ throughout the whole region. When the filament is evolving in time, the process of deuterium fractionation becomes less efficient in the outskirts due to the reduced density compared to the inner part. At around one free-fall 
time the deuteration fraction is readily enhanced in the innermost part of the filament, again matching quite well the overall density structure. In addition, the deuteration fraction appears to be almost homogeneous throughout the filament. However, individual 
overdensities can be identified. At late times, deuterium fractionation becomes also efficient in the more tenuous gas surrounding the filament. The inner ridge of the filament has now significantly increased deuteration fractions 
of the order of $0.01-0.3$. In this sense, regions of strongly increased density show the highest values, as is expected. Interestingly, the previously mentioned striations also show an enhanced deuterium fractionation, giving rise to accretion of highly deuterated 
gas onto regions within the filament. If one compares the filament at $t=180$~kyr and $t=200$~kyr, there appear interesting features. Although the time-evolution of the overall column density between these stages is rather small, the deuteration fraction 
evolves significantly. In detail, the gaps that formed due to mainly turbulent fluctuations appear to be smoother when looking at the deuteration fraction at these times. Although the column density is greatly reduced in these gaps, the density is still high 
enough to promote a rather efficient deuteration process.\\

\subsection{Dependence of the deuterium fractionation on core properties}
\begin{figure*}
	\begin{tabular}{ccc}
	$t=0.5\,t_\mathrm{ff,cyl}$	&$t=t_\mathrm{ff,cyl}$	&$t=2\,t_\mathrm{ff,cyl}$\\
	\includegraphics[width=0.25\textwidth,angle=-90]{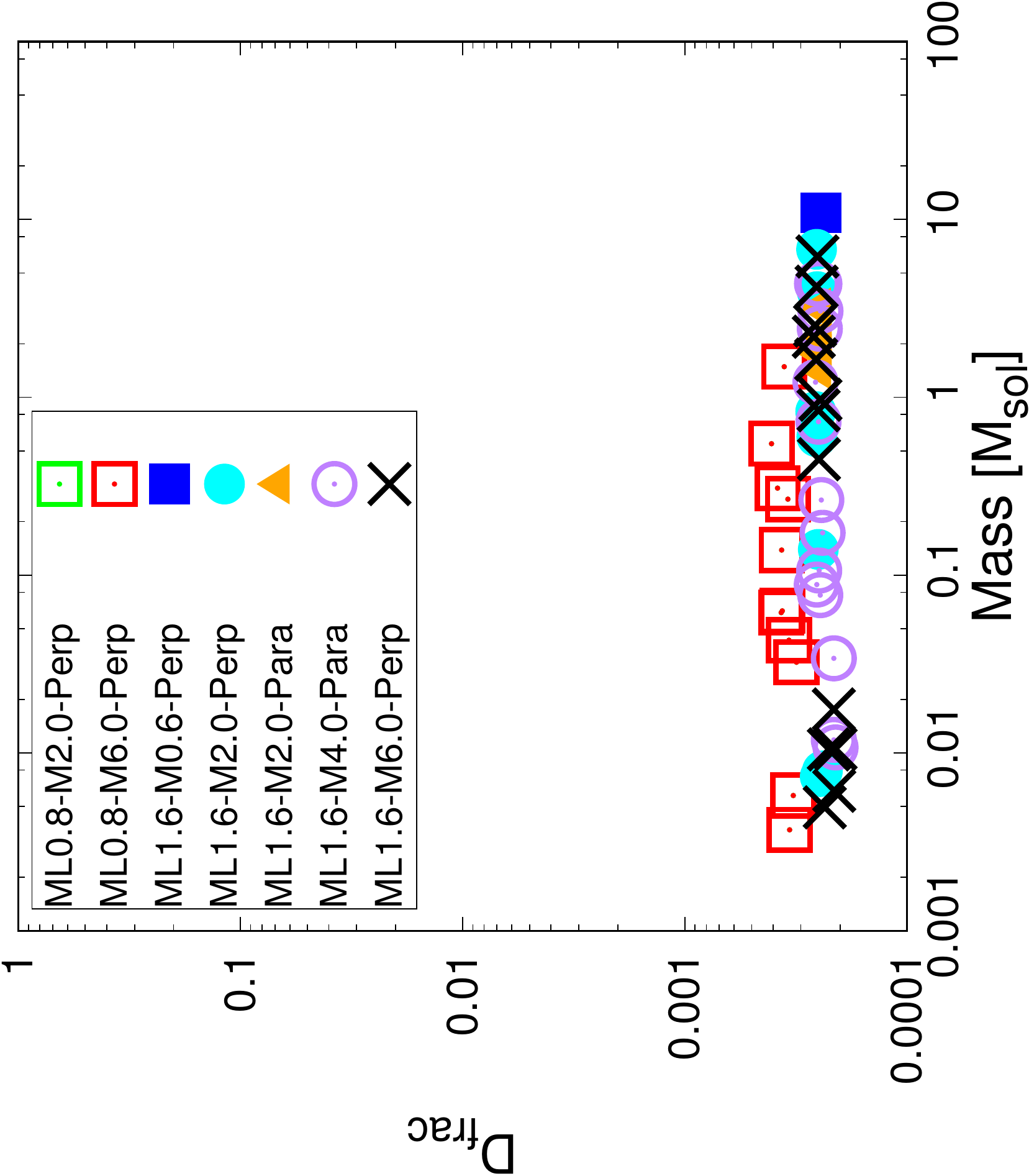} &\includegraphics[width=0.25\textwidth,angle=-90]{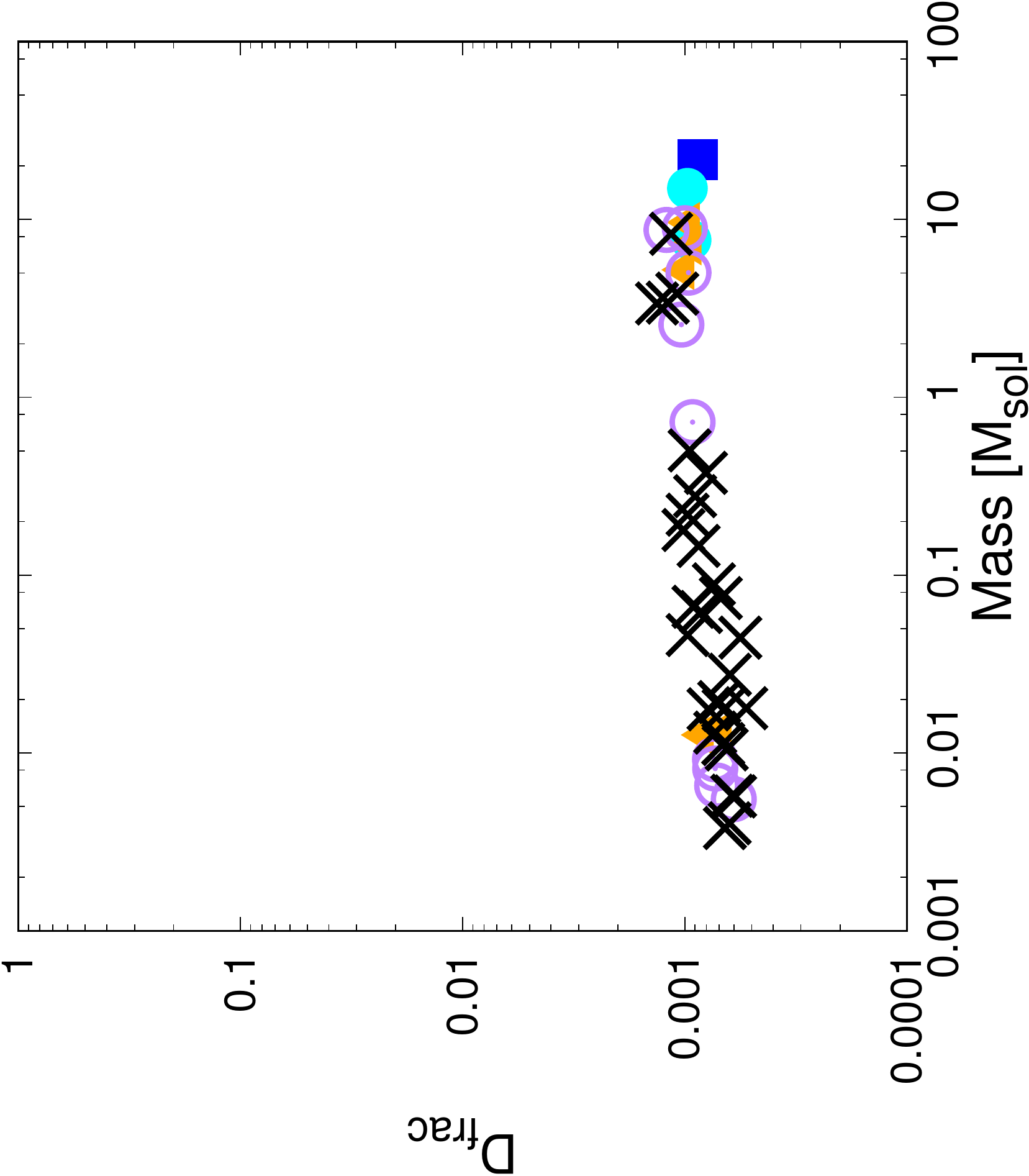} &\includegraphics[width=0.25\textwidth,angle=-90]{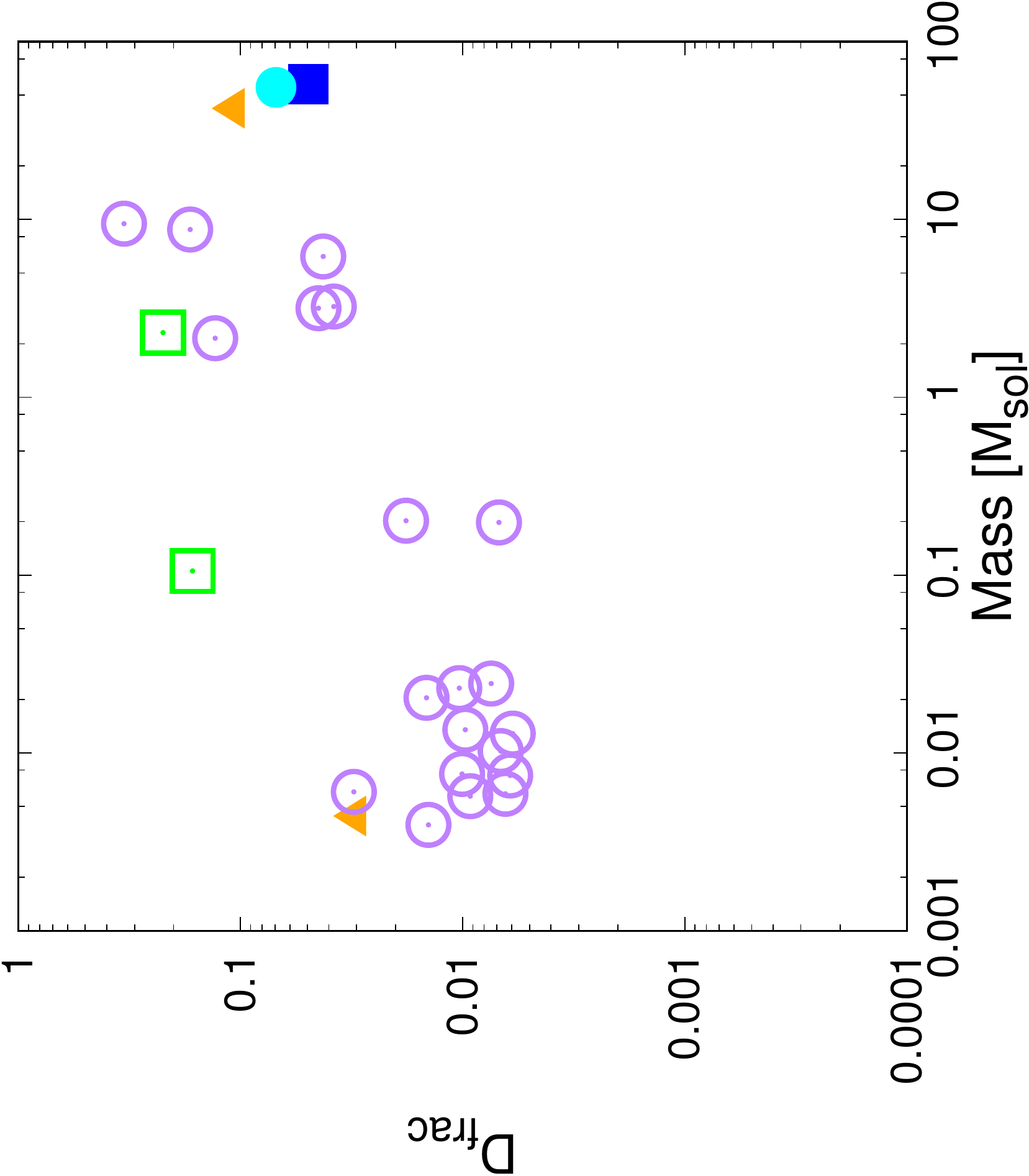} \\
	\includegraphics[width=0.25\textwidth,angle=-90]{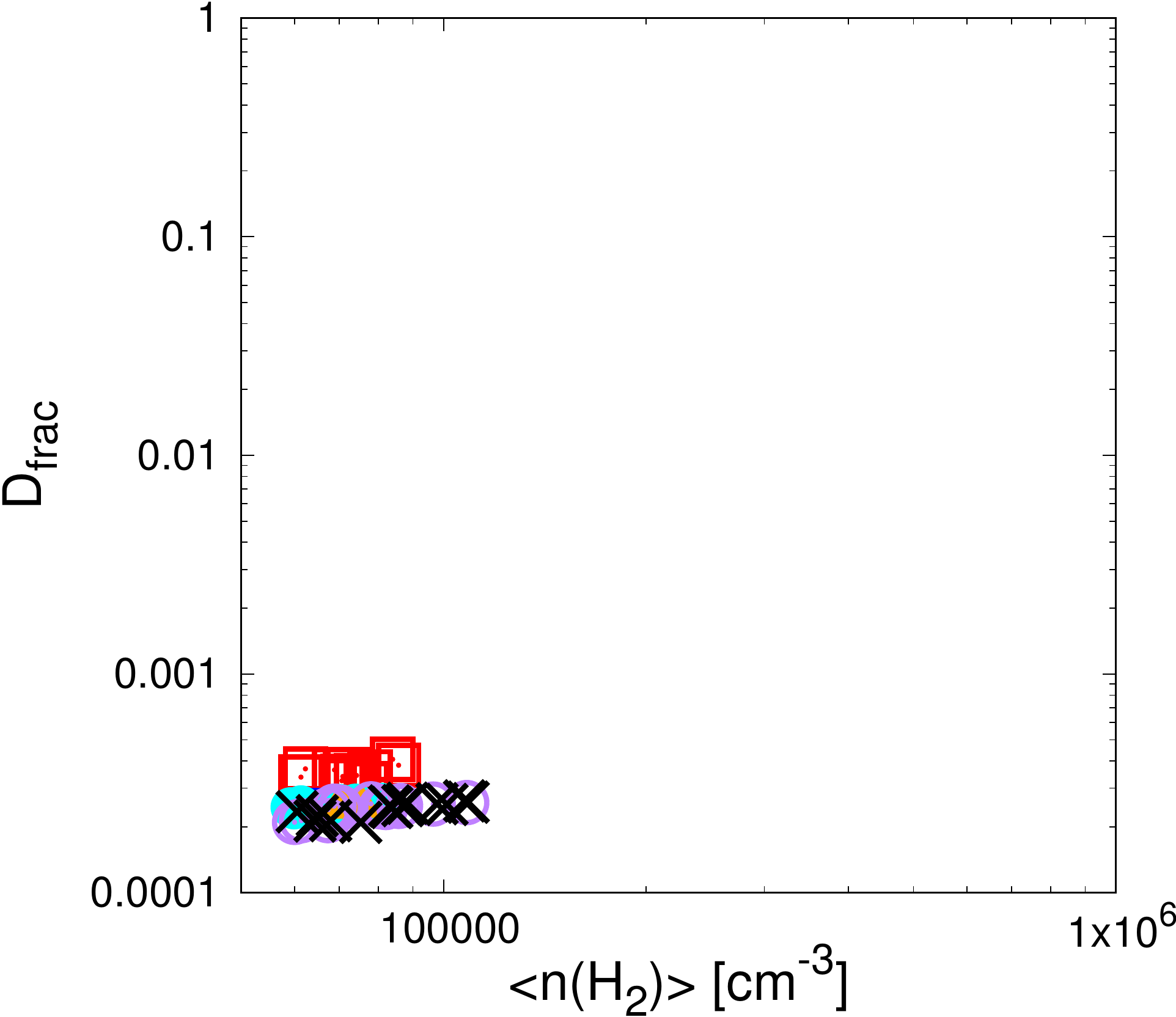} &\includegraphics[width=0.25\textwidth,angle=-90]{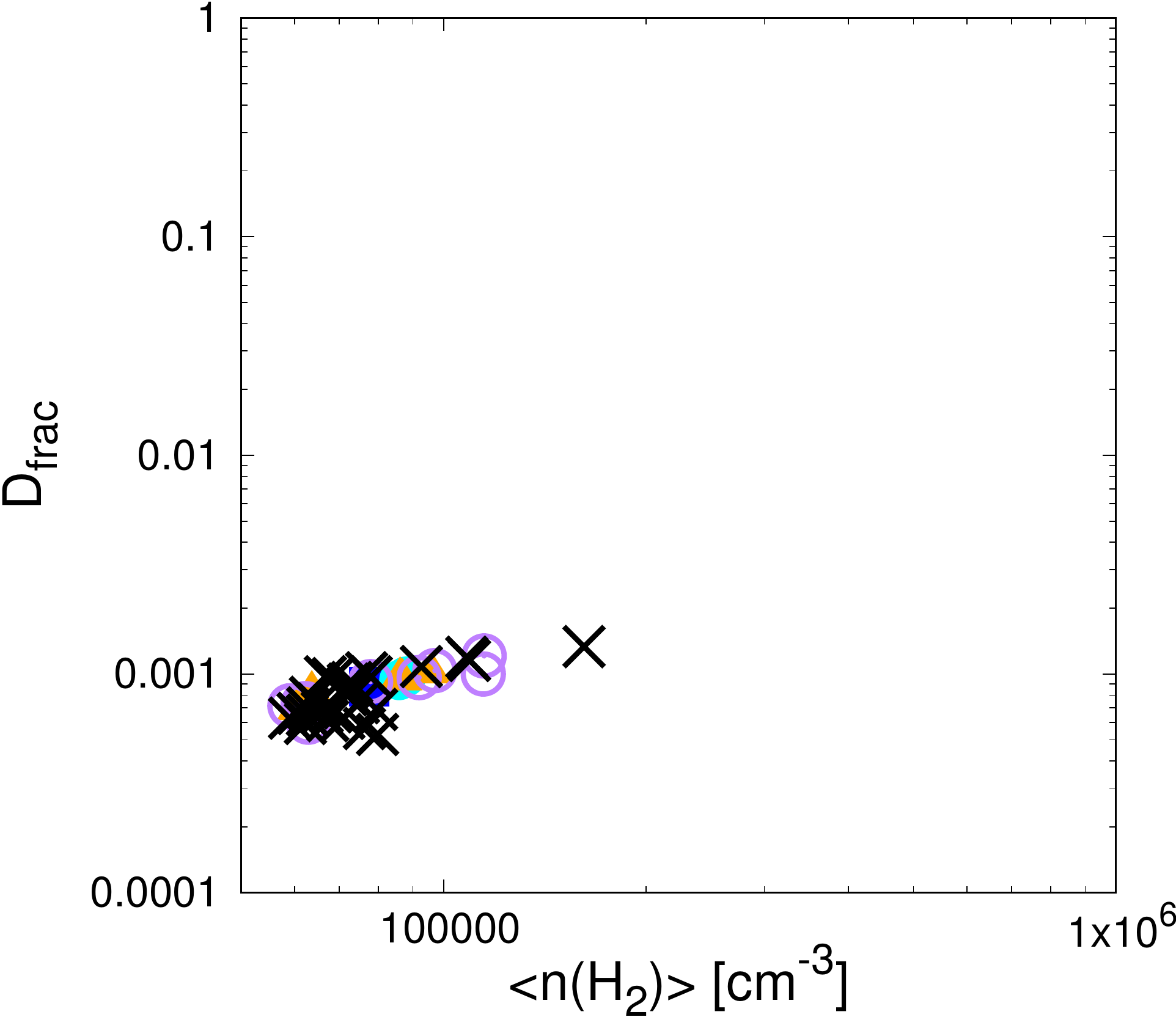} &\includegraphics[width=0.25\textwidth,angle=-90]{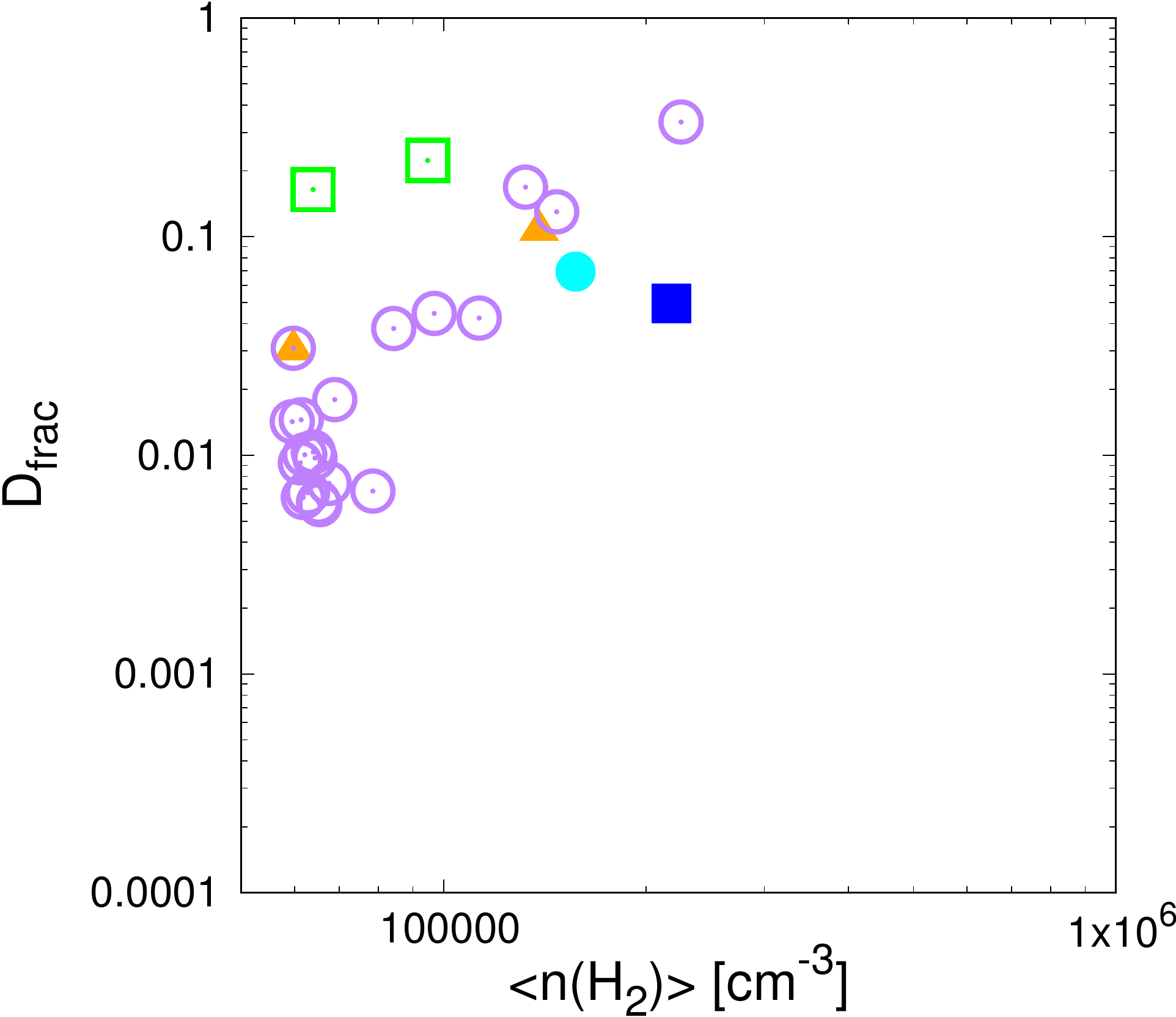} \\
		\includegraphics[width=0.25\textwidth,angle=-90]{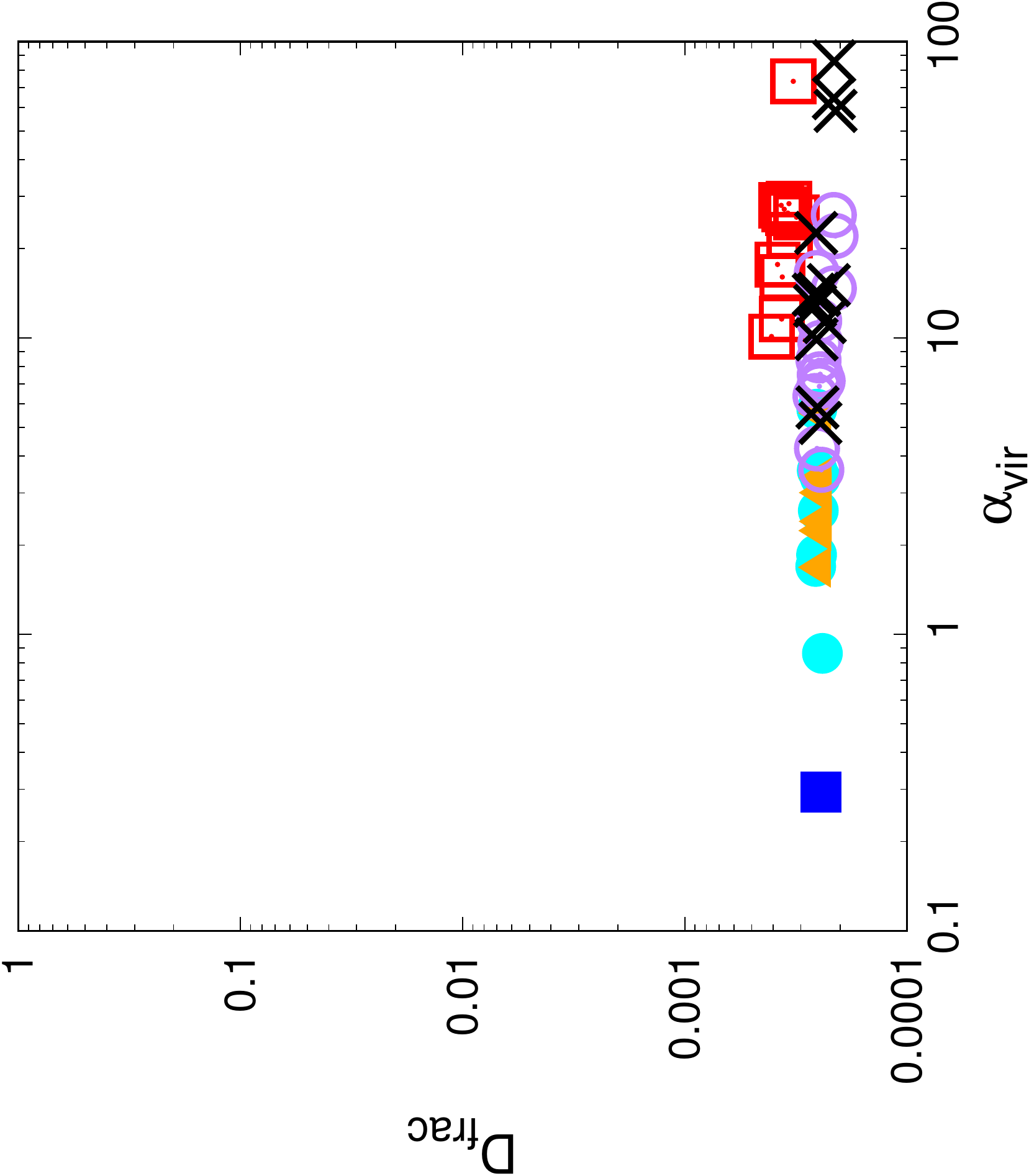} &\includegraphics[width=0.25\textwidth,angle=-90]{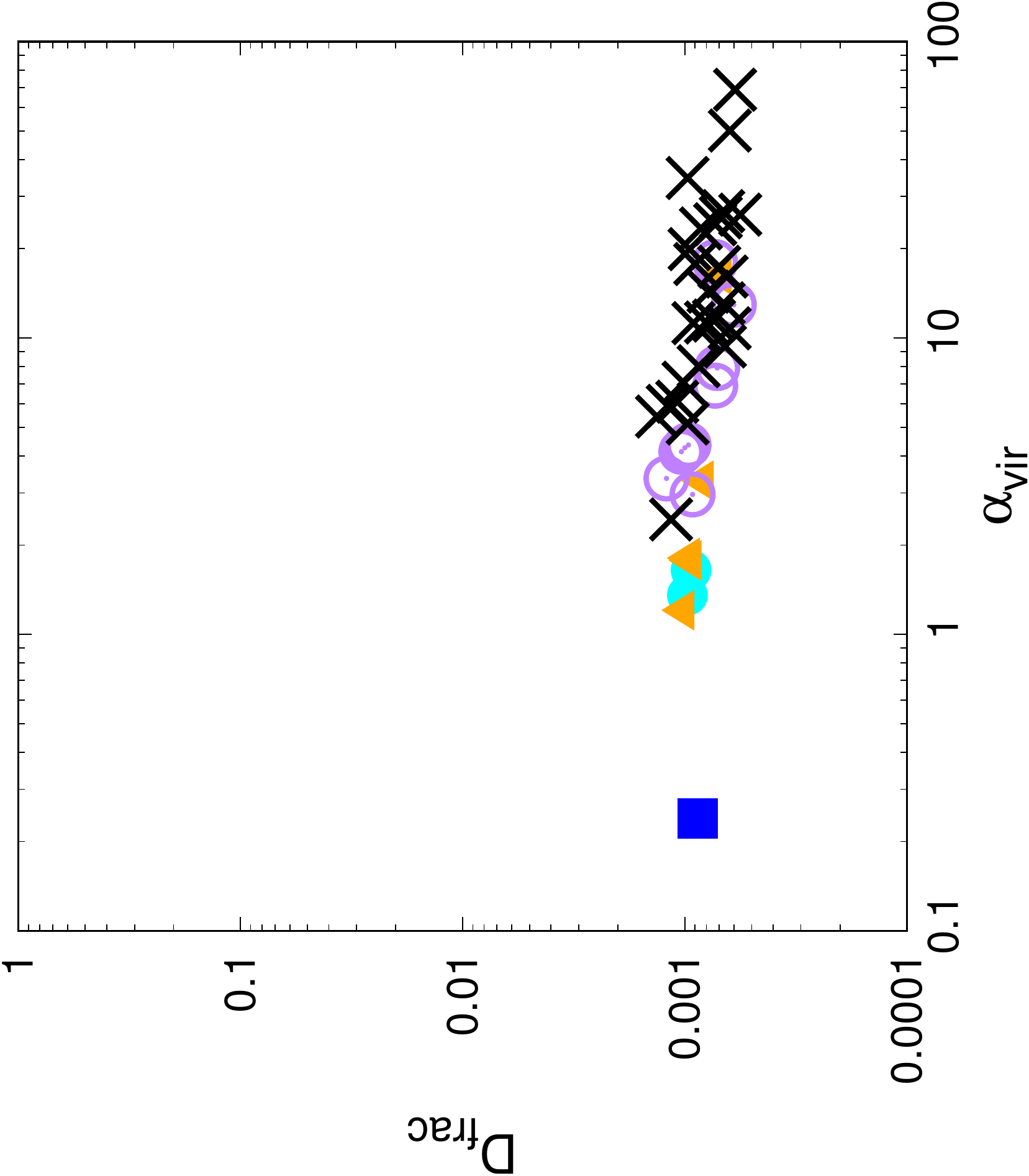} &\includegraphics[width=0.25\textwidth,angle=-90]{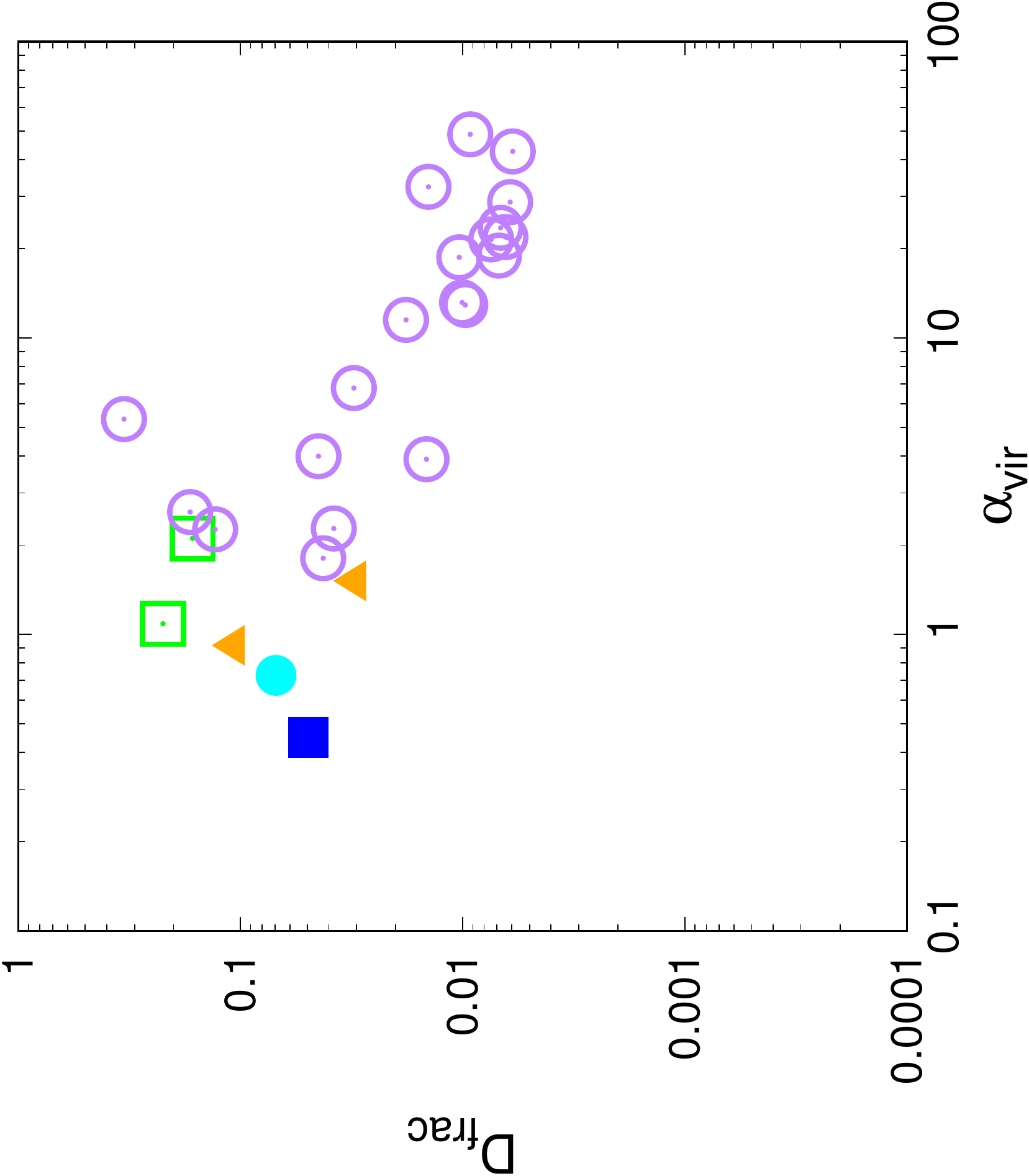} \\
	\end{tabular}
	\caption{Statistical properties of the cores found in the different simulations. We plot the deuteration fraction as a function of core mass (top panel), average H$_2$ density (mid panel) and virial parameter (bottom panel) at three characteristic times in the simulations, corresponding to $0.5$, $1$ and $2$ free-fall times as defined for cylindrical systems. For times $t<t_\mathrm{ff,cyl}$ the chemical evolution is dominated by the average density of the filament and appears to be independent of the dynamical 
	properties of the filament. At late stages, trends emerge, which can be assigned directly to the properties of the cores, but indirectly also to their formation time. In this respect, low-D$_\mathrm{frac}$ cores shown at later stages should have formed later. This is also in agreement with Fig.~\ref{filstruct}, where the major part of the filament shows D$_\mathrm{frac}\sim0.01$. }
	\label{figStatistics}
\end{figure*}

Of particular interest in our investigation is how the properties of the cores evolve over time, and if there are statistical correlations between different properties of the core. We are particularly interested in potential correlations of the deuteration fraction with core mass, H$_2$ density and the virial parameter, which are plotted in Fig.~\ref{figStatistics} at $0.5$, $1$ and $2$  free-fall times as defined for cylindrical systems. As shown in Table~\ref{ini}, the value of the filament free-fall time varies between $92$ and $129$~kyrs for our simulations. The individual cores are defined as spatially connected (i.e. neighboring cells with similar density) structures with a lower threshold density of \mbox{$n_\mathrm{min}=8\times10^4\,\mathrm{cm}^{-3}$}. However, we caution here 
that the term \ita{core} is somewhat ill-defined and mostly based on whether the object can be distinguished from the background. \\
For all three quantities, we find an initially flat relation at $0.5$ free-fall times. This corresponds to the stage where the cores have initially formed, and where their chemical properties are thus dictated by the average evolution within the filament. Therefore, we see no dependence on the parameters of the core. This is different at one free-fall time, where a mild increase of deuteration fraction is visible both as a function of core mass and H$_2$ number density. The correlation with H$_2$ density is expected from a chemical point of view, as it enhances the deuteration rate, and it appears that (with some scatter), the density is correlated with the core mass. In this context, it is consistent that the correlation appears more pronounced when looking at $D_{\rm frac}$ vs density. After two free-fall times, these trends become even more clear and are particularly well visible in our reference run, which includes a larger number of cores. We however also see that there is a dependence on the initial conditions of the filament, as for instance the simulation ML0.8-M2.0-Perp has two cores that follow the same trend, but with enhanced values of $D_{\rm frac}$ compared to the reference run, while ML1.6-M0.6-Perp shows lower values. While it seems plausible that we can expect typical correlations, their normalization may vary from filament to filament.\\
In the plots of $D_{\rm frac}$ as a function of the virial parameter, we find the inverse correlation, i.e. decreasing deuteration fractions for increasing virial parameters. We note that many of the cores that are defined via density thresholds have virial parameters larger than $1$, and are thus not gravitationally bound. This may at first seem surprising, but also observed samples of cores show high virial parameters of order $20$ \citep[][and references therein]{Kauffmann13,Kirk17}. The inverse correlation is then not surprising, as a core with higher virial parameter will typically have lower density, which reduces the efficiency of chemical reactions including deuterated species. These cores might also have higher velocity dispersions, which tend to dissolve these 
regions, thereby reducing the density and hence again the efficiency of chemical reactions. This trend is still weak at about one free-fall time, but pronounced at two free-fall times. 

\subsection{Time evolution of core properties}
\begin{figure*}
	\begin{tabular}{cc}
		\includegraphics[width=0.4\textwidth,angle=-90]{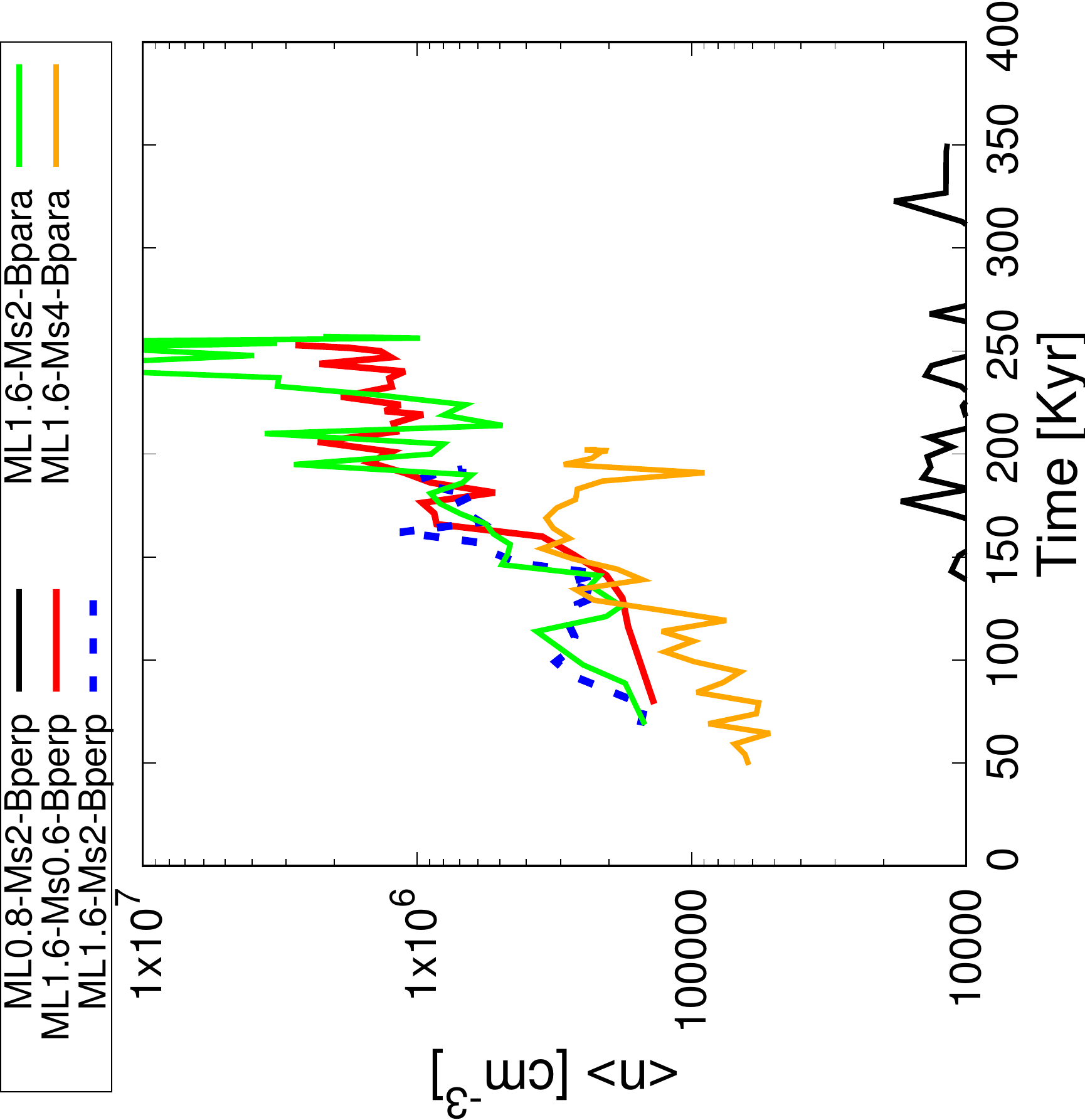} &\includegraphics[width=0.4\textwidth,angle=-90]{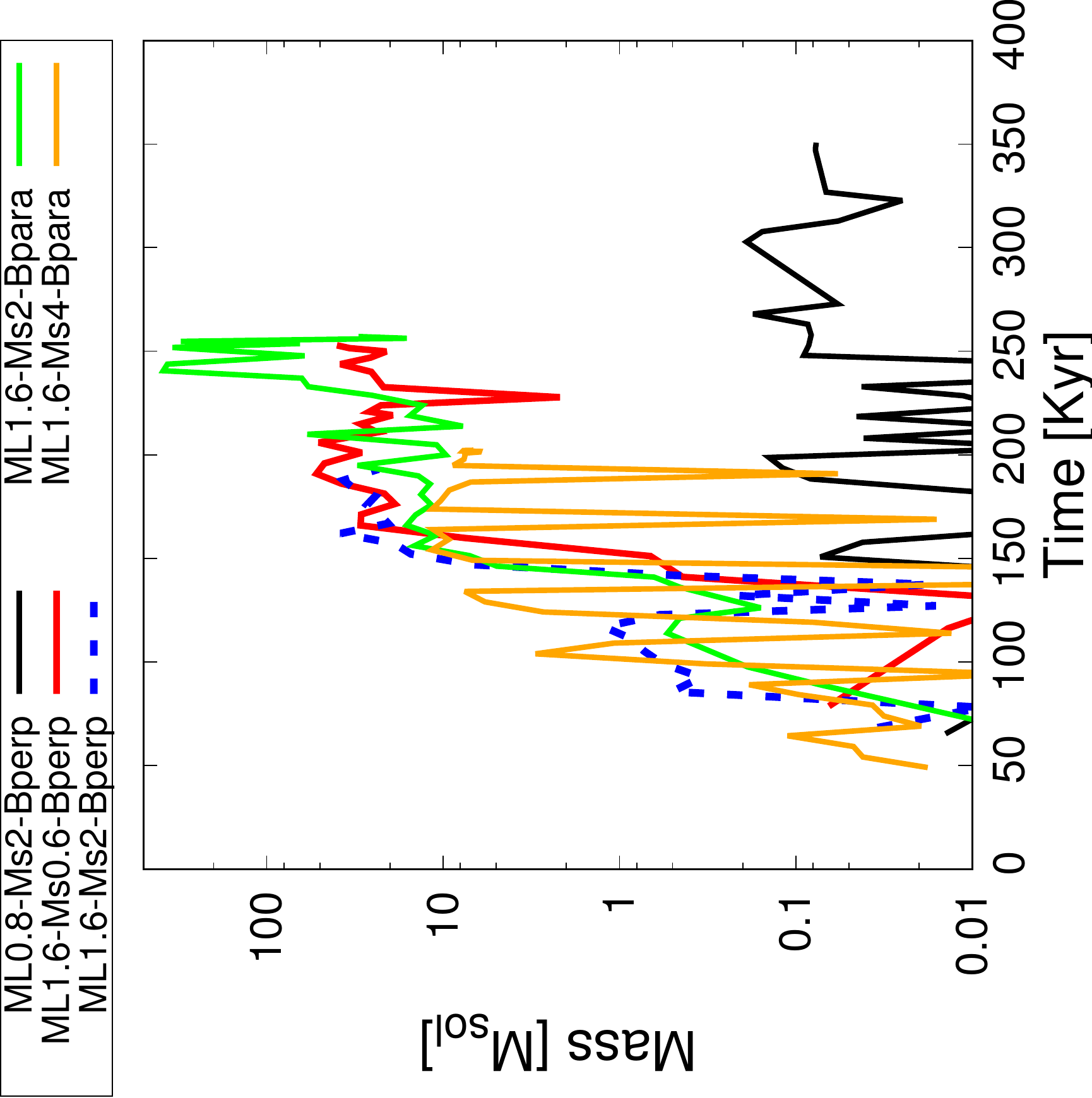} 
	\end{tabular}
	\caption{\ita{Left:} Evolution of average volume number density of the cores given in Table~\ref{tabCoreParams}. When the filament is initially sub-critical the core density is roughly $10^4$~cm$^{-3}$ and hence on the verge of applicability of our model. \ita{Right:} Time evolution of core mass. It is evident that the current setup of filaments provides a large enough mass reservoir that the core can become massive in a rather short time. The scatter in the evolution comes from the highly dynamic environments 
	of the cores, but also due to our somewhat arbitrary definition of a core. }
	\label{figCoreTime1}
\end{figure*}

\begin{figure*}
	\begin{tabular}{cc}
		\includegraphics[width=0.4\textwidth,angle=-90]{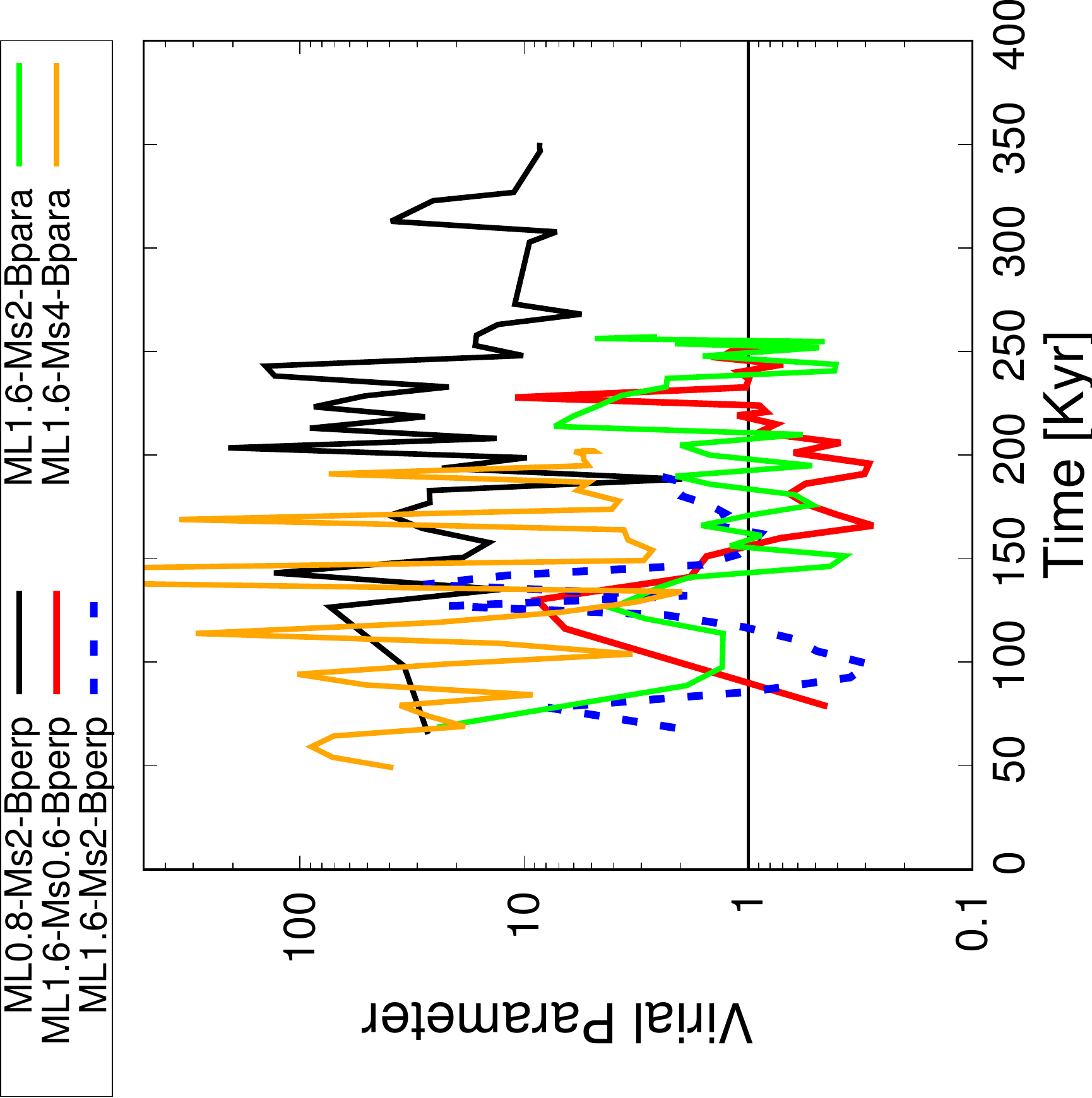} &\includegraphics[width=0.4\textwidth,angle=-90]{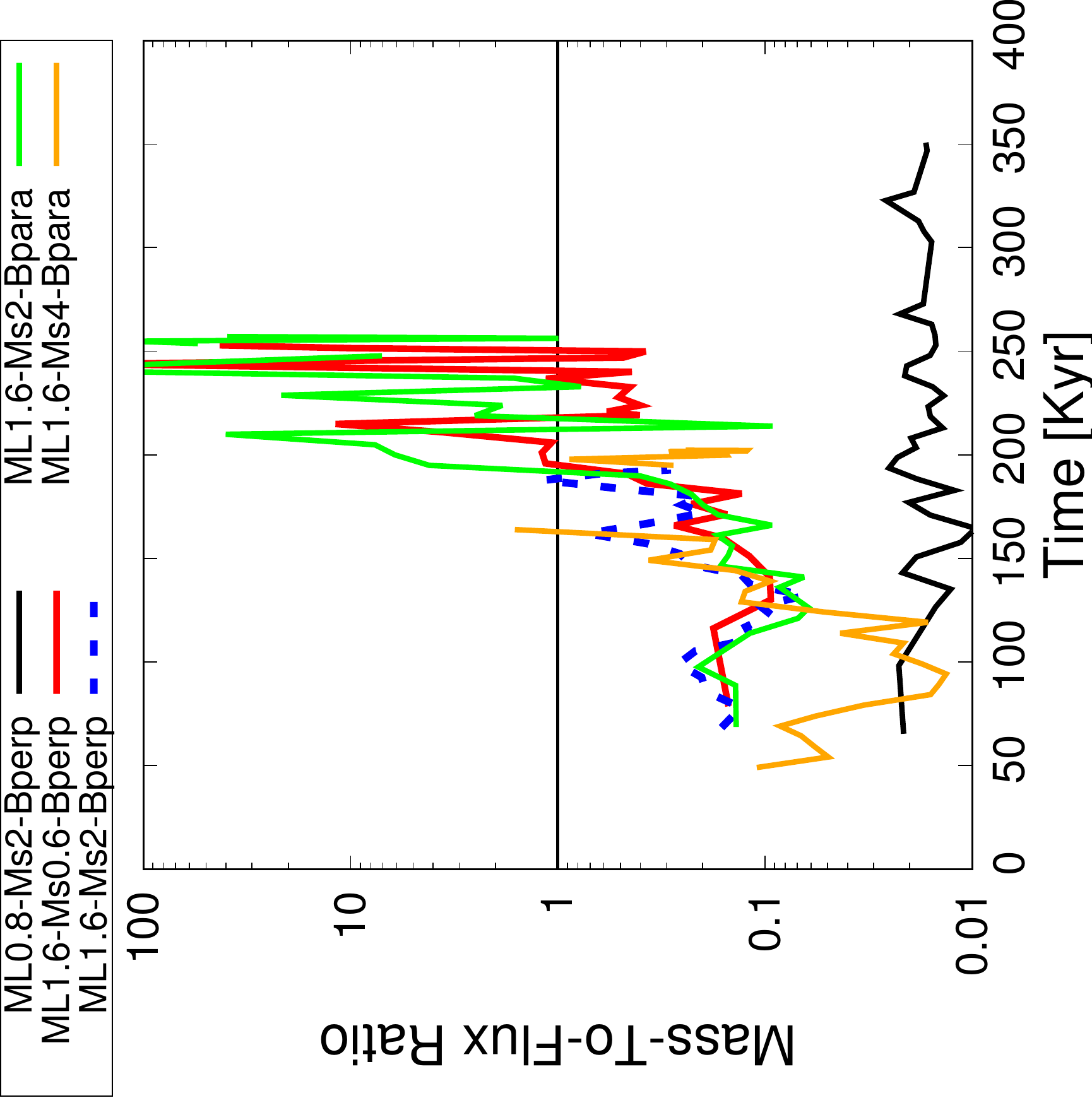} 
	\end{tabular}
	\caption{Evolution of virial parameter (left, only turbulent velocity taken into account) and mass-to-magnetic flux ratio (right) as function of time for the cores given in Table~\ref{tabCoreParams}.}
	\label{figCoreTime2}
\end{figure*}

\begin{figure*}
	\begin{tabular}{cc}
		\includegraphics[width=0.4\textwidth,angle=-90]{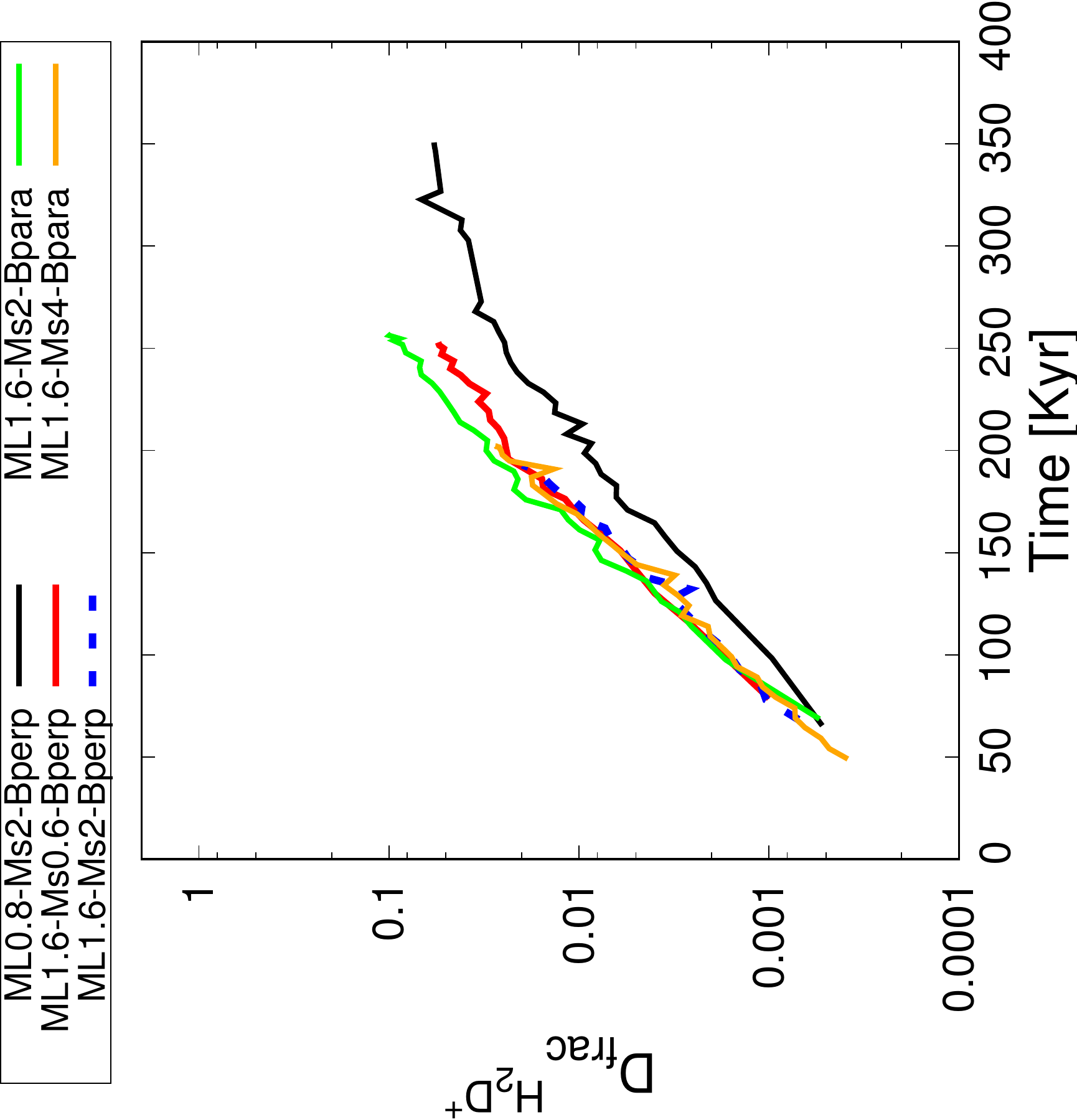} & \includegraphics[width=0.4\textwidth,angle=-90]{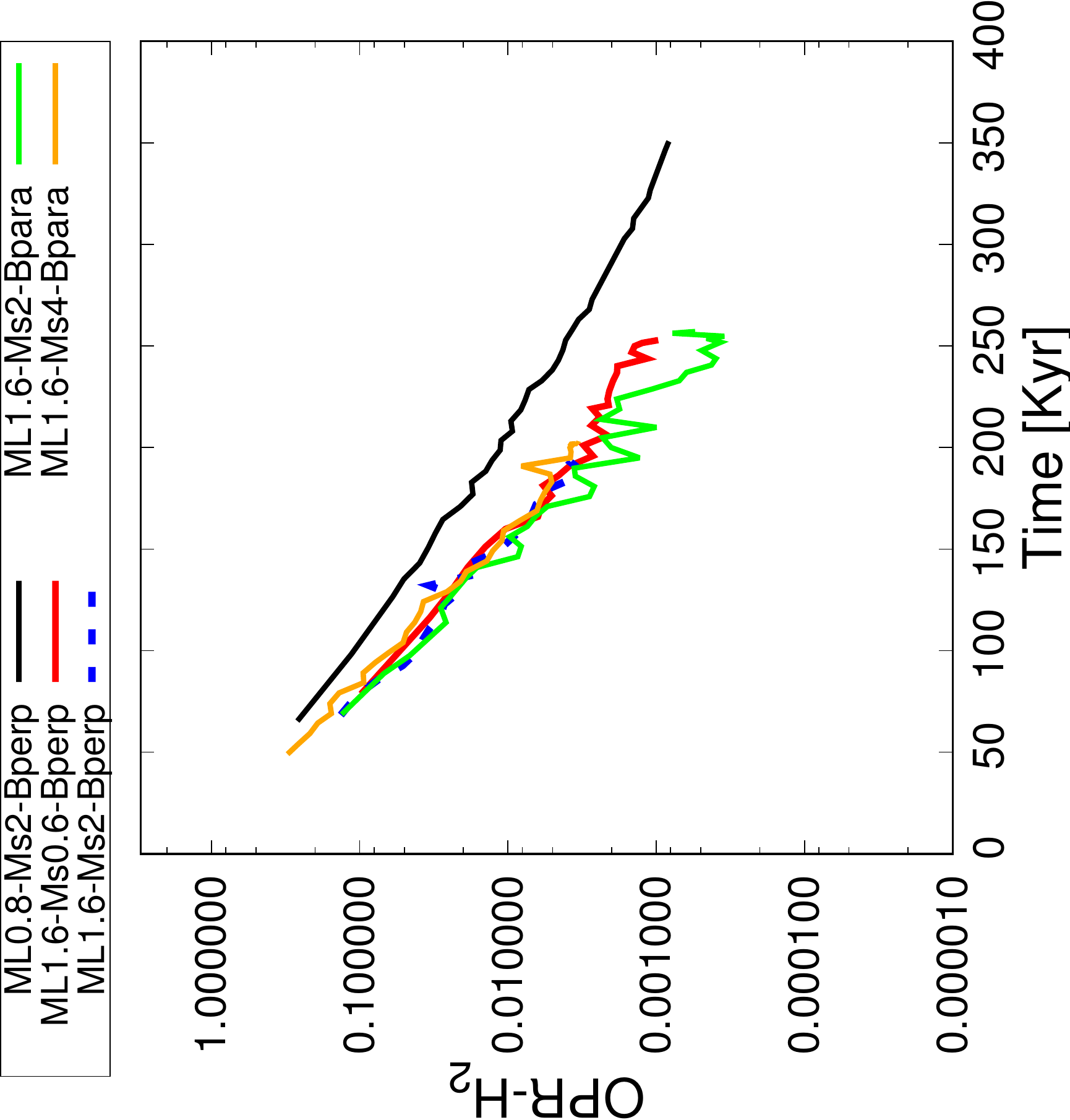} 
	\end{tabular}
	\caption{Time evolution of average deuteration fraction (left) and OPR-H$_2$ (right) in the cores given in Table~\ref{tabCoreParams}. Although the initial OPR-H$_2$ is set to the conservative value of 3, the actual OPR at core formation is about an order of 
	magnitude lower already after only 50~kyr ($\sim0.5\,t_\mathrm{ff,cyl}$) of evolution.}
	\label{figCoreTime3}
\end{figure*}

\begin{figure*}
	\begin{tabular}{cccc}
		\rotatebox[origin=l]{90}{ML1.6-M0.6-Perp}&\includegraphics[width=0.3\textwidth]{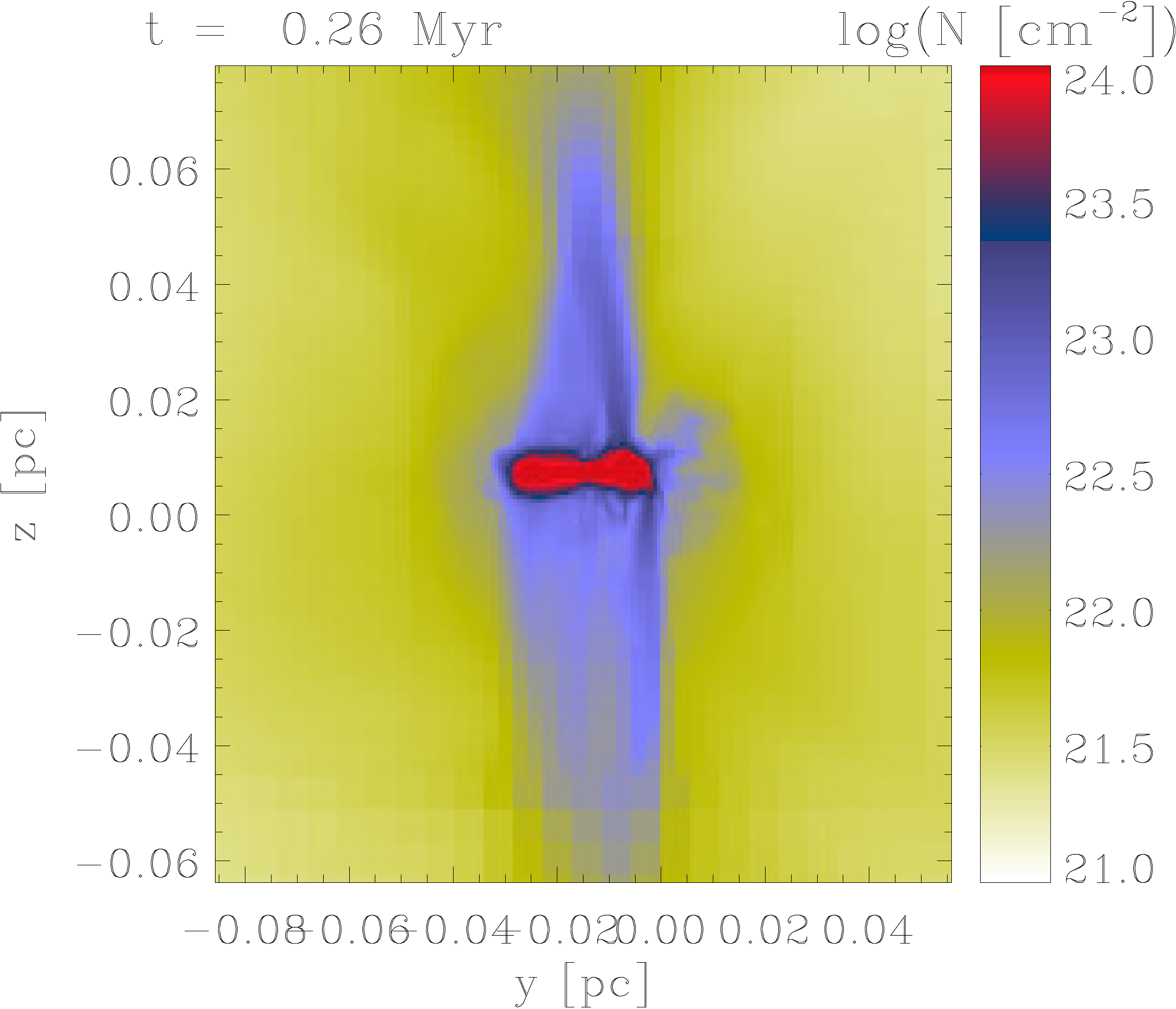} &\includegraphics[width=0.3\textwidth]{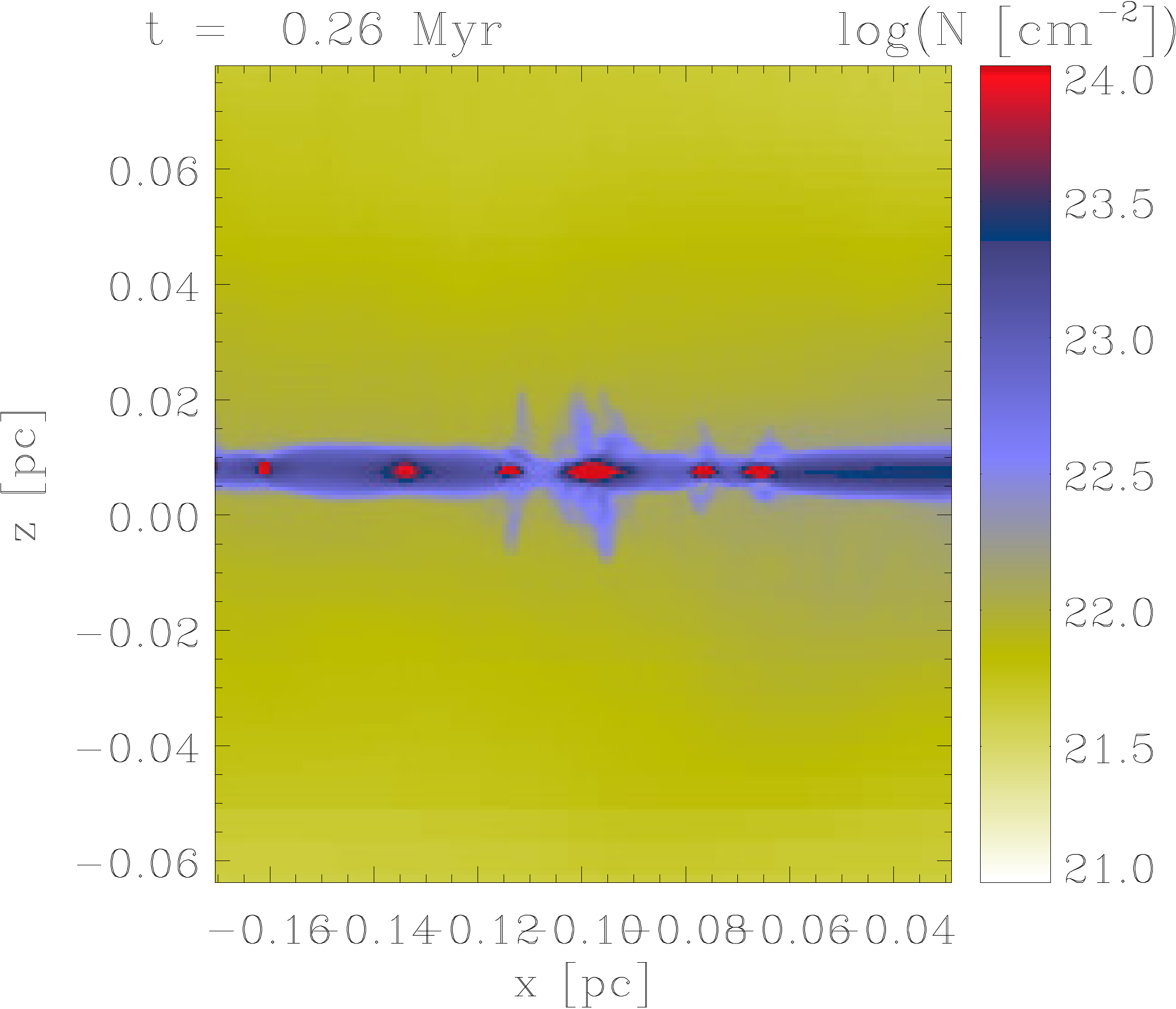} &\includegraphics[width=0.3\textwidth]{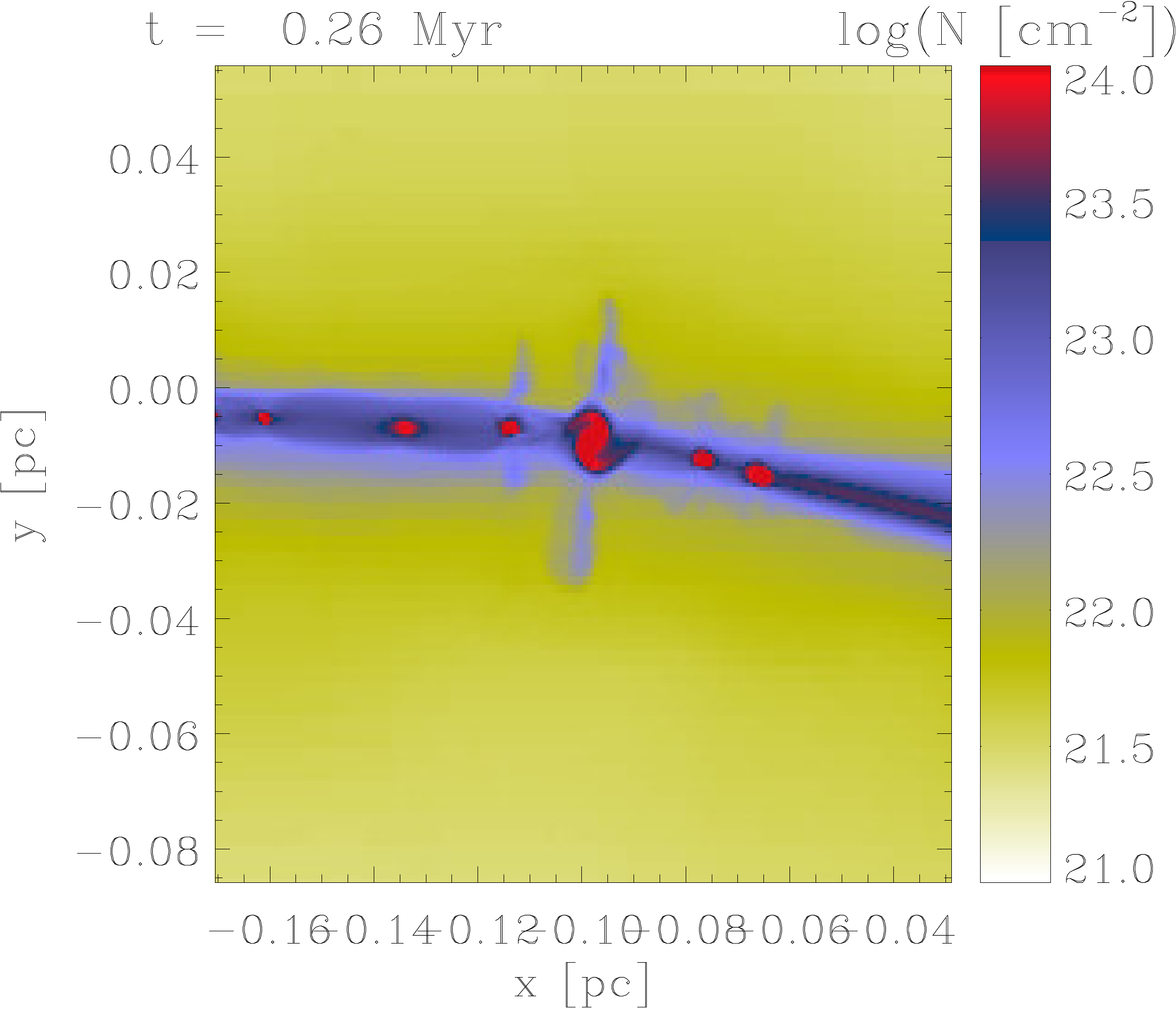} \\
		\rotatebox[origin=l]{90}{ML1.6-M2.0-Perp}&\includegraphics[width=0.3\textwidth]{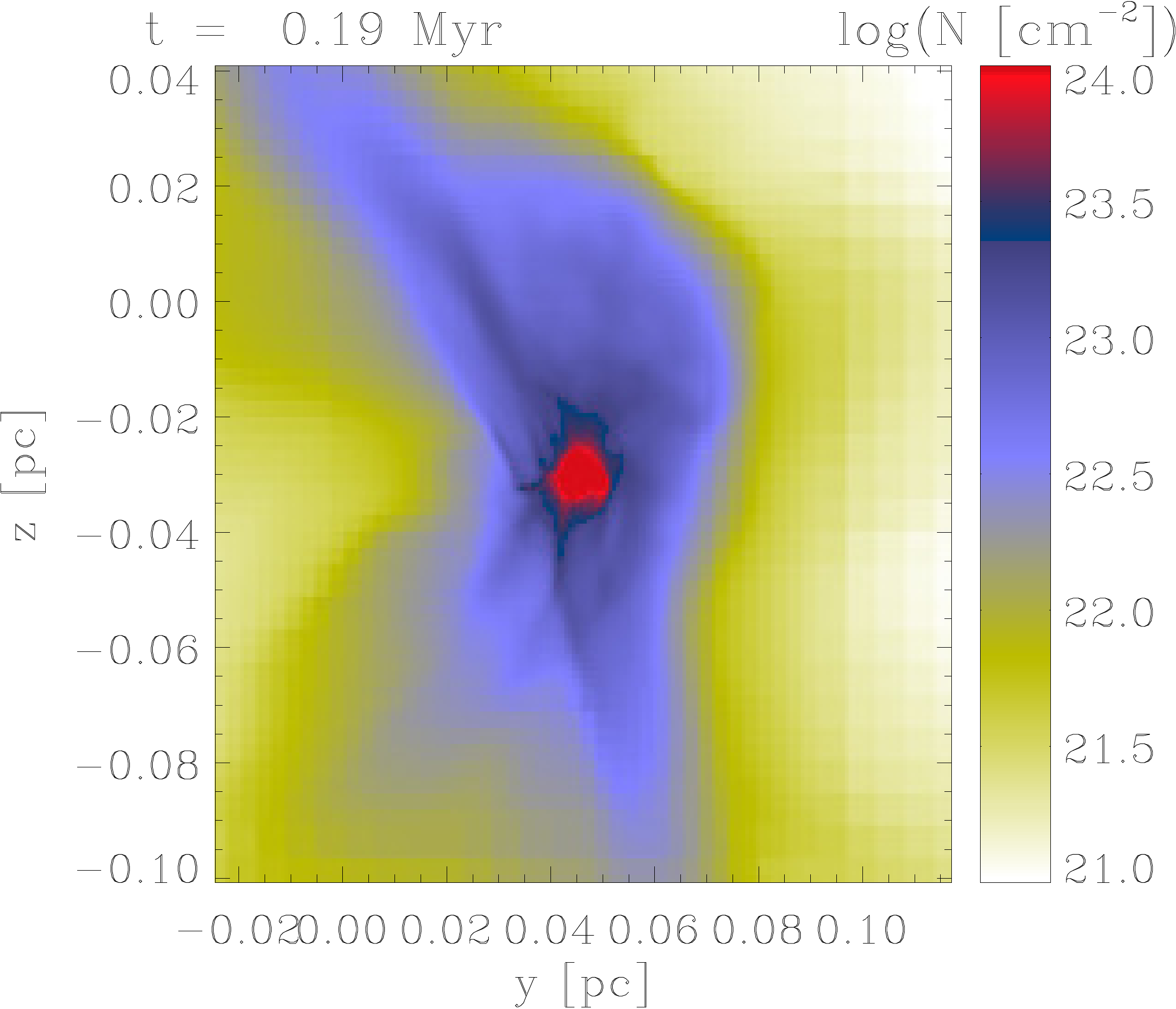} &\includegraphics[width=0.3\textwidth]{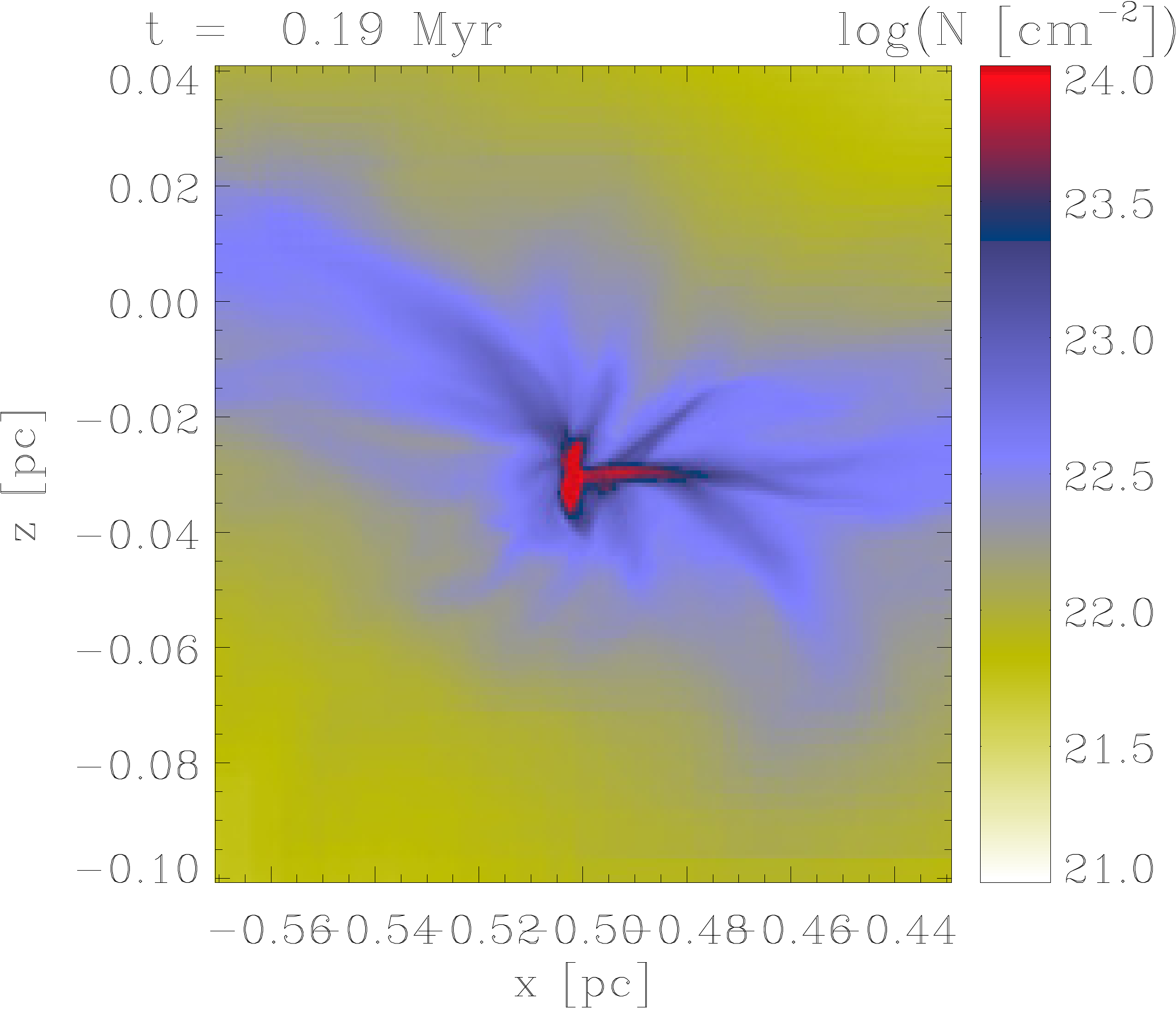} &\includegraphics[width=0.3\textwidth]{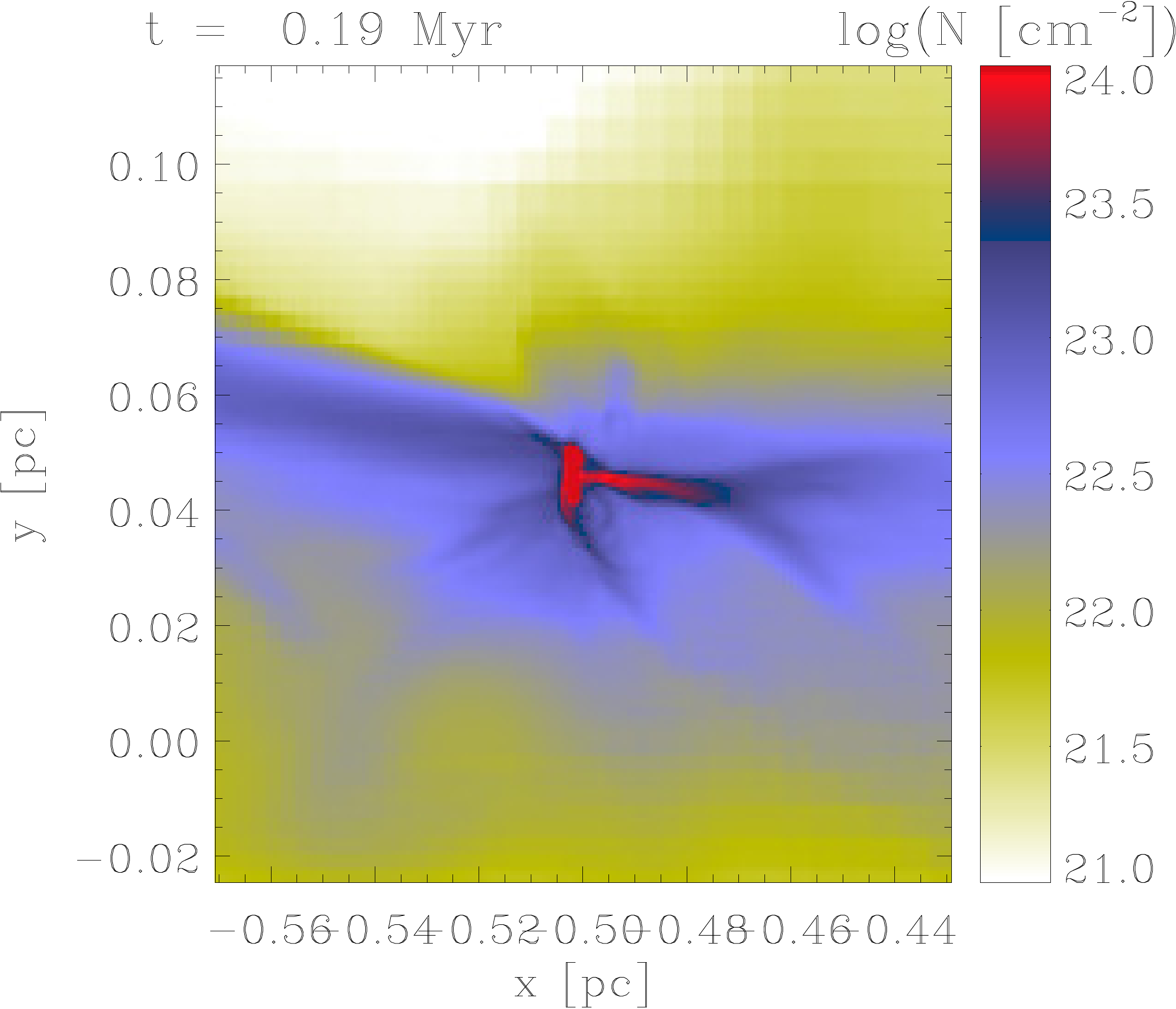} \\
		\rotatebox[origin=l]{90}{ML1.6-M2.0-Para}&\includegraphics[width=0.3\textwidth]{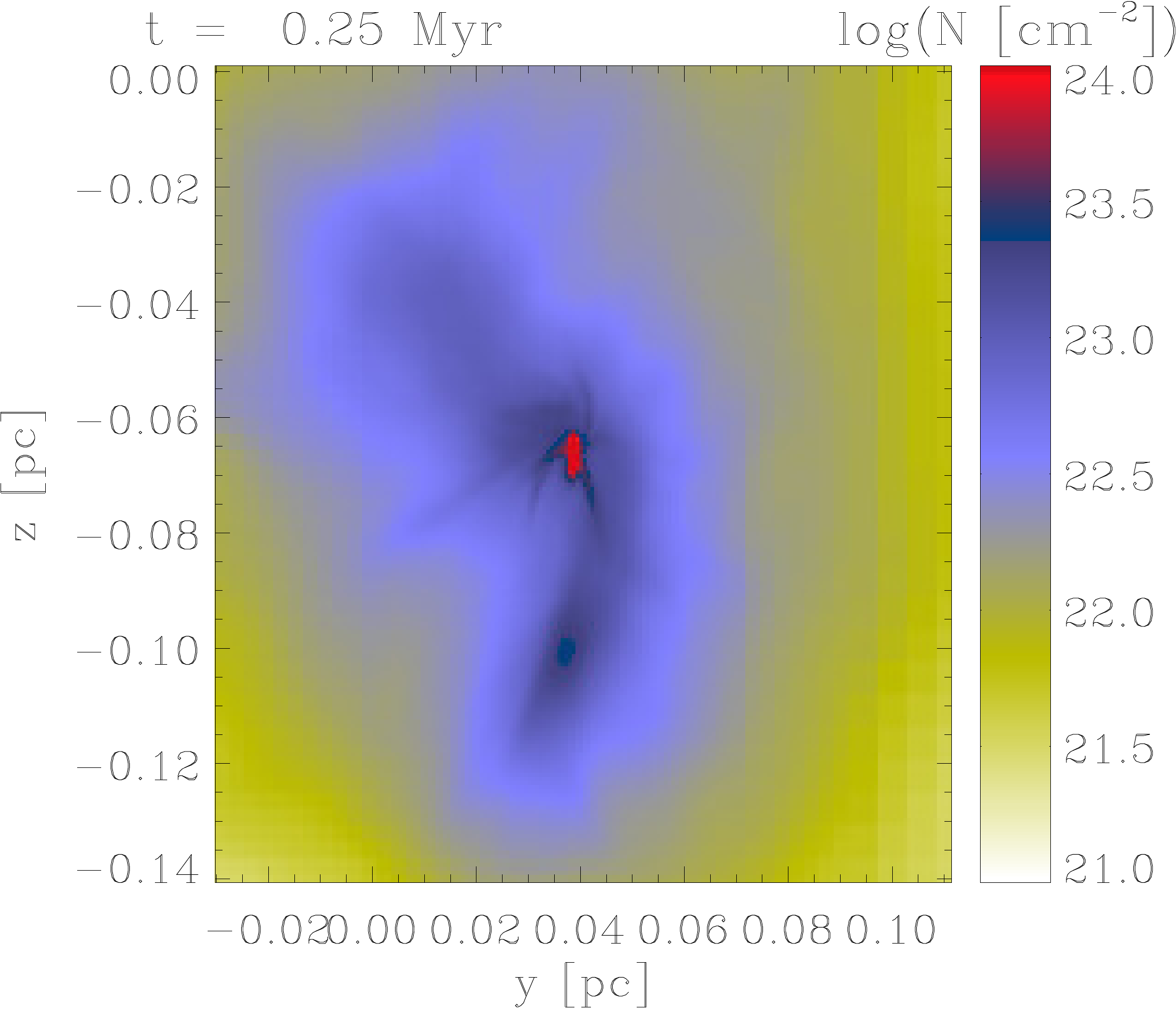} &\includegraphics[width=0.3\textwidth]{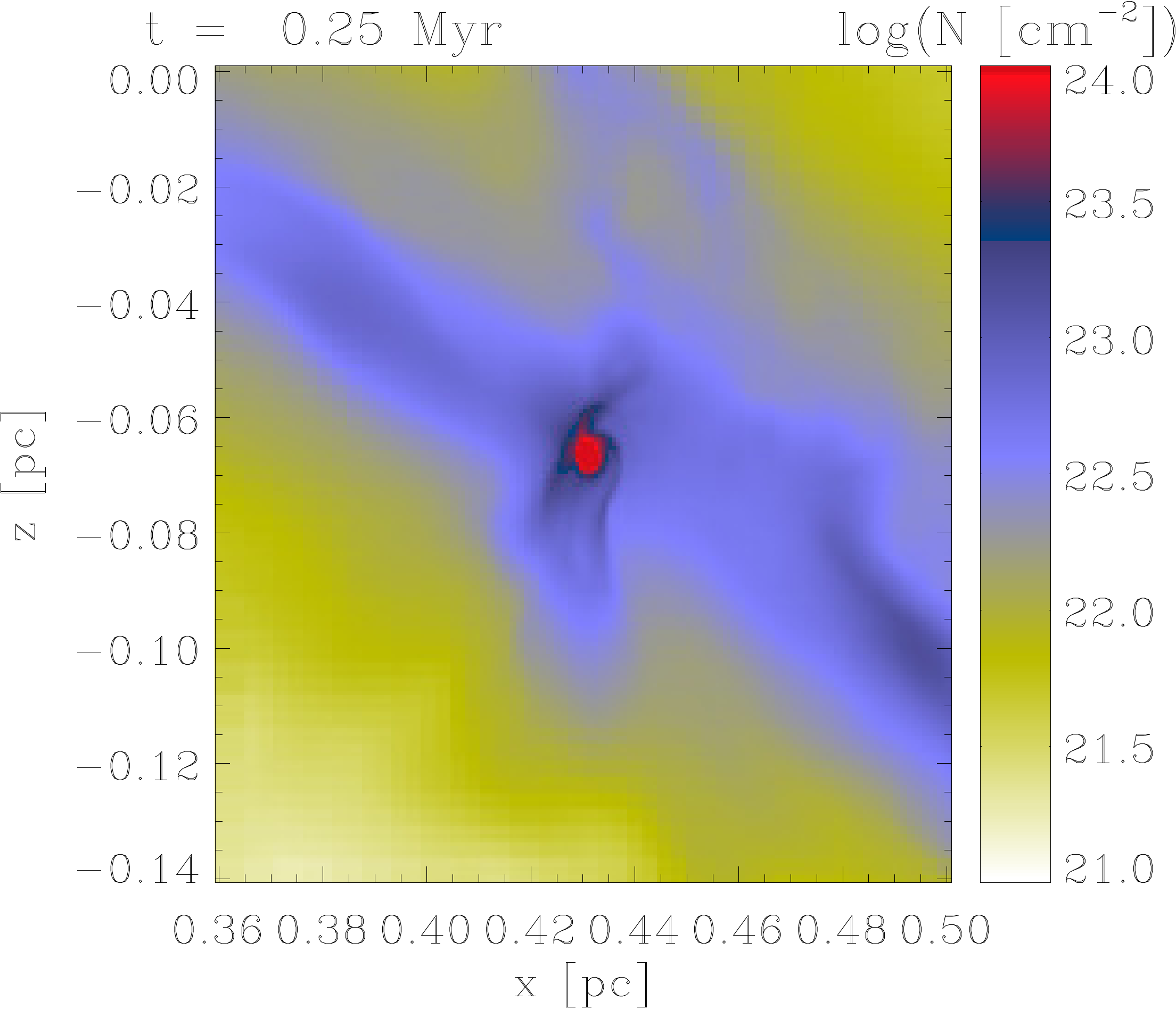} &\includegraphics[width=0.3\textwidth]{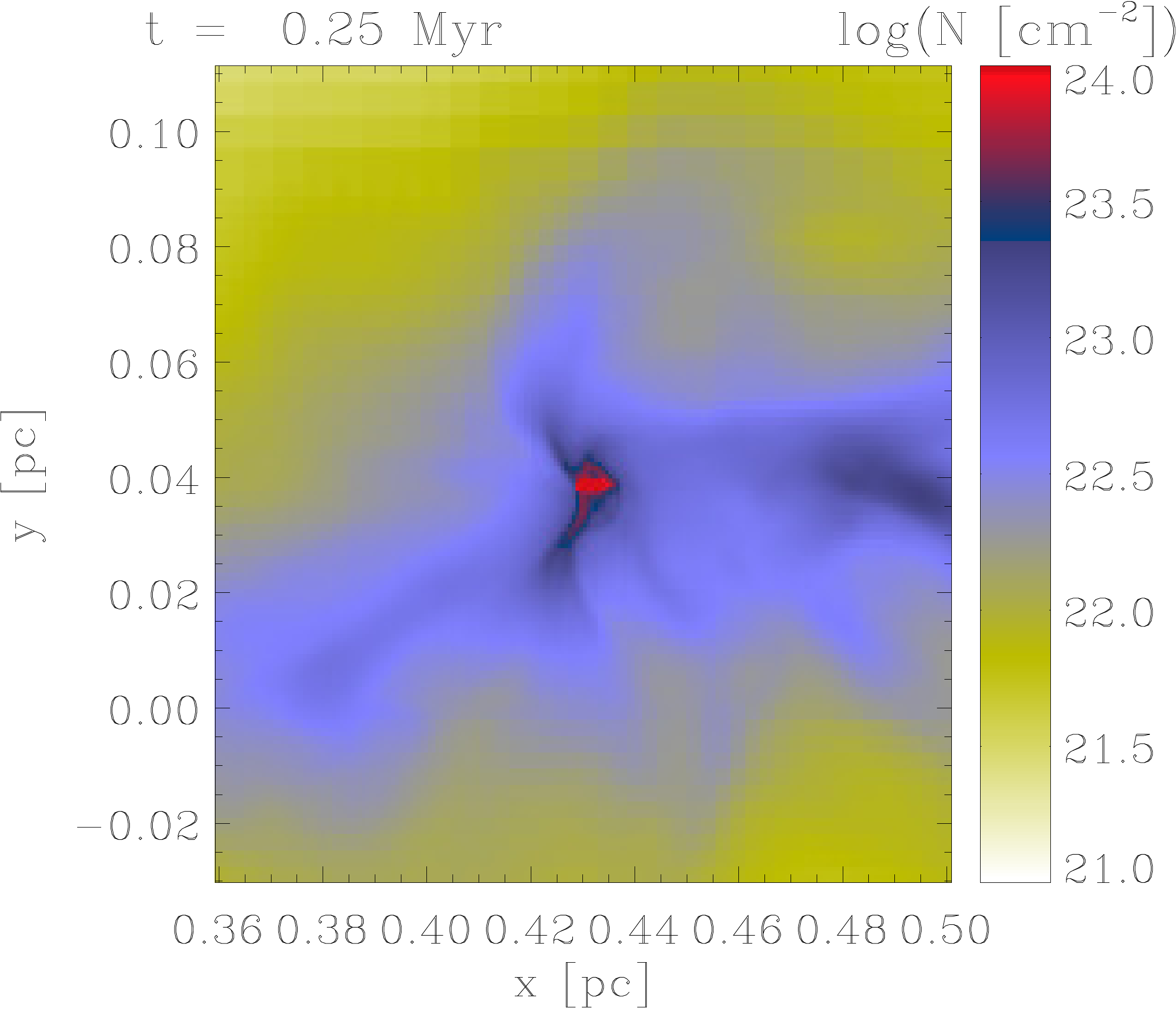} \\
		\rotatebox[origin=l]{90}{ML1.6-M4.0-Para}&\includegraphics[width=0.3\textwidth]{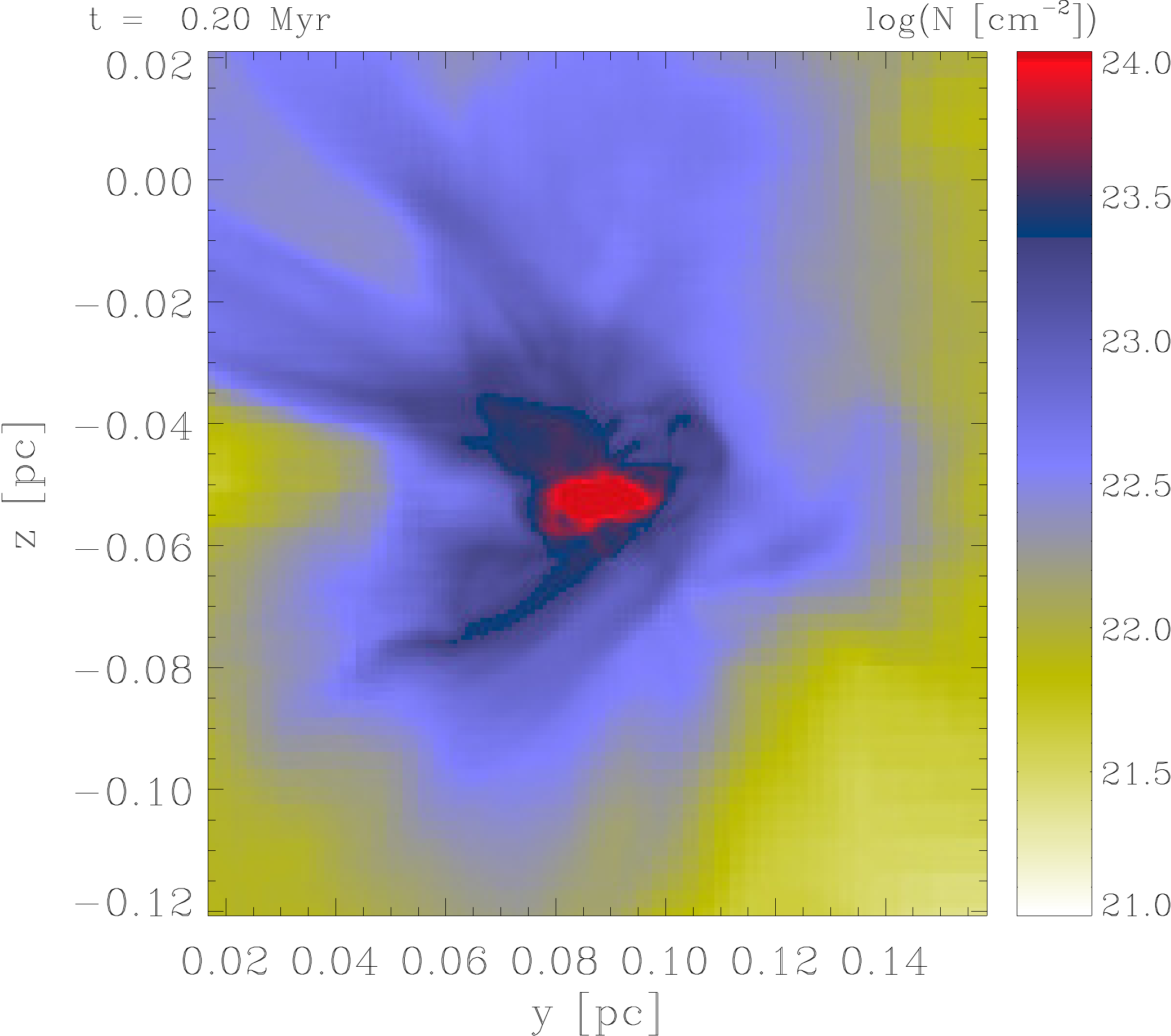} &\includegraphics[width=0.3\textwidth]{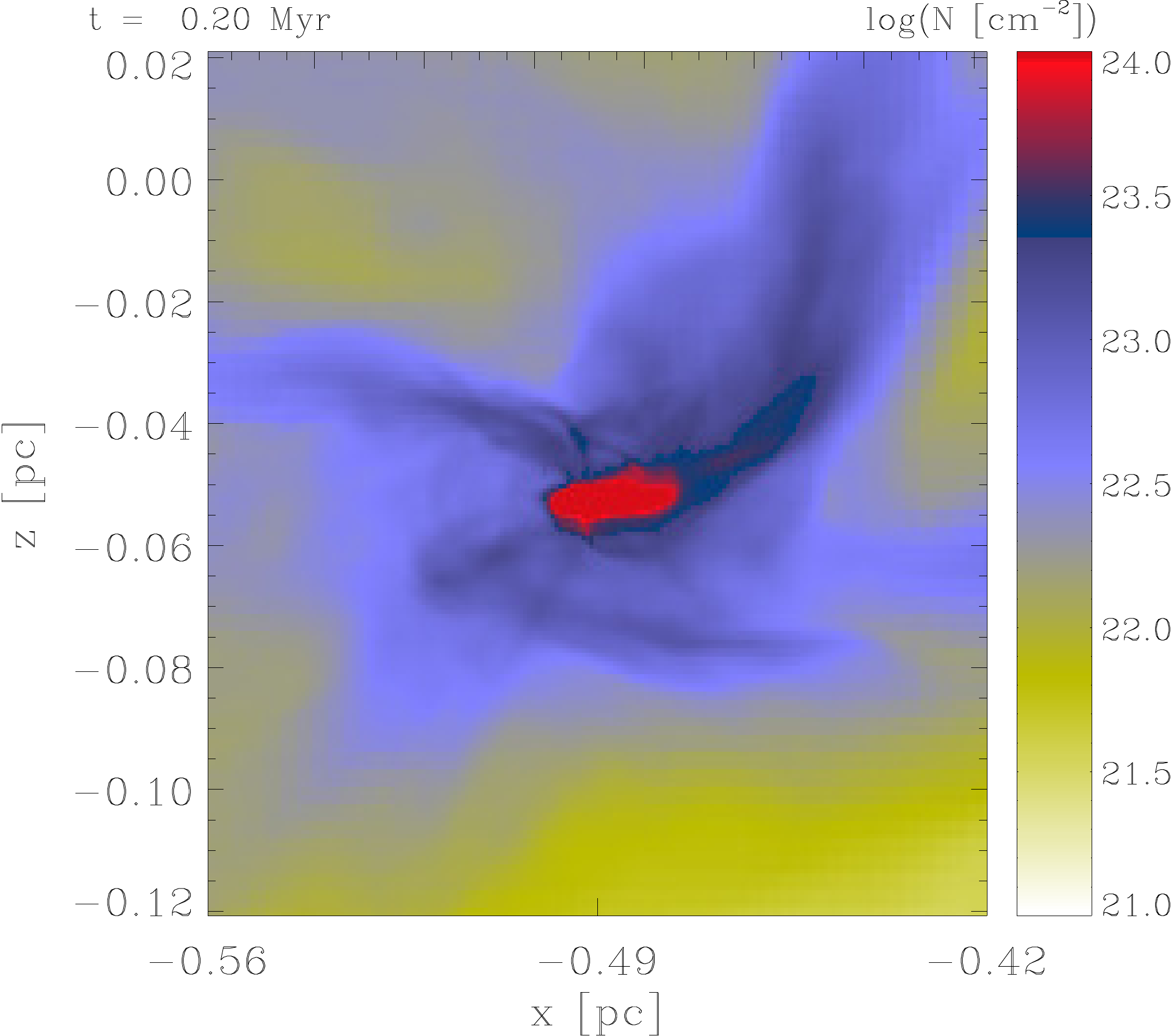} &\includegraphics[width=0.3\textwidth]{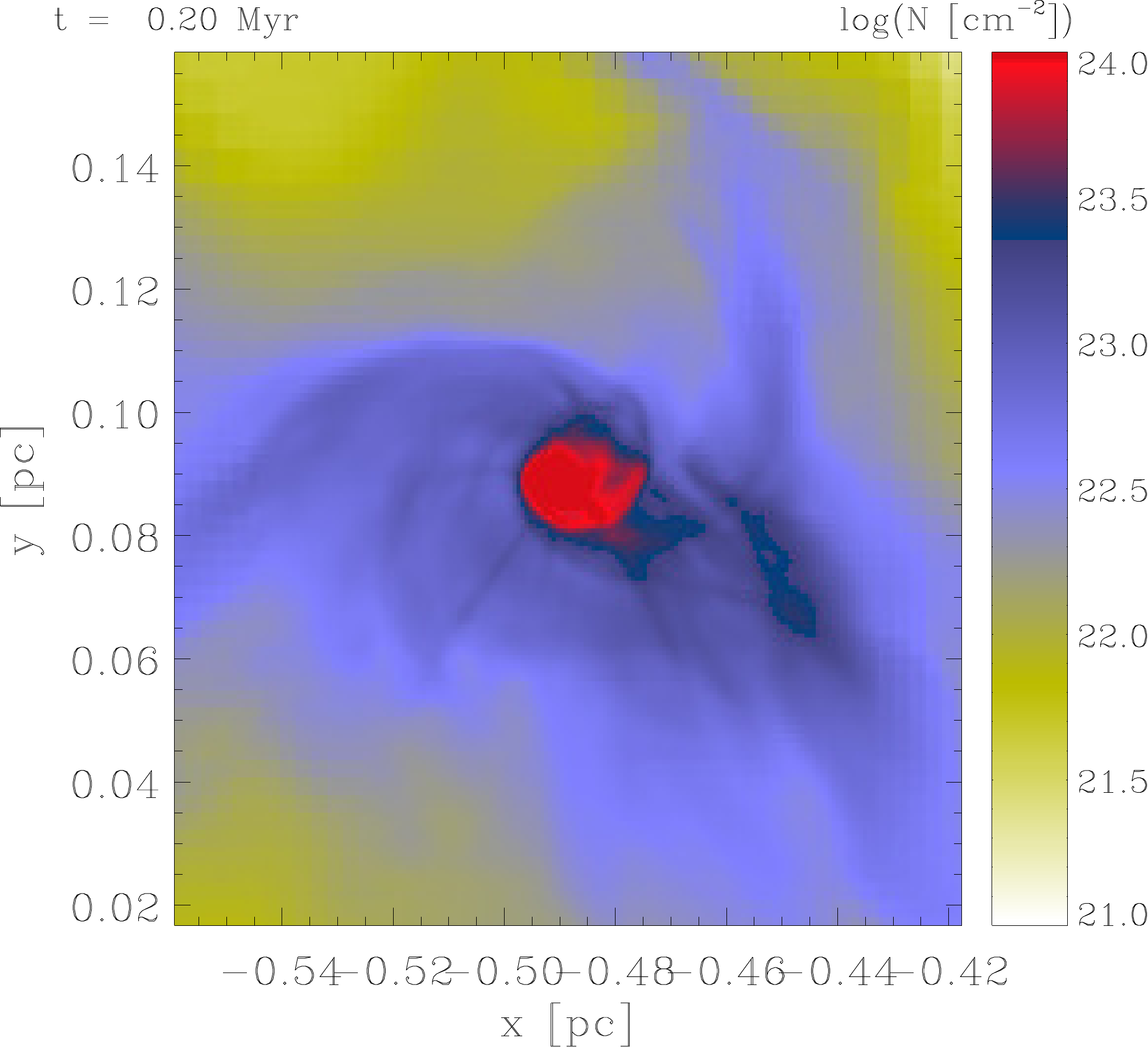} \\
	\end{tabular}
	\caption{Zoom-in to selected cores in different filaments at the end of the simulation. Shown is the column density along different directions. Integration length in all cases is $l=0.2\,\mathrm{pc}$. All cores reveal an oblate geometry. Only with initially supersonic turbulence, the core environment becomes 
	highly irregular and dynamic. Note the finger-like structures of enhanced column density near the cores.}
\label{clumpstruct_cdens}
\end{figure*}

\begin{figure*}
	\begin{tabular}{cccc}
		\rotatebox[origin=l]{90}{ML1.6-M0.6-Perp}&\includegraphics[width=0.3\textwidth]{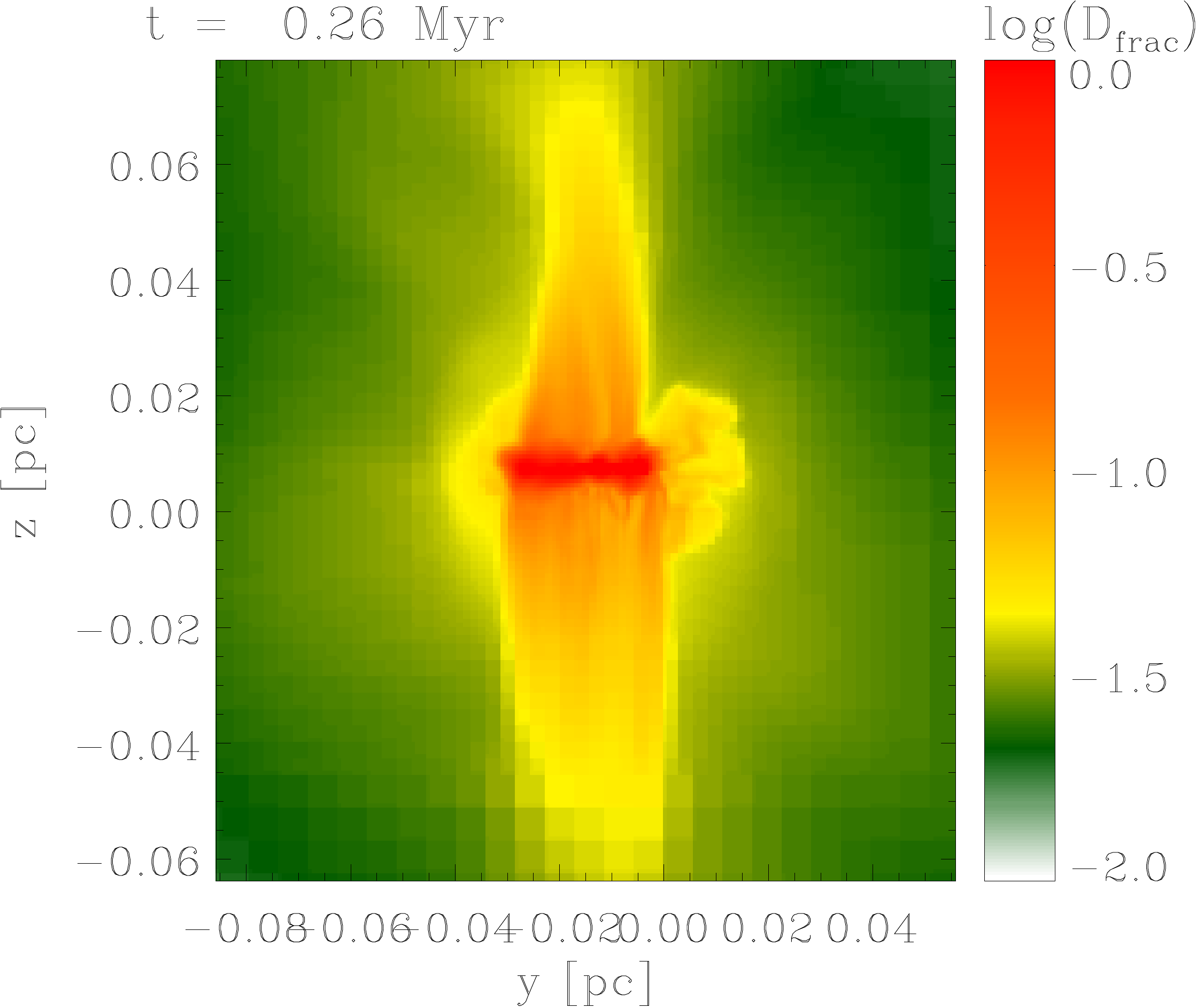} &\includegraphics[width=0.3\textwidth]{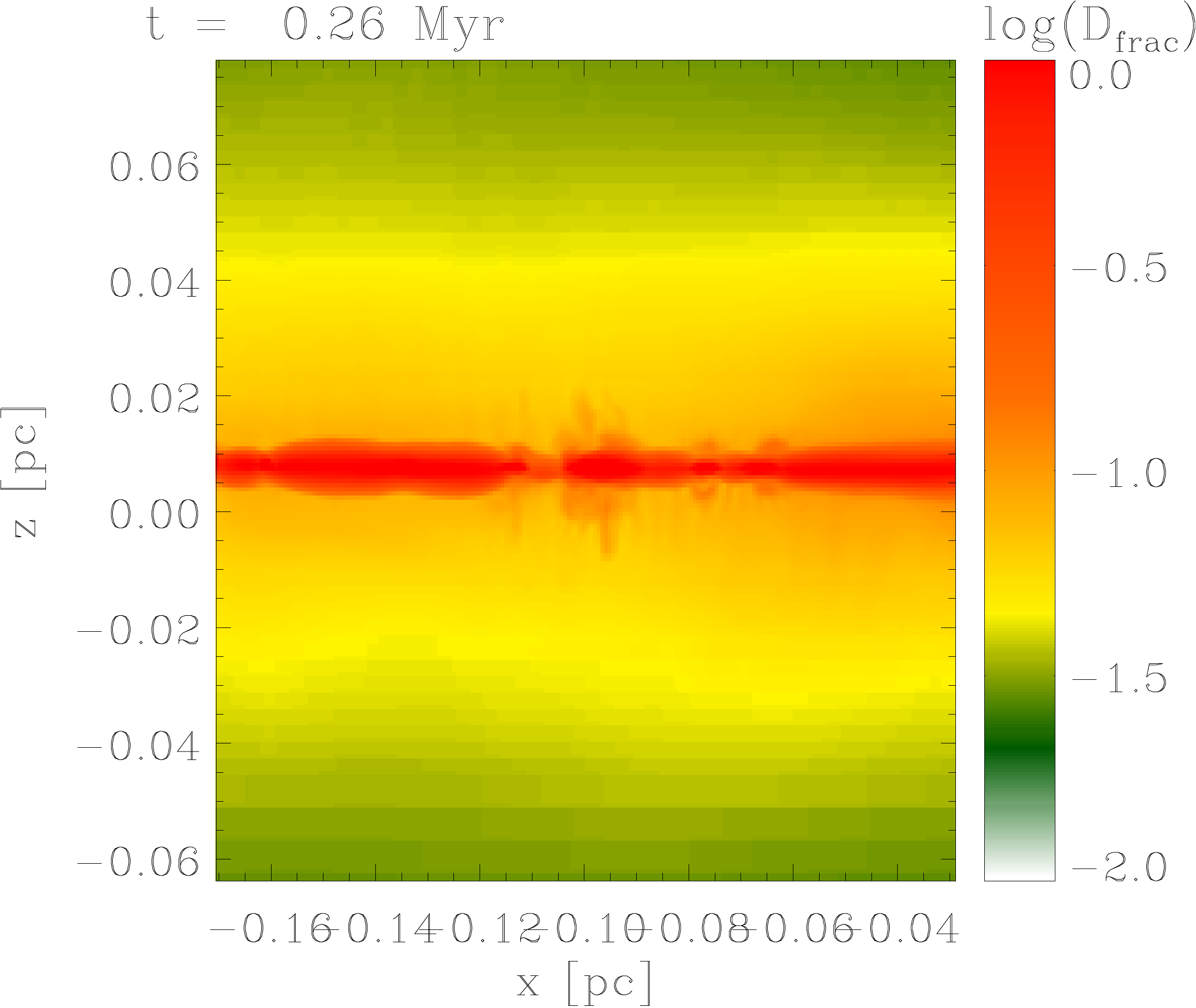} &\includegraphics[width=0.3\textwidth]{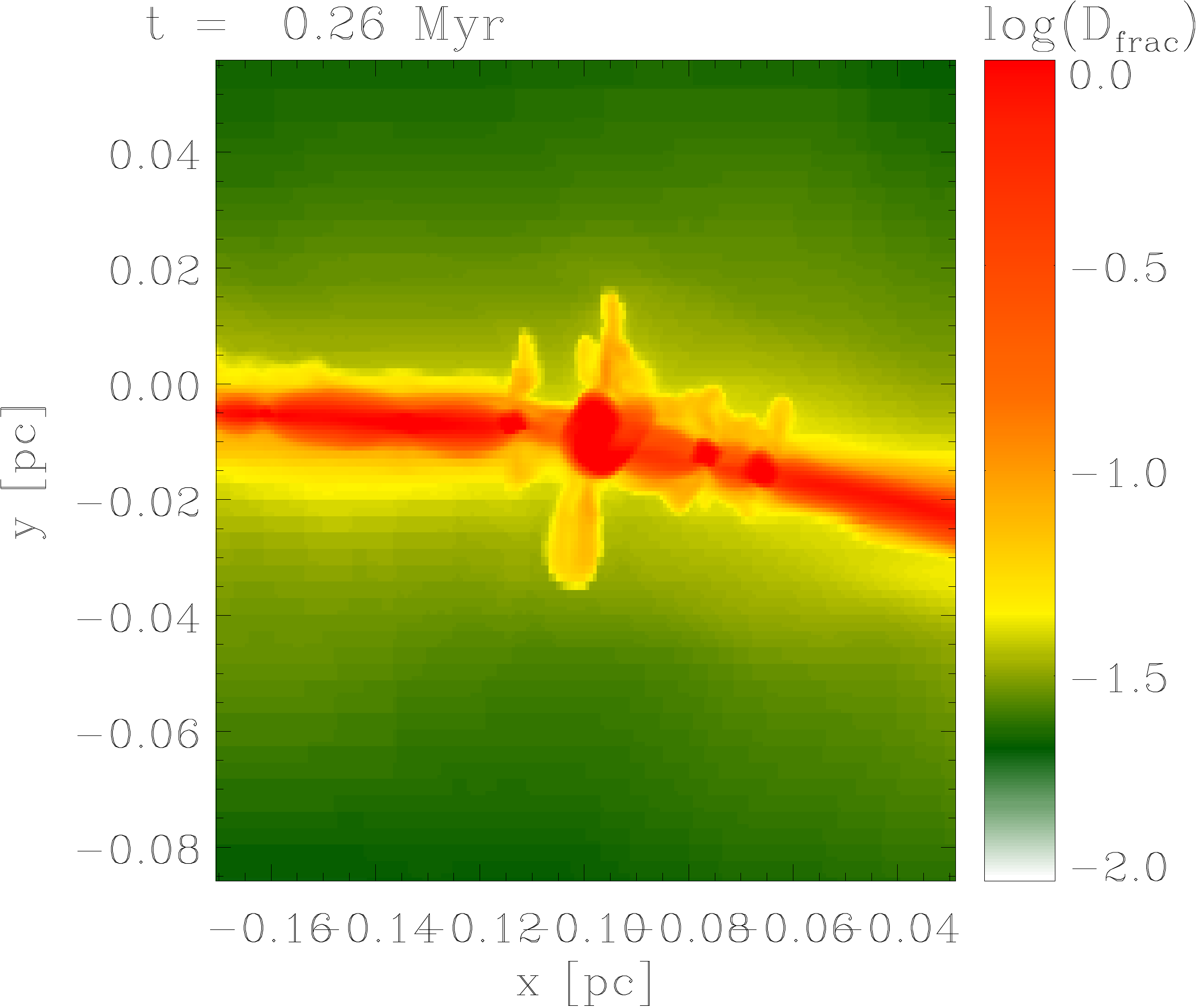} \\
		\rotatebox[origin=l]{90}{ML1.6-M2.0-Perp}&\includegraphics[width=0.3\textwidth]{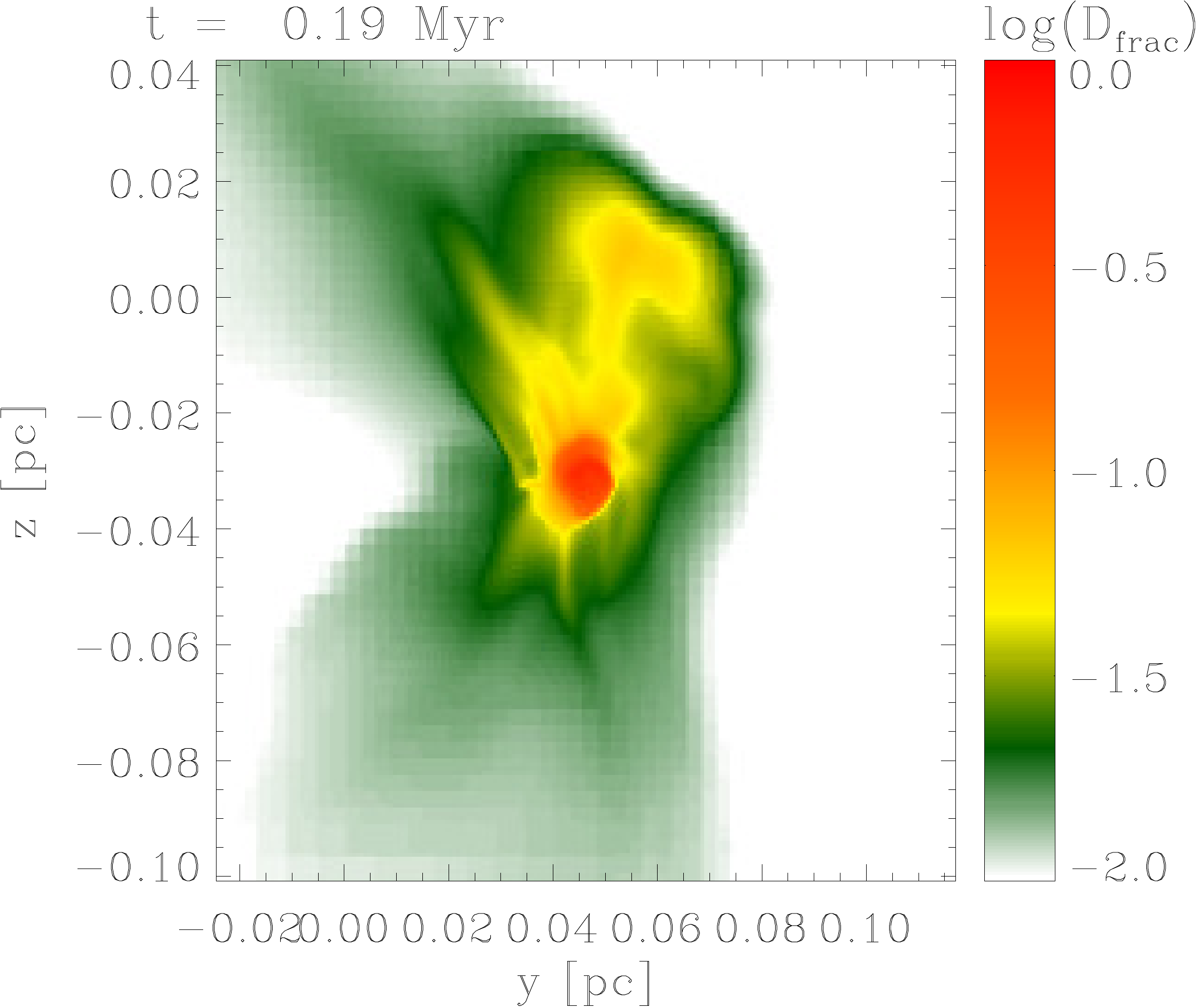} &\includegraphics[width=0.3\textwidth]{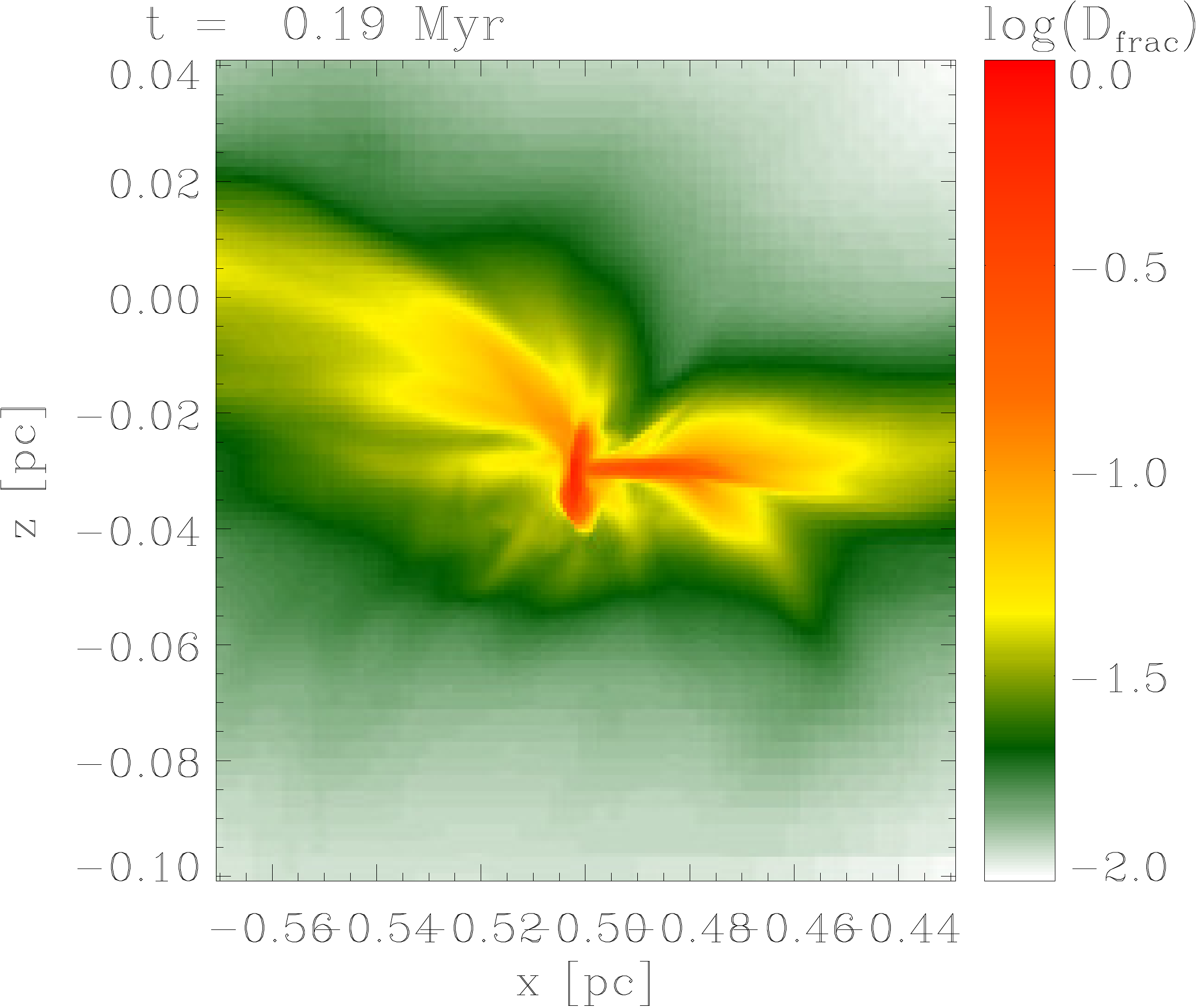} &\includegraphics[width=0.3\textwidth]{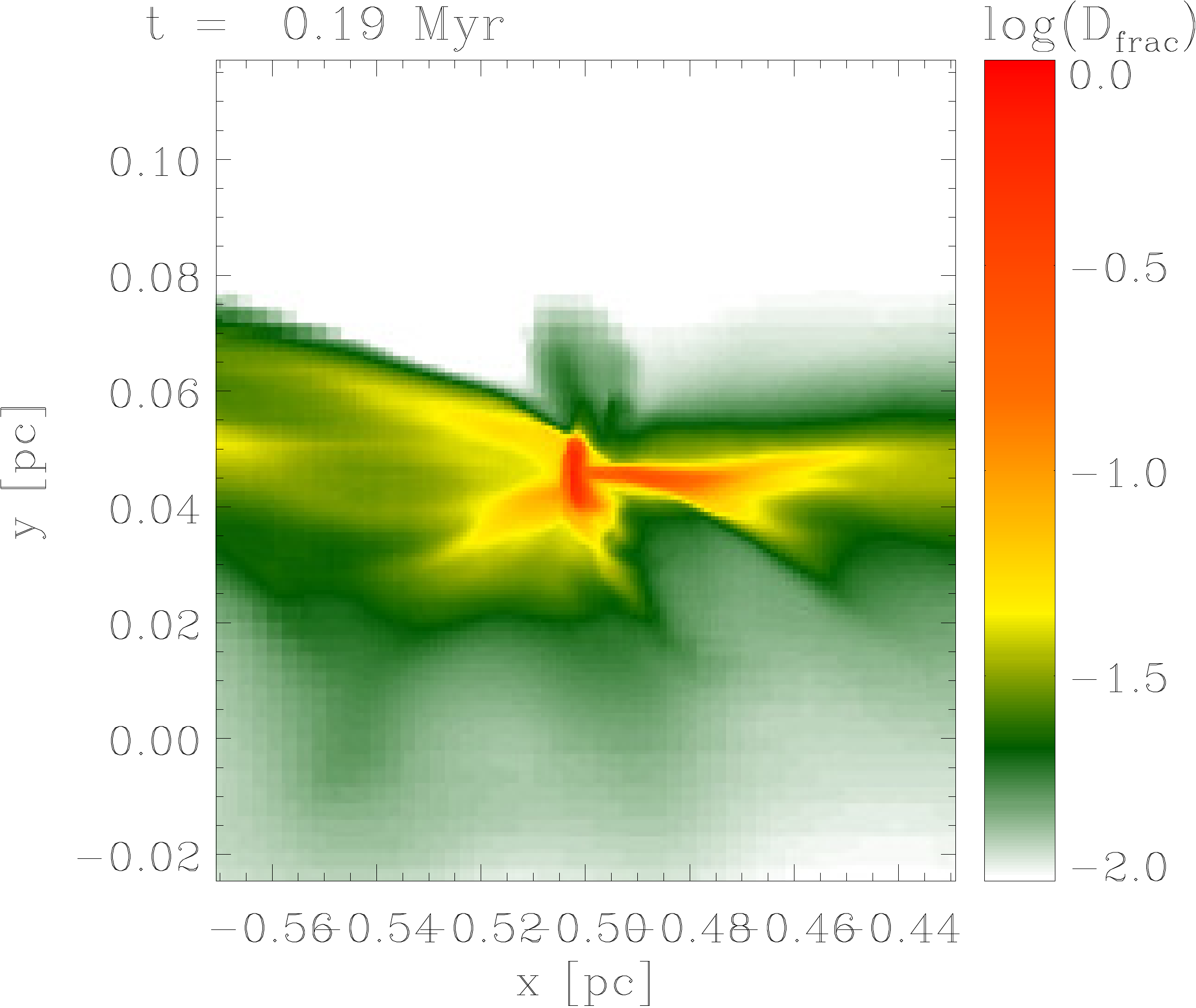} \\
		\rotatebox[origin=l]{90}{ML1.6-M2.0-Para}&\includegraphics[width=0.3\textwidth]{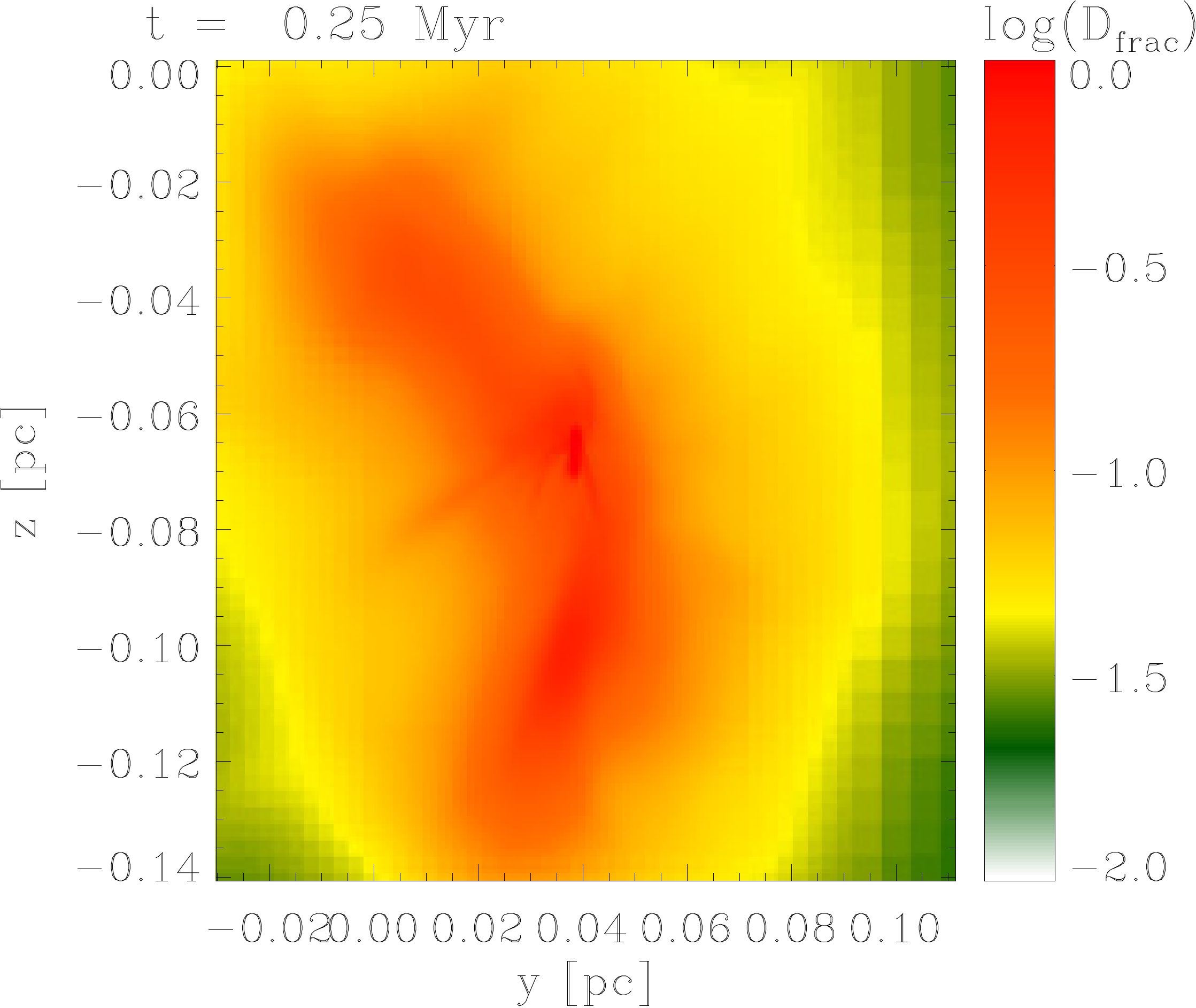} &\includegraphics[width=0.3\textwidth]{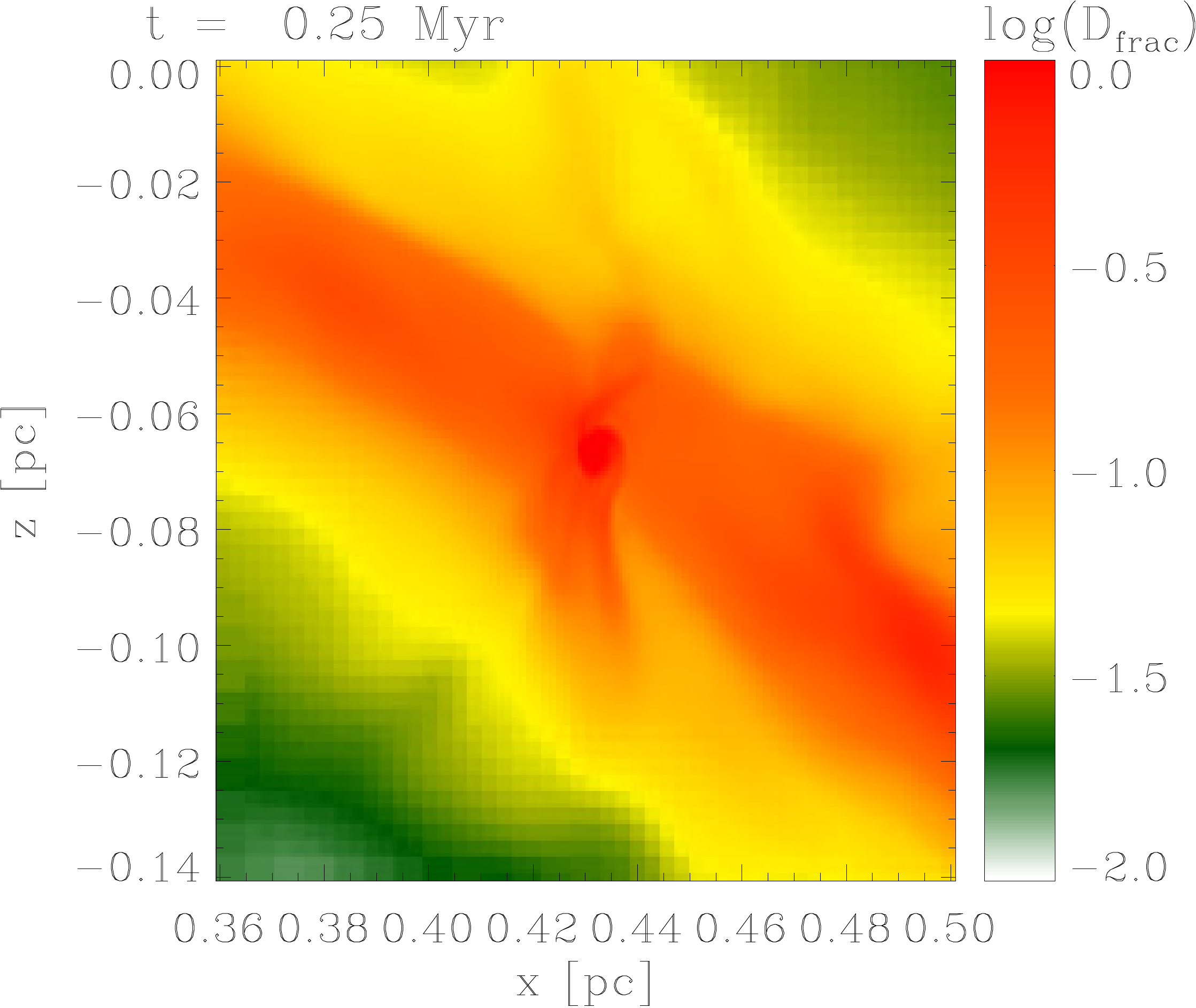} &\includegraphics[width=0.3\textwidth]{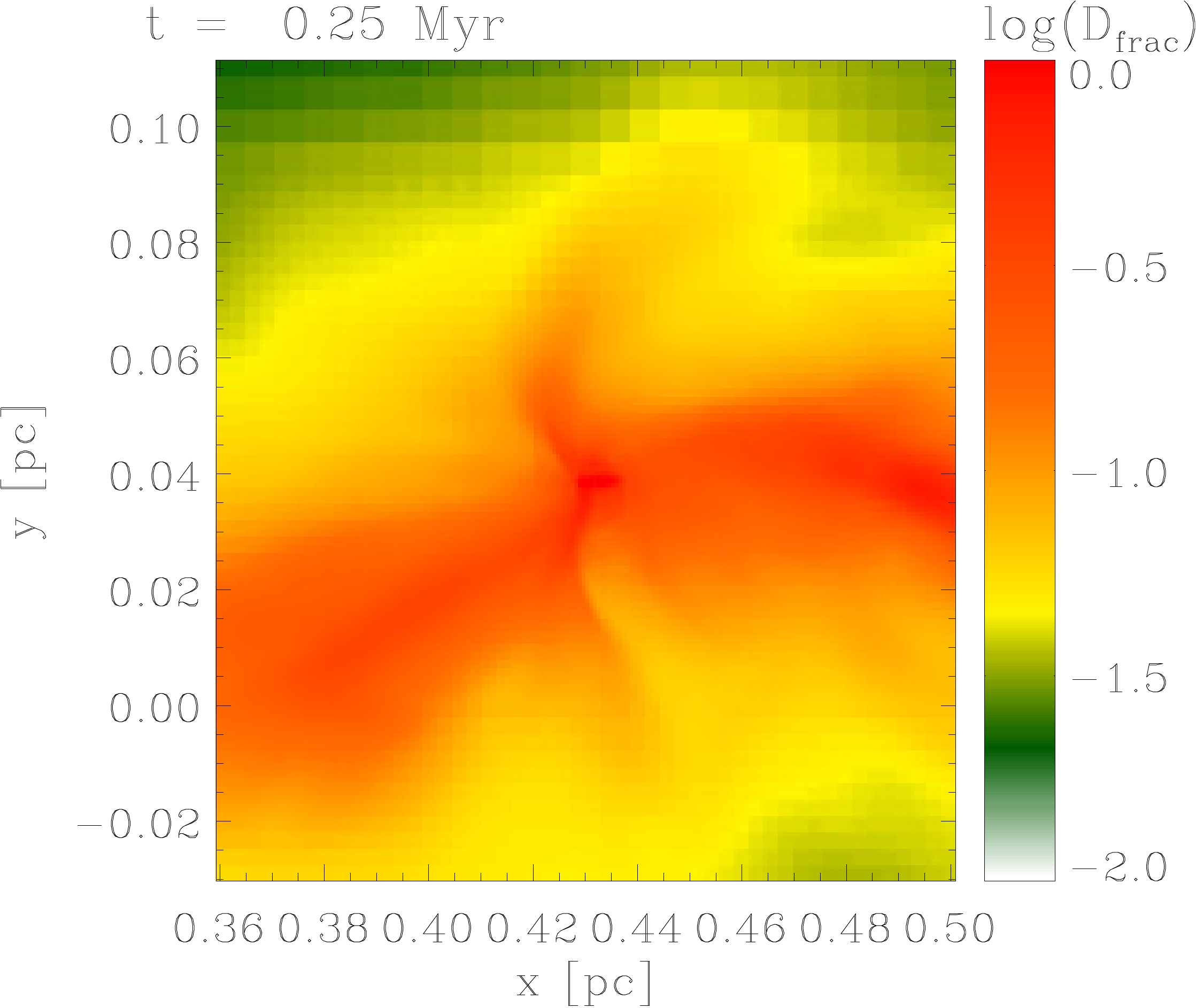} \\
		\rotatebox[origin=l]{90}{ML1.6-M4.0-Para}&\includegraphics[width=0.3\textwidth]{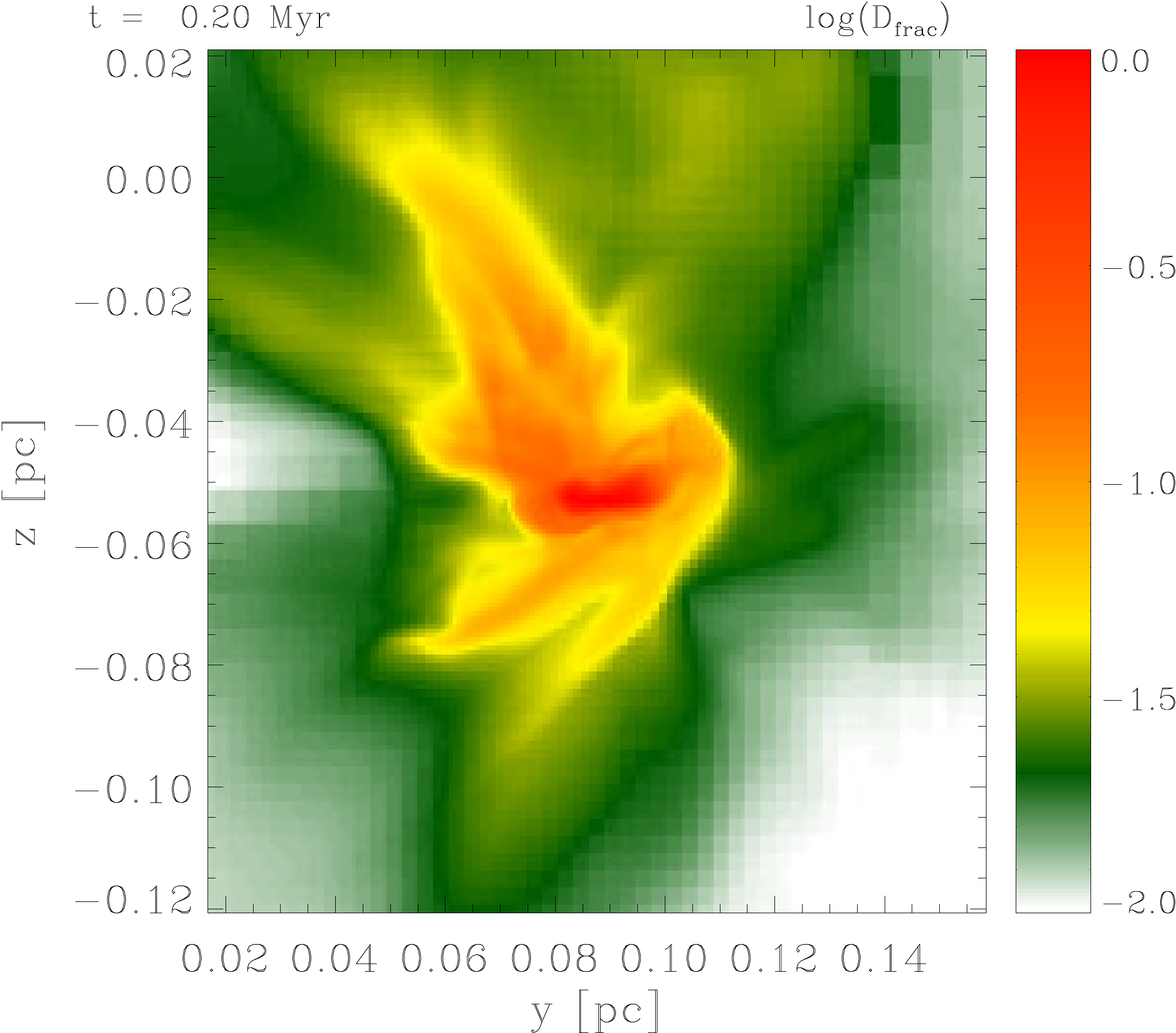} &\includegraphics[width=0.3\textwidth]{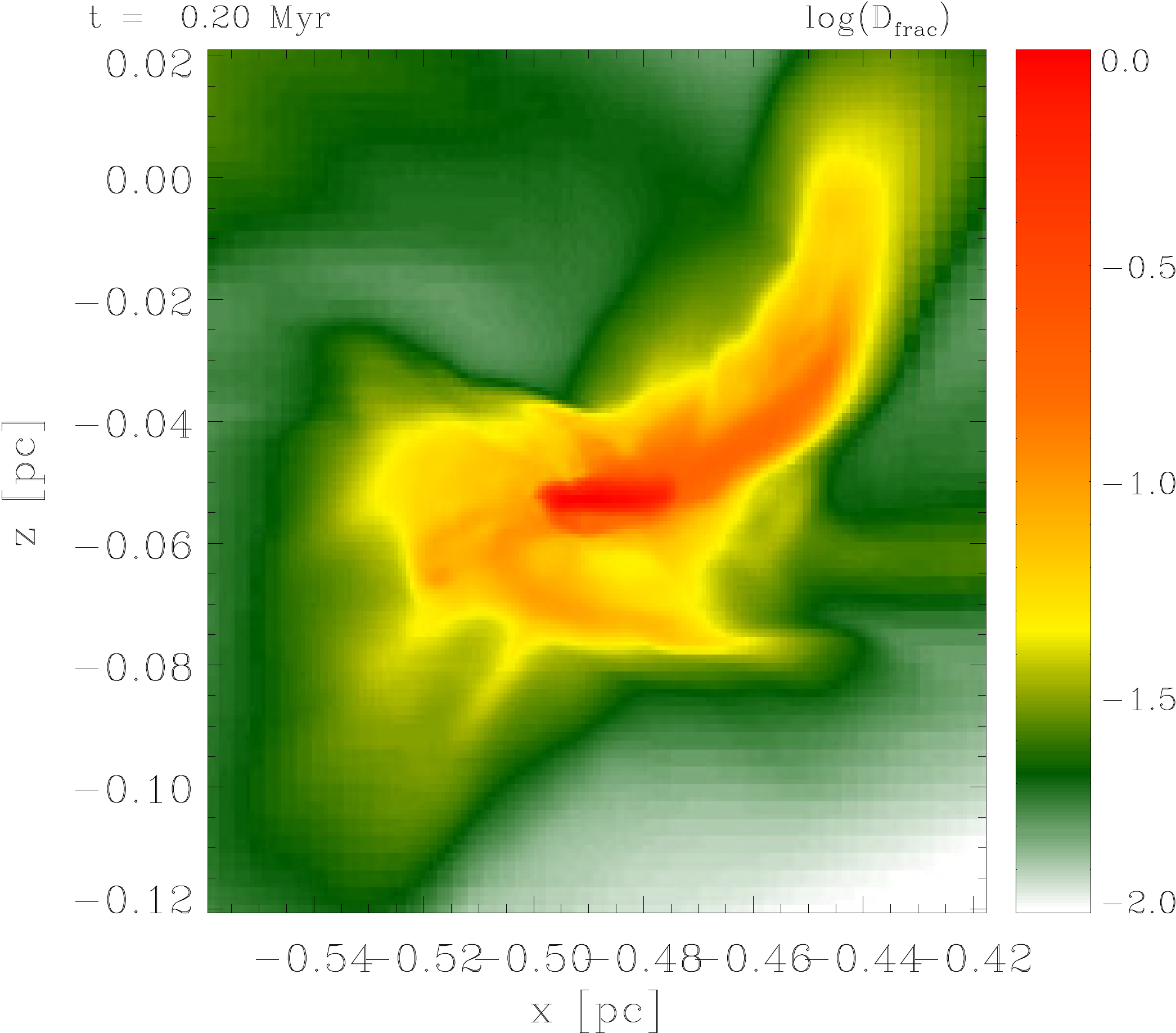} &\includegraphics[width=0.3\textwidth]{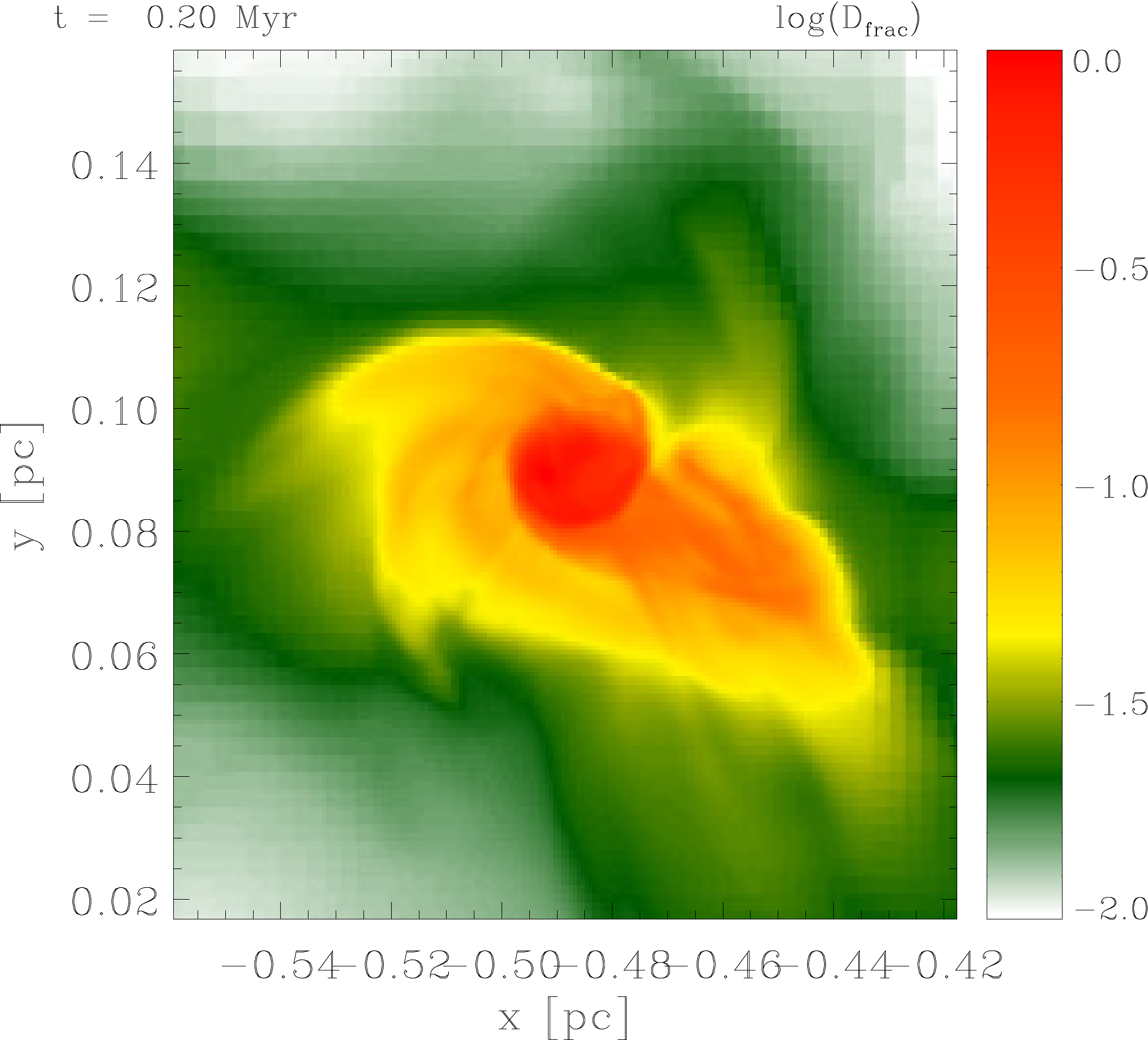} \\
	\end{tabular}
	\caption{Deuteration fraction of selected cores in different filaments at the end of the simulation. Shown is the deuterium fraction along different directions. Integration length in all cases is $l=0.2\,\mathrm{pc}$. The previously mentioned high-density channels are also regions of greatly 
	enhanced deuterium fraction, giving rise to accretion of highly deuterated material.}
\label{clumpstruct_dfrac}
\end{figure*}

To illustrate how the cores in our simulations evolve over time, we subsequently focus on a set of randomly selected cores that are summarized in Table~\ref{tabCoreParams}. We only consider cores from filaments that have been evolved for at 
least $190$~kyrs in order to have a suitable amount of time evolution, which covers the initial as well as the more evolved stages of the filament.\\
\begin{table*}
	\caption{Overview of timescales and density of selected cores for the different filaments. The second column denotes the time when the core is being identified for the first time. The third and fourth column give average core density and free-fall time at first identification point. The following two columns denote the end of the simulation and the evolution of the core in number of free-fall times (calculated using average density from col. 3). The last two columns state the maximum average deuteration fraction in 
	the cores and the number of free-fall times to reach this value.}
	\begin{tabular}{lccccccc}
	\hline
	\hline
	\fat{Run name}	&\fat{Identification}	&\fat{Average}	&\fat{Free-fall}	&\fat{Ending}	&\fat{Number of}	&\fat{Maximum mean}	&\fat{Free-fall times}\\
				&\fat{Time} &\fat{Density} &\fat{Time }	&\fat{Time}	&\fat{Free-fall times}			&D$_\mathrm{frac}$	&\fat{to} D$_\mathrm{frac,max}$\\
				&$\left[\mathrm{kyr}\right]$&$\left[\mathrm{cm}^{-3}\right]$&$\left[\mathrm{kyr}\right]$	&$\left[\mathrm{kyr}\right]$ & & &\\
	\hline
	ML0.8-M2.0-Perp	&65.4	&$7.8\times10^3$	&376.66	&351		&0.76	&0.068	&0.68\\
	ML1.6-M0.6-Perp	&130		&$2.7\times10^5$	&63.90	&253		&1.92	&0.054	&1.92\\
	ML1.6-M2.0-Perp	&55.5	&$1.0\times10^5$	&103.50	&193		&1.33	&0.019	&1.33\\
	ML1.6-M2.0-Para	&68.7	&$1.5\times10^5$	&86.50	&257		&2.18	&0.099	&2.16\\
	ML1.6-M4.0-Para	&49.1	&$6.2\times10^4$	&133.72	&202		&1.14	&0.027	&1.14\\
	\hline
	\end{tabular}
	\label{tabCoreParams}
\end{table*}
In Fig.~\ref{figCoreTime1}, we show the average number density and mass of the cores as a function of time. Most of them do significantly increase as a function of time, with initial number densities of $10^5$~cm$^{-3}$ and later values of up to $10^7$~cm$^{-3}$, and masses from initially $0.01-0.1$~M$_\odot$ up to later $10-100$~M$_\odot$, during an evolutionary time of about $200$~kyrs. Both quantities show significant scatter, arising from the somewhat arbitrary definition of the cores and the density fluctuations that will naturally occur during the evolution. The latter point indeed emphasizes that the cores within the filaments are highly dynamic and not static objects. We also see that in one case, corresponding to ML0.8-M2.0-Perp, both number density and mass of the clump are not increasing. The latter thus corresponds to an object that is strongly stabilized by magnetic fields (the core's Alfv\'{e}n Mach number is $\sim0.2$ throughout its evolution), and will potentially never lead to star formation.\\
To better understand the subsequent further evolution of the cores, we now turn to Fig.~\ref{figCoreTime2}, showing the virial parameter and the mass-to-flux ratio as a function of time. These quantities vary again strongly between the cores, both in terms of their initial values as well as in their evolution.  In two cases, for ML0.8-M2.0-Perp and ML1.6-M4.0-Para, the virial parameter never drops below 1, suggesting that they may never go through gravitational collapse. The second case is partly reflected in Fig.~\ref{figCoreTime1}, where the density increase over $200$~kyrs corresponded to a factor of $3-4$, implying at best a very slow collapse. Interestingly, we nevertheless see no difference in the deuteration fraction for this case. The other cases drop to values close to or below a virial ratio of $1$ during their evolution, and are thus gravitationally unstable. However, also for these cores, the scatter in the time evolution of the dynamical parameters is rather large.\\
A similar behavior can be seen in the mass-to-flux ratio within the same figure, where all cores start out with mass-to-flux ratios of $0.01-0.1$. Consistent with the results above, ML0.8-M2.0-Perp remains at this low value,  while ML1.6-M4.0-Para at least comes close to a value of $1$ (and crosses the value temporarily). All other cores show a growing mass-to-flux ratio with time, explained by their rapid collapse as visible through the density evolution. We therefore conclude that the initial mass-to-flux ratio of a clump does not necessarily reflect this ratio at later stages of the evolution.\\
We now explore the chemical properties in Fig.~\ref{figCoreTime3}, showing the mass-weighted deuteration fraction and H$_2$ ortho-to-para ratio as a function of time. We find a similar evolution in almost all cases, with slight differences only in the case of ML0.8-M2.0-Perp, where the evolution of mass and density is considerably different from the other cores. The other cores (and even still this one) show a remarkable similarity in their chemical evolution, with the H$_2$ ortho-to-para ratio steadily decreasing as a function of time, with an initial value of $\sim0.3$ at core formation, down to a value of $0.001$ after about $250$~kyrs of evolution. These values are in agreement with the ones assessed via observations. For instance, an ortho-to-para ratio of 0.1 was found in the pre-stellar core L183 \citep{Pagani2009_2}, and 10$^{-4}$ toward IRAS 16293-2422 \citep{Brunken2014}.\\
The decrease of the ortho-to-para ratio of H$_2$ over time in turn also boosts the deuterium fractionation, where this ratio starts at a value of $\sim0.001$ at core formation, and reaches values of $\sim0.1$ after about $250$~kyrs. A similar value is even reached in the non-collapsing core ML0.8-M2.0-Perp, even though on a somewhat longer timescale of $300-350$~kyrs. This is in fact in agreement with the 
observational findings by \citet{Fontani11}, who give an evolutionary sequence for the deuterium enrichment in pre-stellar cores and ultra-compact H\,II regions. In addition, the deuteration fraction in the 
core of run ML0.8-M2.0-Perp starts to saturate, which is due to the fact that the density in the core does not evolve at all.\\
In Fig.~\ref{clumpstruct_cdens} we show column density maps of selected cores in the thermally supercritical filaments at the end of the individual simulation. The filament with initially subsonic turbulence reveals a highly ordered morphology since this filament 
collapsed almost unimpeded along the radial direction. The ambient turbulent fluctuations in the gas still lead to the formation of individual small-scale overdensities with a rather regular spacing. If seen edge-on, these structures seem to accrete material from 
the diffuse environment, thereby generating fingers of increased column density. The selected core can be clearly identified in the leftmost picture as the structure with spiral-features. \\
In contrast, if the initial turbulent fluctuations are supersonic, the filaments become locally highly disordered as can be seen from the other figures. The selected cores now appear to be in isolation as overdense structures surrounded by a high-density halo. 
However, this halo is not spherical and its morphology is linked to the orientation of the ambient magnetic field as can be seen by the morphological difference between the cores with parallel or perpendicular magnetic field. We emphasize that in all cases the 
cores resemble oblate rather than spherical objects due to significant angular momentum in this region. \\
All cores reveal an order of magnitude or larger increase in column density with respect to their ambient medium. However, taking into account their virial parameter and mass-to-flux ratio, as shown in Fig.~\ref{figCoreTime3}, indicates that these objects are still not gravitationally unstable, but are in an equilibrium configuration, where any significant increase in density/mass will drive them unstable. \\
The deuteration fraction of the selected cores and their environment is shown in Fig.~\ref{clumpstruct_dfrac}. Similar to the column density maps, the filament in run ML1.6-Ms0.6-Bperp reveals a thin region of greatly enhanced deuteration fraction with values of 
D$_\mathrm{frac}\sim0.5-0.9$. Interestingly, the deuteration appears to be widespread and enhanced also in the more diffuse gas surrounding this filament, as can be seen from the middle panel. However, a closer look to the left panel shows that the area of 
enhanced deuteration fraction only extends along the z-direction. The magnetic field in this filament is initially oriented along the y-direction (i.e. perpendicular to the major axis) and hence this increase can be associated with overdensities formed by converging motions along the field lines 
towards the inner part with a deeper gravitational potential well. The only other filament with enhanced deuterium fractionation in the environment is the one with initial Mach number of $\mathcal{M}=2$ and a parallel oriented magnetic 
field. In this scenario, the field provides support against radial contraction, thus giving the gas enough time to build up a sufficiently large amount of H$_2$D$^+$ compared to H$_3^+$. In addition, turbulent mixing also leads to high deuterium 
fractionation in the more diffuse gas. The other two filaments ML1.6-M2.0-Perp and ML1.6-M4.0-Para only show a small region of enhanced deuteration fraction, while the outskirts reveal values of D$_\mathrm{frac}\sim0.01$. A closer look at the latter 
filament highlights that the material surrounding the core has a slightly higher value of D$_\mathrm{frac}$ due to the stabilizing effect of the (parallel) magnetic field and the larger amount of turbulence. Overall, our results thus demonstrate that deuteration is highly efficient and largely independent of the specific evolution within the cores.


\subsection{Radial profiles of selected cores}
\begin{figure*}
	\begin{tabular}{cccc}
	\rotatebox[origin=r]{90}{ML0.8-M2.0-Perp}&\includegraphics[width=0.25\textwidth,angle=-90]{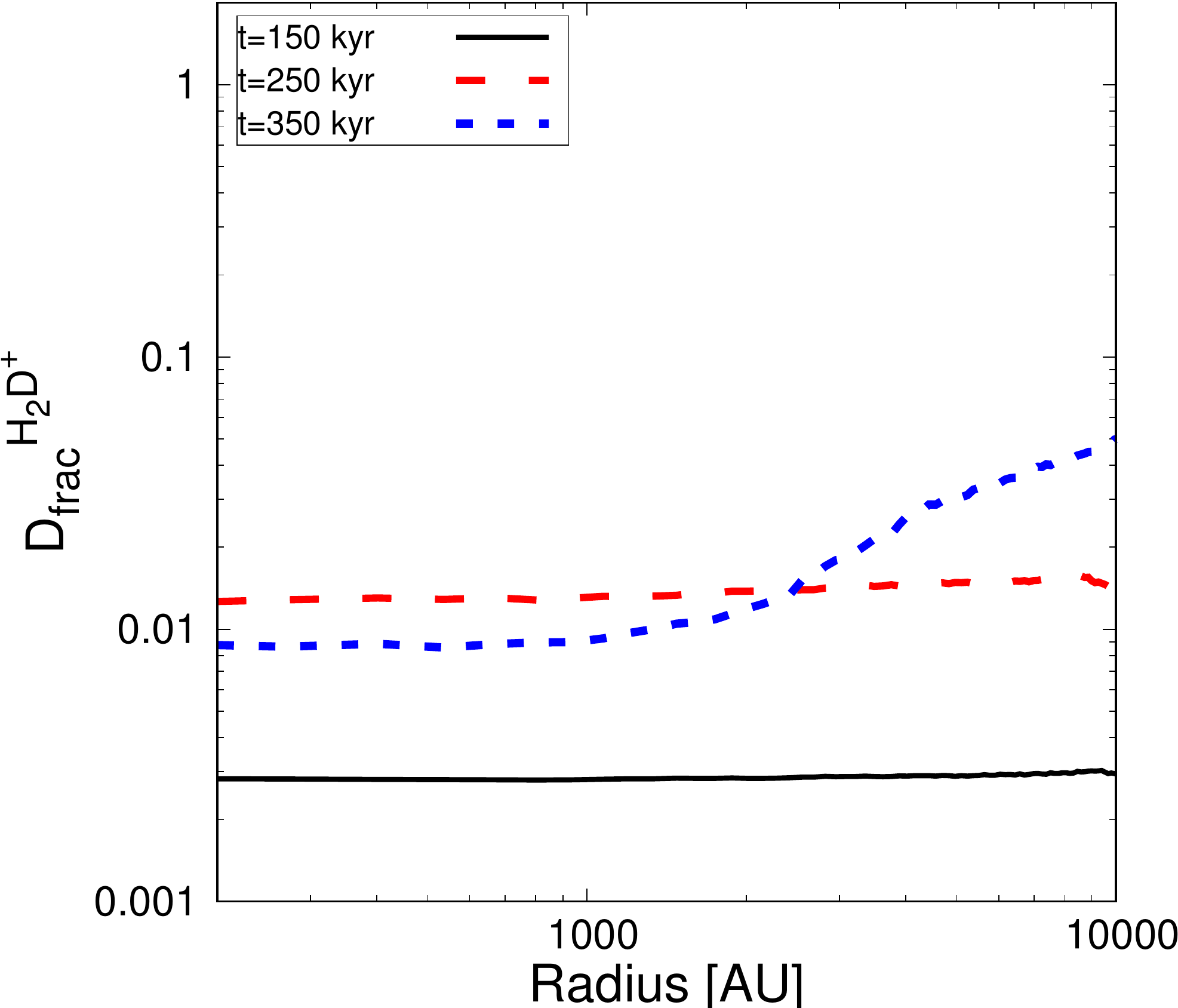}&\includegraphics[width=0.25\textwidth,angle=-90]{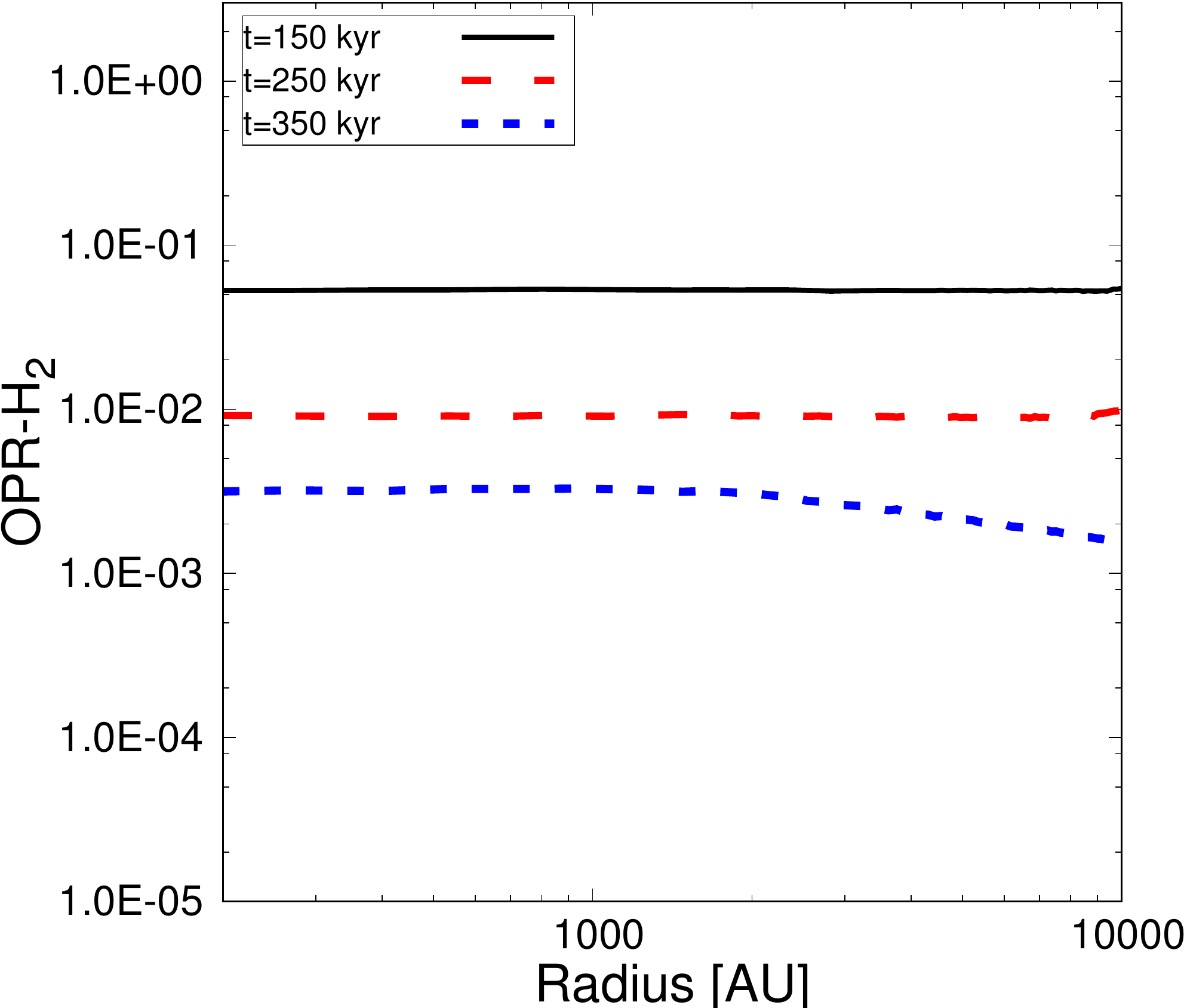}&\includegraphics[width=0.25\textwidth,angle=-90]{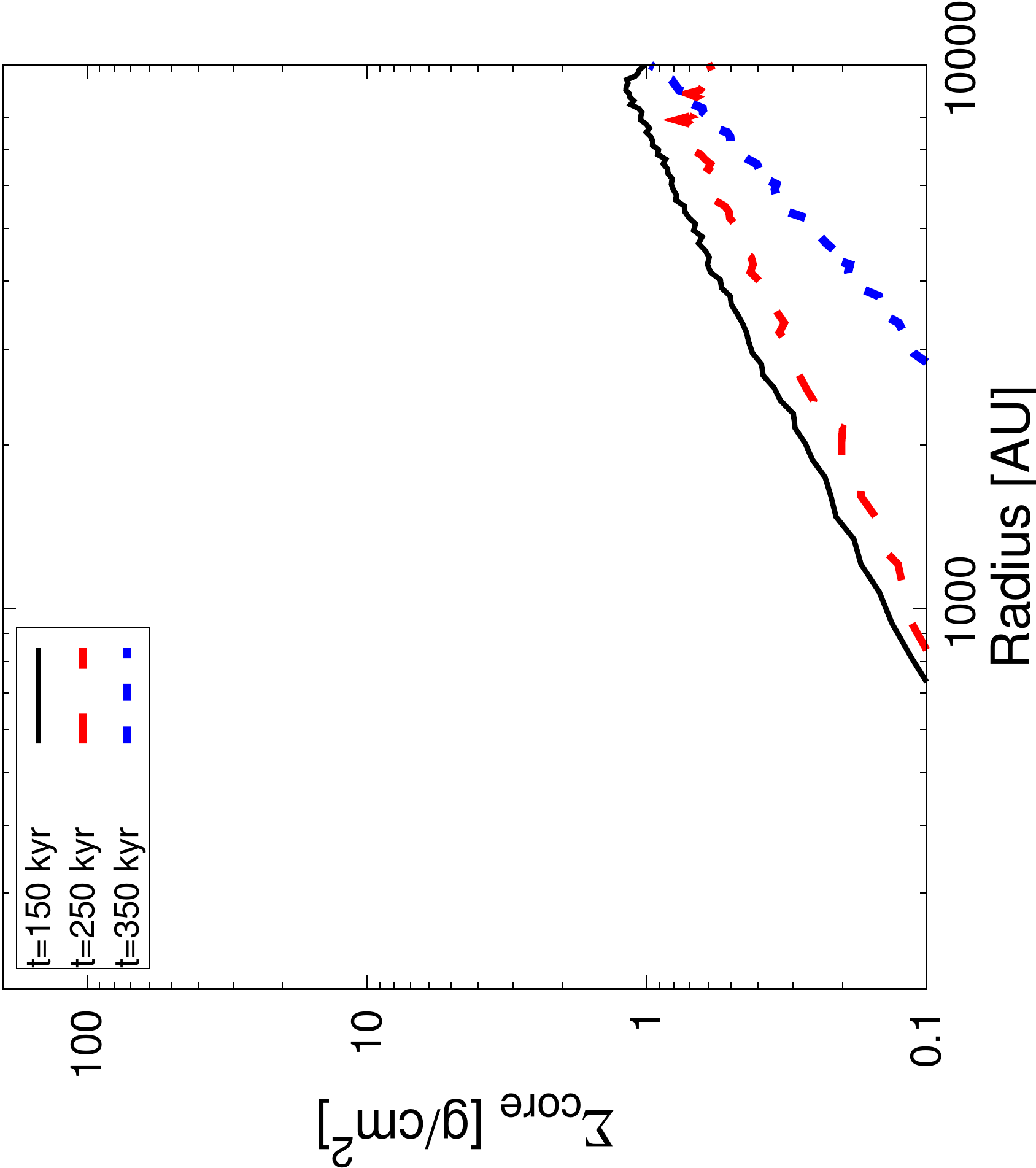}\\
	\rotatebox[origin=r]{90}{ML1.6-M2.0-Perp}&\includegraphics[width=0.25\textwidth,angle=-90]{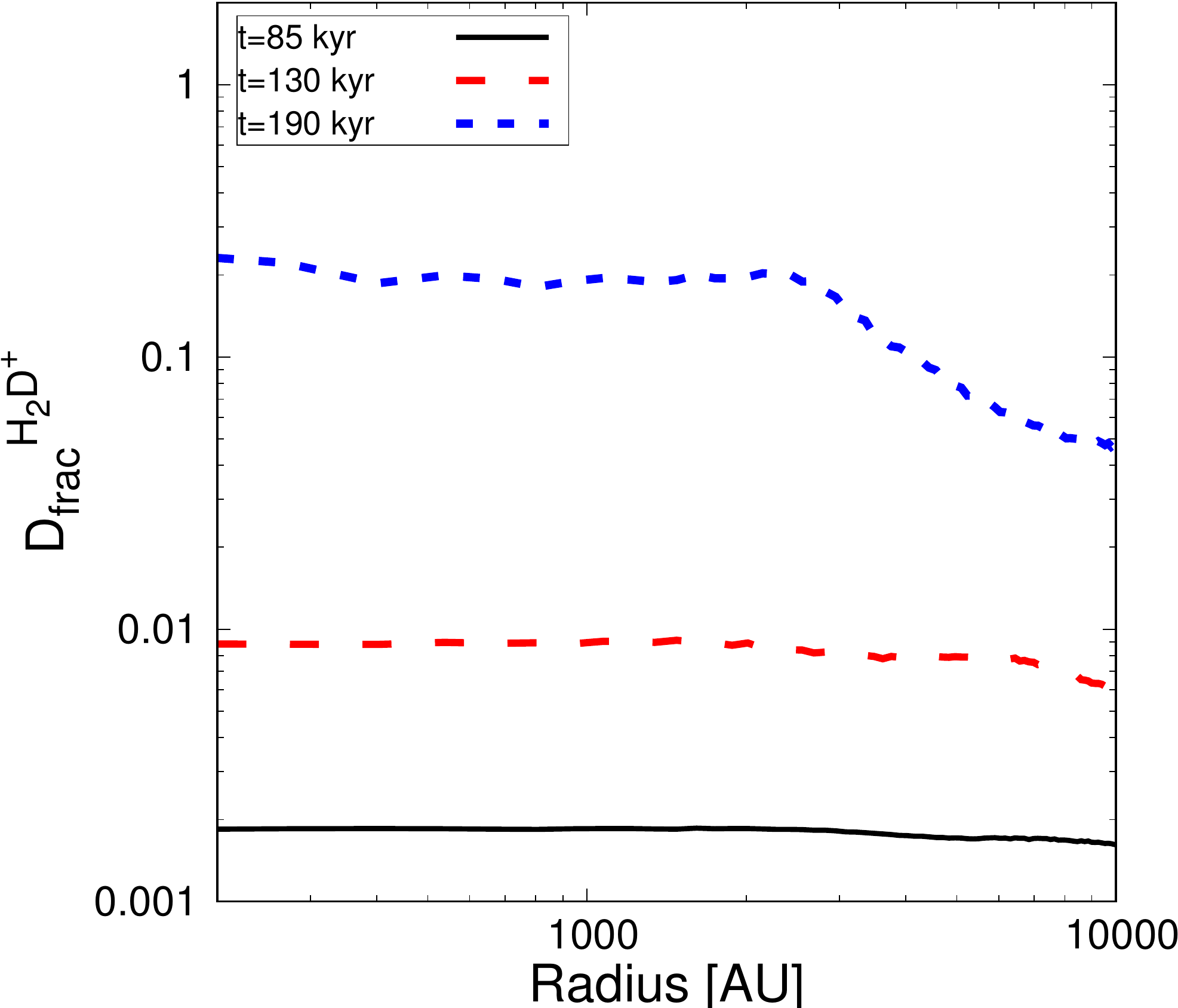}&\includegraphics[width=0.25\textwidth,angle=-90]{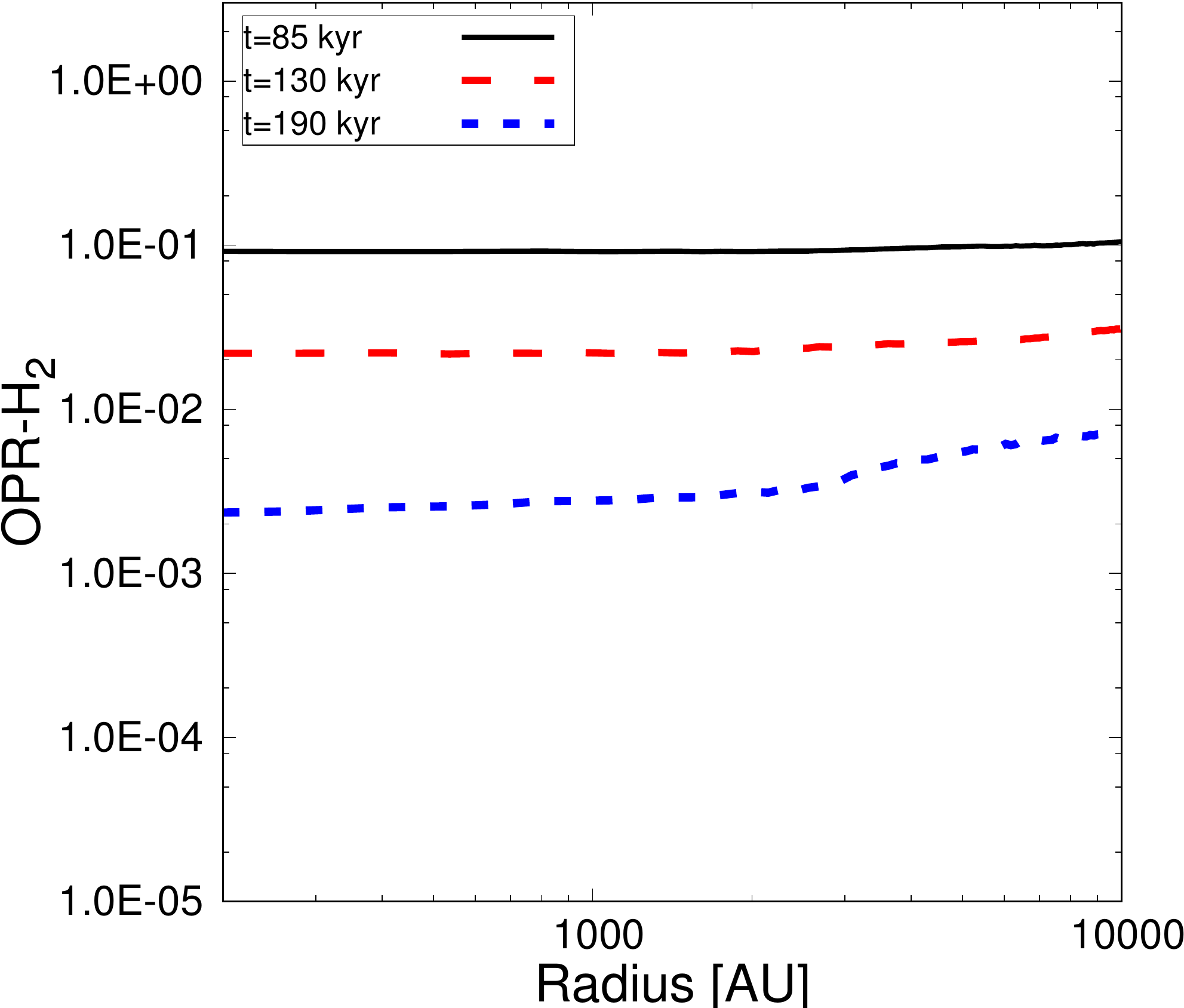}&\includegraphics[width=0.25\textwidth,angle=-90]{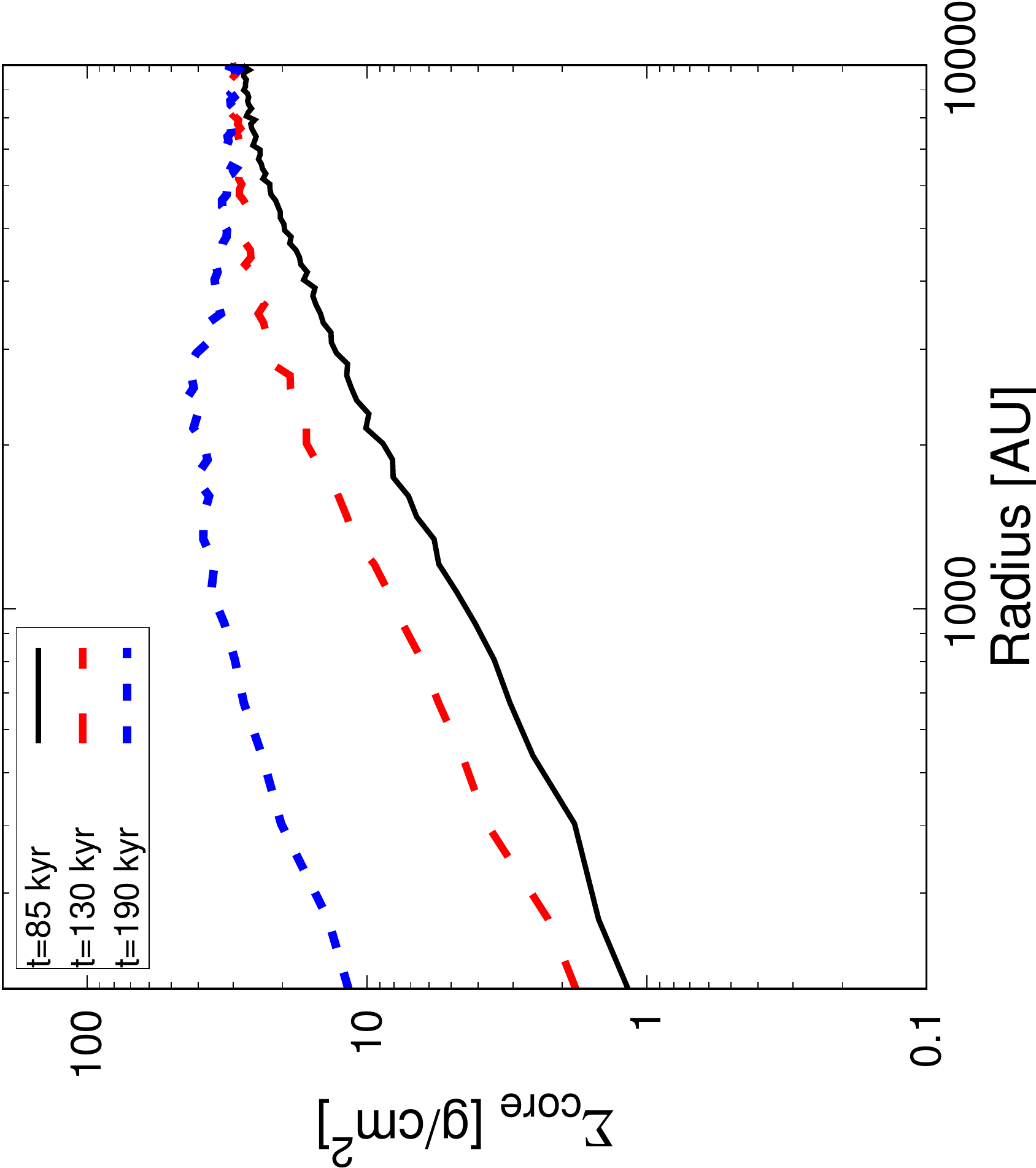}\\
	\rotatebox[origin=r]{90}{ML1.6-M2.0-Para}&\includegraphics[width=0.25\textwidth,angle=-90]{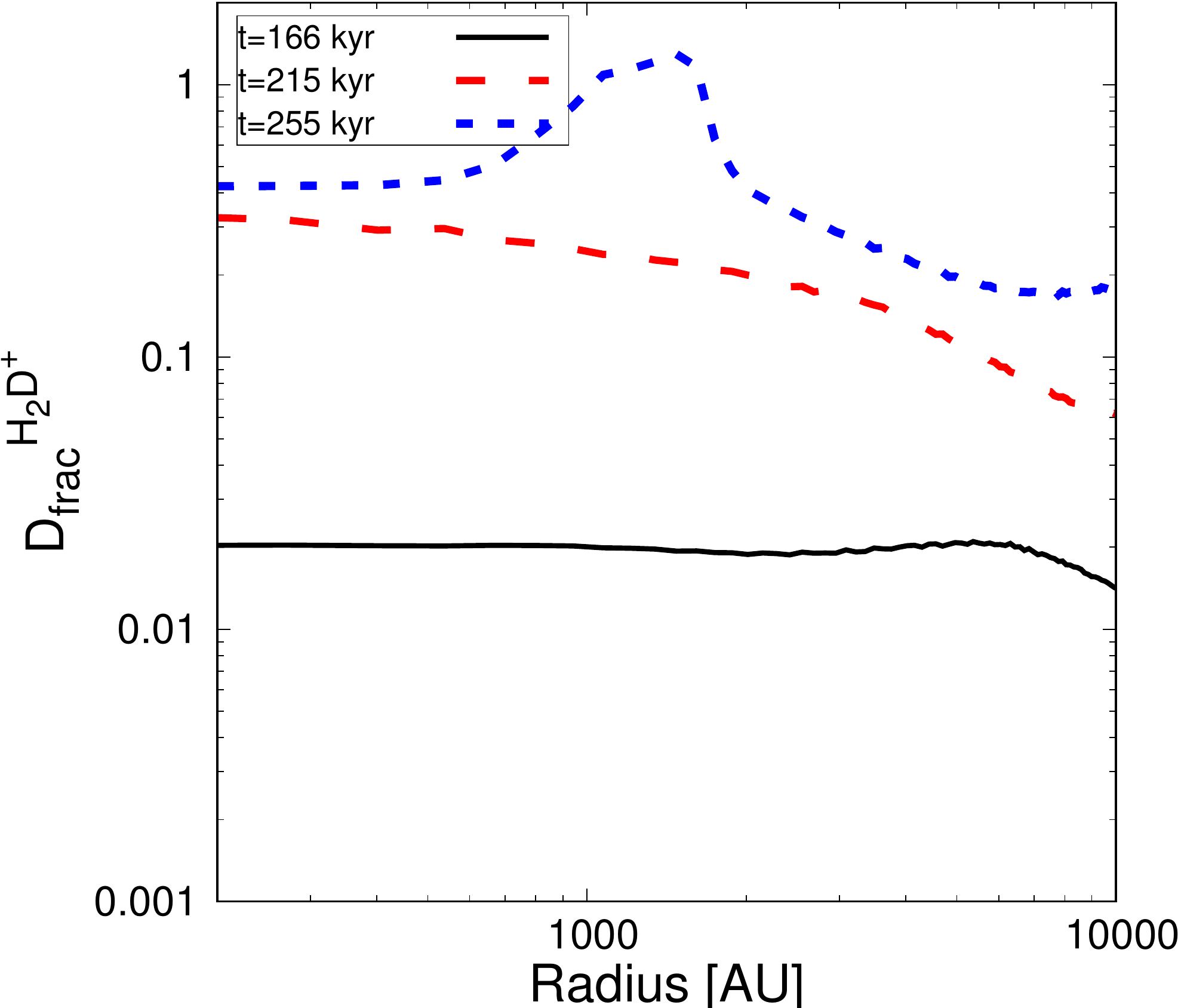}&\includegraphics[width=0.25\textwidth,angle=-90]{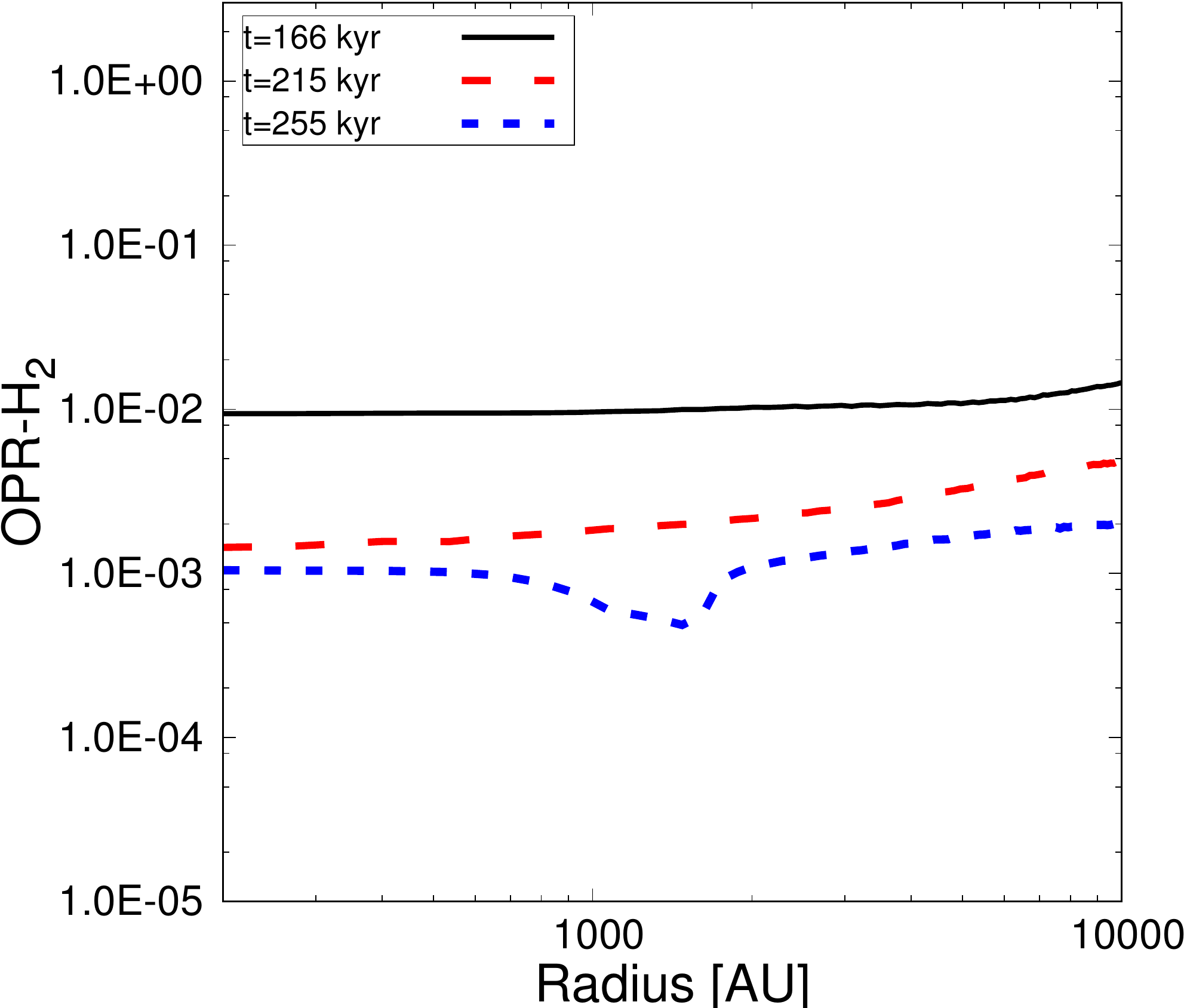}&\includegraphics[width=0.25\textwidth,angle=-90]{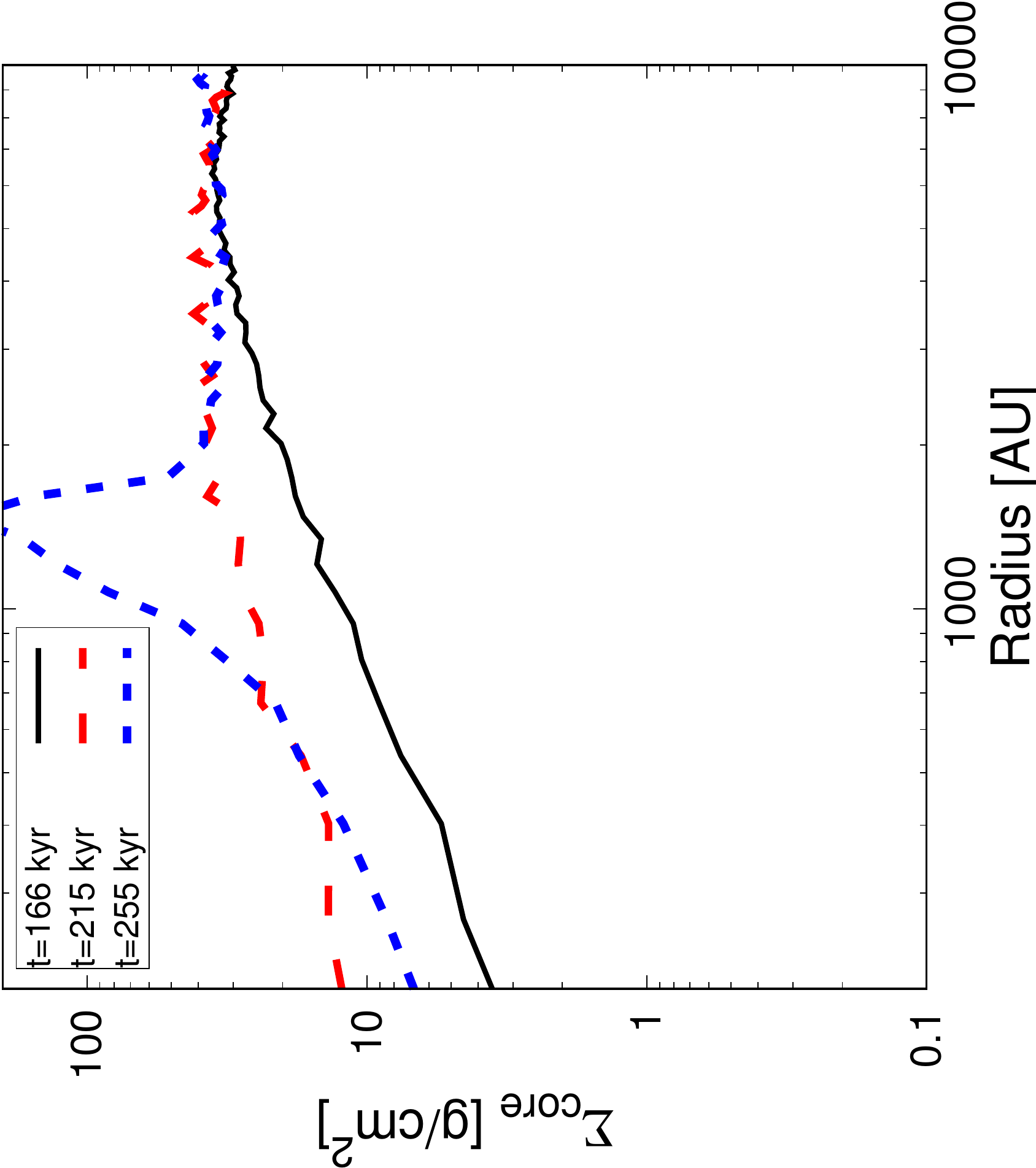}\\\\
	\rotatebox[origin=r]{90}{ML1.6-M4.0-Para}&\includegraphics[width=0.25\textwidth,angle=-90]{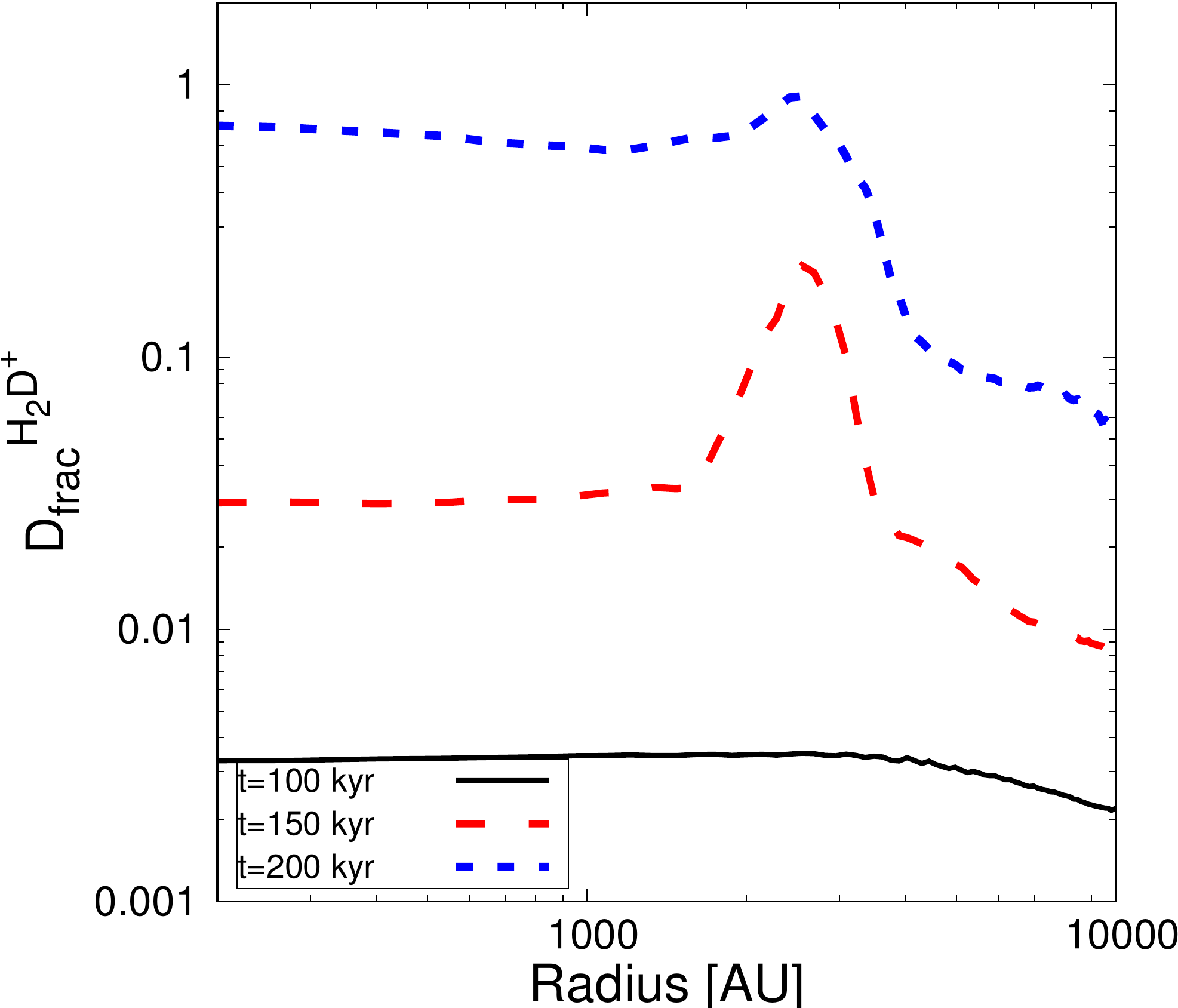}&\includegraphics[width=0.25\textwidth,angle=-90]{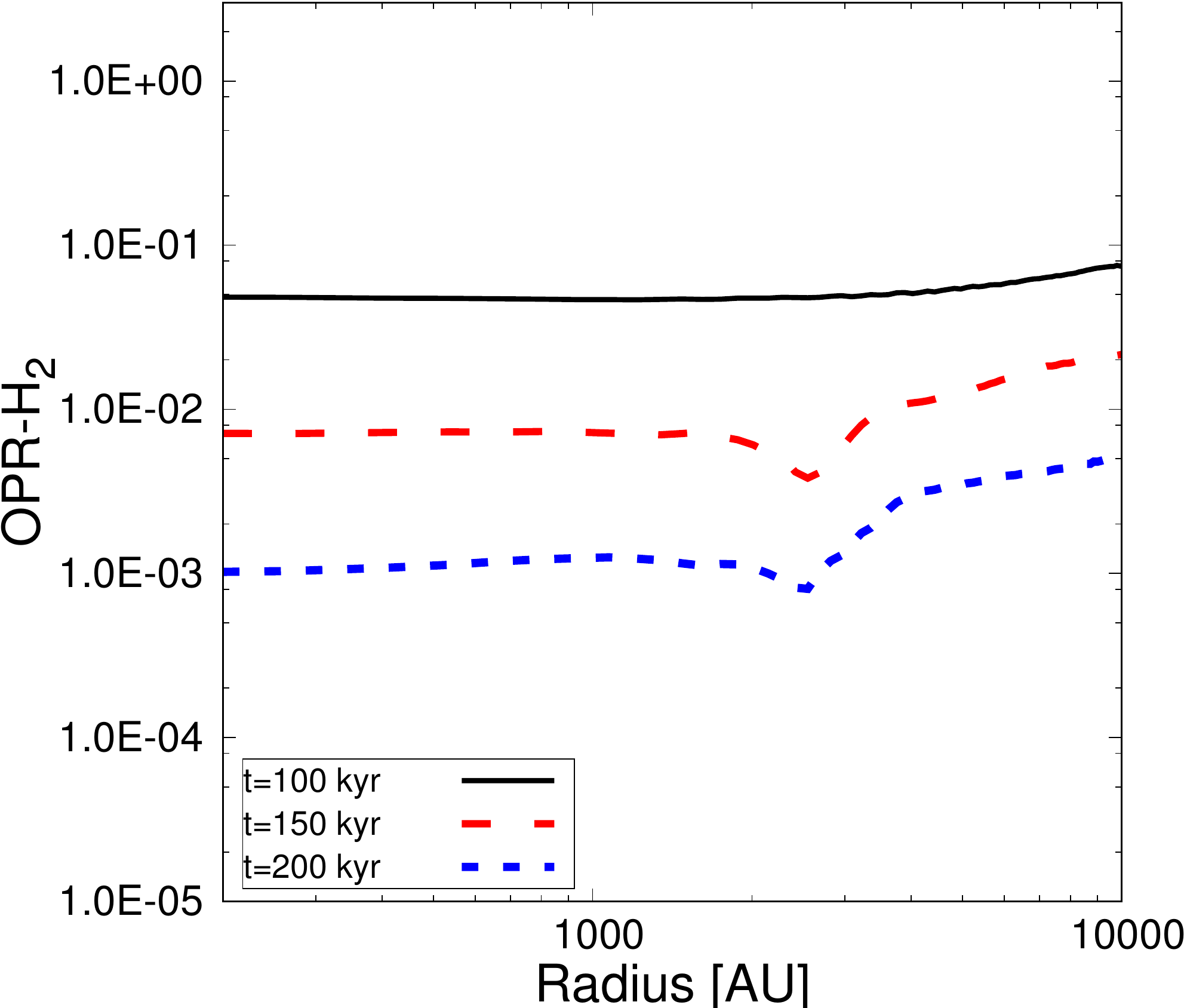}&\includegraphics[width=0.25\textwidth,angle=-90]{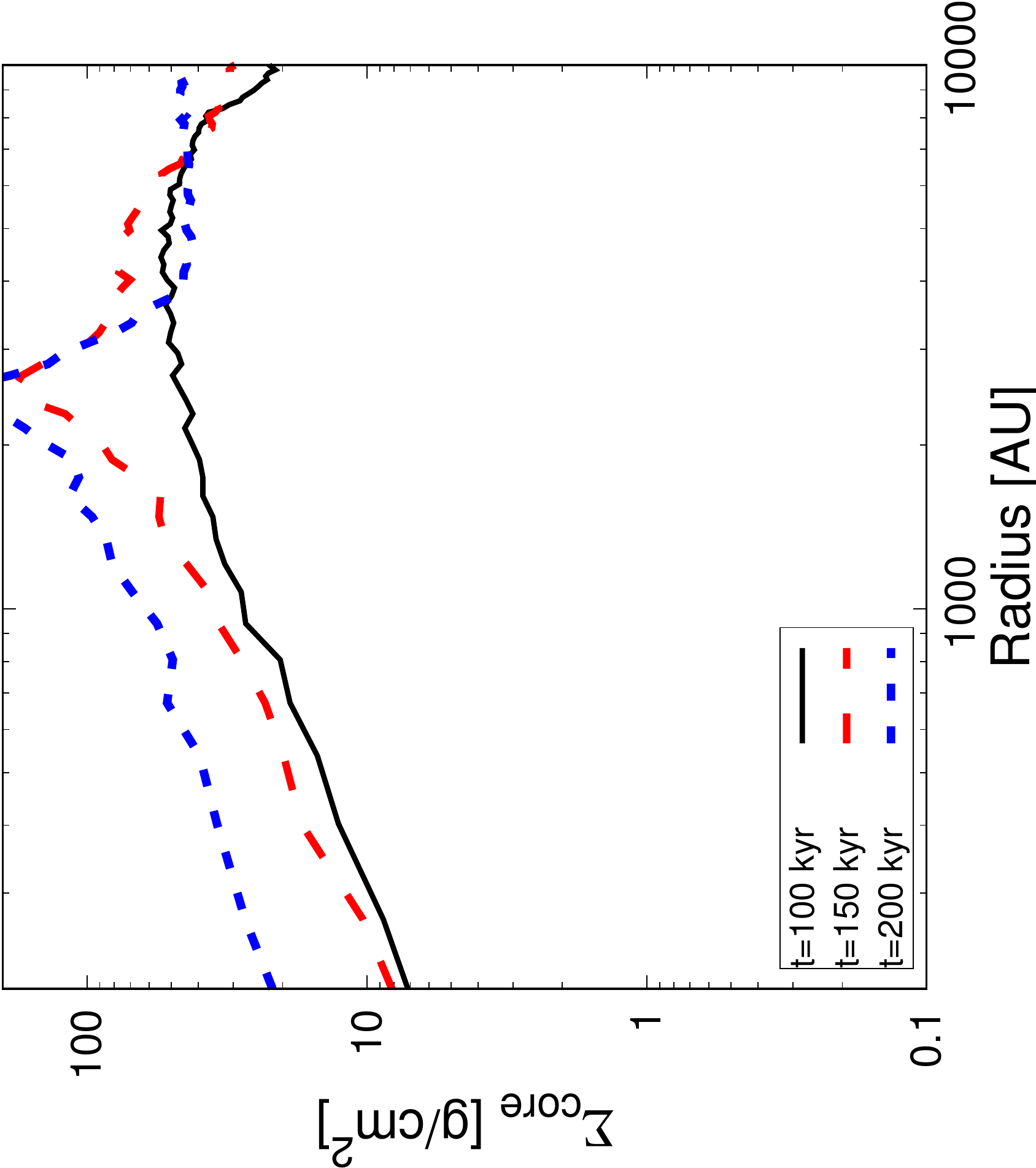}\\\\
	\end{tabular}
	\caption{Radial profiles of deuteration fraction, H$_2$ ortho-to-para ratio and gas surface density for the some of the selected cores at different times. It is observed that the deuteration fraction mainly follows the surface density, as is expected. The 
	OPR-H$_2$ shows almost no variation along the profile, except some scatter collocated with peaks in the surface density. Note, the core center is defined as the center of mass and not as the point of maximum surface density.}
	\label{figCoreProfiles}
\end{figure*}

We finally analyze the radial profiles of the selected cores at different times, focusing on deuteration fraction, surface density as well as H$_2$ ortho-to-para ratio. The corresponding plots are given in Fig.~\ref{figCoreProfiles} for the different cores. We use the 
center of mass to define the core center, which leads to surface density peaks not associated with the center of the profile.\\
 In our special case ML0.8-M2.0-Perp, which is stable against collapse, the surface density is hardly evolving and remains at low values with $0.1-1$~g~cm$^{-2}$.  This is different in all other runs, where the surface density is significantly increasing and reaching values of $10-100$~g~cm$^{-2}$ within $200$~kyrs. The runs with magnetic fields parallel to the filament show a particularly prominent peak of the surface density at $1000-2000$~AU, and a somewhat milder peak in ML1.6-M2.0-Perp with perpendicular field structure. Overall, many of the surface density structures appear consistent with the results found for collapsing cores by \citet{Koertgen17}, suggesting that the characteristic peak is moving outward with increasing turbulent Mach number.
Again similar to the \citet{Koertgen17} results, the H$_2$ ortho-to-para ratio has only a weak dependence on the underlying density structure, and appears as approximately flat within the radial profile, with small dips in the places of the surface density peak, as the ratio decreases more efficiently at high densities. Apart from such specific features, the results for the ortho-to-para H$_2$ appear quite comparable between the different cores. \\
The deuteration fraction is again reflecting the density structure more strongly, with pronounced peaks in the runs with parallel fields, and somewhat weaker / reduced peaks in runs with perpendicular field structures. The latter may be indicative of a different mode of fragmentation and core formation, as also suggested by \citet{Seifried15}, and therefore explain the differences in the structure. Even given these differences, quite high deuteration fractions up to values $\sim1$ are reached in almost all cases through a rather monotonic evolution. The only partial exception is ML0.8-M2.0-Perp, the non-collapsing core, where the deuteration fraction even slightly decreases between $250$ and $350$~kyrs, and never exceeds values of $0.05$. We attribute this to the overall particular evolution in this case, where also the surface density decreases over time, suggesting a strong interaction and exchange between the core and its environment. However, it is important to say that values of about 10$^{-3}$--10$^{-2}$ already represent enhanced deuteration and have been reported in different observational works \citep[e.g.][]{Lackington2016,Barnes16}.

\section{Discussion and conclusions}\label{discussion}
We have pursued here the first 3D magneto-hydrodynamical simulations of filaments including a detailed chemical model to explore deuterium fractionation. For this purpose, we have adopted the network by \citet{Walmsley2004}, assuming full depletion, valid in the inner $0.2$~pc of our filaments, which is the main region we are interested in. We studied both the chemical evolution of the filament as a whole, as well as the formation and chemical evolution of individual cores. When investigating deuteration fraction as a function of core mass and density, we find an initially flat relation right after the formation of the cores, which increasingly steepens at more advanced evolutionary stages. The slope of this relation is thus potentially indicative of the evolutionary stage of the filament, even though we note that the normalization of the relation may vary from filament to filament depending on the initial conditions.\\
We subsequently considered the time evolution of a number of selected cores (based on the requirement that we have sufficient data to follow at least $200$~kyrs of evolution), finding a remarkably similar chemical evolution in all cores, including one that is strongly stabilized by turbulence and magnetic field and therefore not going through gravitational collapse. Assuming a more or less chemically homogeneous initial condition, the chemical evolution of the cores, particularly regarding deuteration effects, is thus very similar. Of course, environmental effects such as the local cosmic ray ionization rate could induce potential differences, by changing the ionization degree. Similarly, metallicity effects at lower densities may depend on the local conditions. At least within one filament, it is however plausible that the deuteration fraction is indeed indicative of chemical age. We find here that about two free-fall times \citep[as defined for cylindrical systems, see][]{Toala12} are sufficient to reach core deuteration fractions of $\gtrsim 0.1$.
We finally investigated also the radial structure of the core, finding overall similar properties as in the isolated collapsing cores studied by \citet{Koertgen17}. The H$_2$ ortho-to-para ratio appears to be approximately flat and only weakly dependent on radius.  Both the deuteration fraction and the gas surface density show a peak on scales of about $1000-2000$~AU, which is particularly pronounced in the case of parallel magnetic fields. As found previously, this peak moves outward with increasing turbulent Mach number, indicating the amount of support against gravity. The difference in the visibility of the peak may result from the difference in the fragmentation mode in both cases, as previously described by \citet{Seifried15}, thus potentially affecting the structure of the resulting cores. Overall, our results have shown that the observed high deuteration fraction in prestellar cores can be readily reproduced in simulations of turbulent magnetized filaments. We further found that deuteration fractions of order $0.1$ can be produced independent of the specific history of the cores, both for high and low virial parameters. The latter suggests that deuteration is potentially very efficient. \\
While the main purpose of this study was to explore this feasibility in general terms, it is crucial that future investigations assess these results in further detail, by e.g. using more improved chemical networks \citep[][]{Vastel12,Majumdar17}. In addition, considering grain-surface effects may further enhance deuteration as for instance the ortho-to-para conversion process \citep{Bovino17}. But also include a dependence on additional free parameters like CO freeze-out and cosmic ray flux would improve our knowledge on the entire process. We believe that such calculations will become feasible in the near future, and thus help to address the important question regarding the timescale to achieve high deuteration fractions.

\section*{Acknowledgement}
BK, SB, DRGS, and RB thank for funding through the DFG priority program `The Physics of the Interstellar Medium' (projects BO 4113/1-2, SCHL 1964/1-2, and BA 3706/3-2). Furthermore RB acknowledges additional funding from the DFG for this project via the grants BA 3706/4-1 and BA 3706/15-1. DRGS thanks for funding through Fondecyt regular (project code 1161247), the ''Concurso Proyectos Internacionales de Investigaci\'on, Convocatoria 2015'' (project code PII20150171), ALMA Conicyt (project 3116000001) and the BASAL Centro de Astrof\'isica y Tecnolog\'ias Afines (CATA) PFB-06/2007. The software used in this work was developed in part by the DOE NNSA
 ASC- and DOE Office of Science ASCR-supported Flash Center for Computational Science at the University of Chicago. The simulations were performed on the local GPU cluster \ita{HUMMEL} at the University of Hamburg as well as on HLRN--III (www.hlrn.de) under project--ID hhp00022. BK further gratefully acknowledges the Gauss Centre for Supercomputing e.V. (www.gauss-centre.eu) for funding this project (project--ID pr92pu) by providing computing time on the GCS Supercomputer SuperMUC at Leibniz Supercomputing Centre (LRZ, www.lrz.de).

\bibliography{mybib_D} 

\bibliographystyle{mn2e}
\end{document}